\documentclass[11pt,a4paper]{JHEP3}
\usepackage{graphicx}
\usepackage{amssymb}
\newcommand{\al}{\alpha}
\newcommand{\bt}{\beta}
\newcommand{\gm}{\gamma}
\newcommand{\dl}{\delta}
\newcommand{\ep}{\epsilon}

\newcommand{\et}{\eta}
\newcommand{\kp}{\kappa}
\newcommand{\lm}{\lambda}
\newcommand{\rh}{\rho}
\newcommand{\sg}{\sigma}
\newcommand{\ta}{\tau}

\newcommand{\ph}{\phi}

\newcommand{\ps}{\psi}
\newcommand{\om}{\omega}

\newcommand{\Gm}{\Gamma}

\newcommand{\Sg}{\Sigma}

\newcommand{\half}{\frac{1}{2}}
\newcommand{\quart}{\frac{1}{4}}

\newcommand{\unity}{1\!\! 1}

\newcommand{\eela}[1]{\label{#1}\end{equation}}
\newcommand{\eeala}[1]{\label{#1}\end{eqnarray}}
\newcommand{\be}{\begin{equation}}
\newcommand{\ee}{\end{equation}}
\newcommand{\bea}{\begin{eqnarray}}
\newcommand{\eea}{\end{eqnarray}}

\newcommand{\at}{\tilde a}

\newcommand{\tc}{t_{\rm c}}
\newcommand{\mc}{m_{\rm c}}

\newcommand{\vrel}{v_{\rm rel}}
\newcommand{\Krel}{K_{\rm rel}}
\newcommand{\CEI}{\mbox{\scriptsize CE-I}}
\newcommand{\CEII}{\mbox{\scriptsize CE-II}}
\newcommand{\GCEI}{\mbox{\scriptsize GCE-I}}
\newcommand{\lP}{\ell_{\rm P}}
\newcommand{\mlp}{m\ellP}
\newcommand{\mPl}{m_{\rm P}}

\newcommand{\lmmin}{\lm_{\rm min}}
\newcommand{\kpmin}{\kp_{\rm min}}
\newcommand{\rmin}{r_{\rm min}}
\newcommand{\re}{{\rm Re}}
\newcommand{\im}{{\rm Im}}
\newcommand{\Emin}{E_{\rm min}}

\newcommand{\mE}{\mathcal{E}}
\newcommand{\ab}{\bar{a}}
\newcommand{\rb}{\bar{r}}
\newcommand{\Lb}{\bar{L}}
\newcommand{\sbar}{\bar s}
\newcommand{\amin}{a_{\rm min}}
\newcommand{\aB}{a_{\rm B}}
\newcommand{\Eb}{E_{\rm b}}
\newcommand{\Ebzero}{E_1}
\newcommand{\Ebone}{\Delta E_1}

\newcommand{\rL}{r_{\rm s}}
\newcommand{\tG}{\tilde G}

\newcommand{\rmax}{r_{\rm max}}

\newcommand{\rs}{r_{\rm s}}

\newcommand{\mO}{\mathcal{O}}

\newcommand{\mP}{m_{\rm P}}
\newcommand{\ellP}{\ell_{\rm P}}

\preprint{}

\title{
Three models of non-perturbative quantum-gravitational binding}
\author{Jan Smit\\
Institute for Theoretical Physics, University of Amsterdam, \\
Science Park 904, P.O.\ Box 94485, 1090 GL, Amsterdam, the Netherlands.\\
}

\abstract{
Known quantum and classical perturbative long-distance corrections to the Newton potential are extended into the short-distance regime using evolution equations for a `running' gravitational coupling, which is used to construct examples non-perturbative potentials for the gravitational binding of two particles.
Model-I is based on the complete set of the relevant Feynman diagrams. Its potential has a singularity at a distance below which it becomes complex and the system gets black hole-like features.
Model-II is based on a reduced set of diagrams and its coupling approaches a non-Gaussian fixed point as the distance is reduced.  Energies and eigenfunctions are obtained and used in a study of time-dependent collapse (model-I) and bouncing (both models) of a spherical wave packet.
The motivation for such non-perturbative `toy' models stems from a desire to elucidate the mass dependence of binding energies found 25 years ago in an explorative numerical simulation within the dynamical triangulation approach to quantum gravity.
Models I \& II suggest indeed an explanation of this mass dependence, in which the Schwarzschild scale plays a role. An estimate of the renormalized Newton coupling is  made by matching with the small-mass region. Comparison of the dynamical triangulation results for mass renormalization with `renormalized perturbation theory' in the continuum leads to an independent estimate of this coupling, which is used in an improved analysis of the binding energy data.
}

\keywords{quantum gravity, lattice theory}

\begin{document}

\section{Introduction}
\label{intro}

An explorative numerical computation of two-particle binding was performed within the original time-space symmetrical dynamical triangulation (SDT)\footnote{Customarily known as Euclidean dynamical triangulation (EDT). The acronym SDT was introduced earlier by the author to  emphasise the difference with causal dynamical triangulation (CDT). It is not ideal but we shall keep it here to distinguish it from other versions of EDT mentioned later.} approach to quantum gravity  \cite{deBakker:1996qf}.
The binding energies found were puzzling in their dependence on the masses of the particles.
In the present paper we study models in the continuum with the aim of improving our acquaintance with possible mass dependencies of binding energies and then return to the SDT results.

These continuum models are derived from one-loop perturbative corrections to the Newton potential, which include quantum gravitational contributions \cite{Donoghue:1993eb,Donoghue:1994dn}
as well classical ones in which $\hbar$ cancels \cite{Iwasaki:1971vb,Holstein:2004dn}. At large distances the corrections are independent of the UV regulator.
The calculations were interpreted within effective field theory  \cite{Donoghue:1993eb,Donoghue:1994dn} and were subsequently also carried out by other authors, as discussed in \cite{BjerrumBohr:2002kt,BjerrumBohr:2002ks} which' final results we are using in this work.
In \cite{BjerrumBohr:2002ks} it was observed that the quantum contributions from the subset one-particle-reducible (1PR) diagrams (`dressed one-particle exchange') suggested a `running' gravitational coupling depending on the distance scale, a simple example of a renormalization-group type evolution with a non-Gaussian fixed point \cite{Codello:2008vh}. Later such running was found to be not universally applicable \cite{Anber:2011ut,Donoghue:2019clr}. However, similar running couplings including also the classical contributions are employed here, solely for the construction of non-perturbative (`toy') models of quantum gravitational binding.

The models are specified by a running potential
\be
V_{\rm r} = -\tG m^2 r\,,
\ee
in which a dimensionless running coupling $\tG$ satisfies an evolution equation with an asymptotic condition
\be
- r\frac{\partial \tG}{\partial r} = \beta(\tG,\sqrt{G}\,m)\,,
 \qquad
 \tG\to \frac{G}{r^2}\,, \quad r\to\infty\,,
 \ee
were $G$ is the Newton constant.\footnote{Units in which $\hbar=c=1$; we shall also use a Planck length $\ellP=G^{1/2}$, Planck mass $\mP=G^{-1/2}$ and when convenient units $G=1$. For convenience, we shall call models using the potential $-Gm^2/r$: `Newton models'.} The `beta function' $\bt$ depends on the (equal) mass $m$ of the particles through the classical perturbative corrections.
For large masses the classical terms in $\bt$ tend to dominate and dropping the quantum part leads to simpler `classical evolution models'.

Model-I starts from the long-distance potential including all one-loop corrections \cite{BjerrumBohr:2002kt}.
It leads to an evolution with singularities at a distance $\rs$. We interpret the singularities as {\em distributions}, which enables continuing the running past $\rs$ to zero distance where the potential vanishes. When $r$ passes $\rs$ the potential gets an imaginary part.
For large particle masses $\rs\approx 3 G m$; it is of order of the Schwarzschild radius of the two-particle system and the model has black hole-like features, such as absorbing probability out of the two-particle wave function.
Model-II uses only the 1PR contributions.
Its  evolution of $\tG$ has a non-Gaussian fixed point, the potential is regular and real for all $r\ge 0$ and it has a minimum at a distance $\rmin$. For large masses $\rmin\approx G m$, hence also of order of the Schwarzschild radius.\footnote{The single point particles have no horizon; the long distance corrections and the beta functions derived from them do not contain a `back reaction' of the particles on the geometry.}

For the most part in this work the models are equipped with a non-relativistic kinetic energy operator $K$.
However, a relativistic kinetic energy operator $\Krel$ gives interesting qualitatively different results in model I. For example, with the simpler classical-evolution potential, classical particles falling in from a distance $r>\rs$ obtain the velocity of light when reaching the singularity at $\rs$, as happens for particles approaching a Schwarzschild black hole horizon  \cite{LandauLifshitz:1971}. In model-II such particles' maximal velocity stays below that of light. For brevity the models with $\Krel$ will be dubbed `relativistic models'.\footnote{Also in the relativistic Newton model the particle velocity reaches that of light, but at zero distance. Models with an energy-independent potential and relativistic kinetic energy can sometimes describe interesting physics. For example,
%with a linear potential
such a model can describe the linear relation between spin and squared-mass of hadrons
%as described by Regge trajectories in a Chew-Frautschi plot \cite{Chew:1962eu,Smit:2002ug}.
\cite{Smit:2002ug}. But the relativistic models I and II are qualitative and not intended to describe merging black holes and neutron stars as done in sophisticated Effective One Body models \cite{Buonanno:1998gg,Buonanno:2000ef,Schafer:2018kuf}. The field theoretic introduction of particles in the SDT computation is also relativistic.}

Computations of binding energies lead naturally to more general knowledge of the spectrum of eigenvalues and eigenfunctions of the Hamiltonian. It is fun and instructive to exploit this in studying also the development in time of a spherical wave packet released at a large distance. It shows oscillatory bouncing and falling back in both models, and in model-I during decay.

In the SDT study \cite{deBakker:1996qf}, the binding energy $\Eb$ was found to increase only moderately with $m$ (it had even decreased at the largest mass), a behavior  differing very much from the rapid increase of $\Eb=G^2 m^5/4$ in the Newton model. Finite-size effects, although presumably present, were expected to diminish with increasing mass (the effective extent of the wave function was assumed $\propto 1/m$). An important clue for a renewed interpretation here of the results is suggested by the fact that in models I and II,
%
%For interpreting the binding energy results in SDT, an important clue comes from the fact that
at relatively large masses, the bound state wave function
%in models I and II
is maximal near $\rs$ respectively $\rmin$. Since these scales grow with $m$ this suggests that finite-size effects become {\em larger} with increasing mass.
We estimate the renormalized $G$ by matching to Newtonian behavior in the small mass region. This is helped by renormalized perturbation theory, which provides independent estimates of $G$ from the SDT results for mass renormalization.

Mentioning some aspects of DT may be useful here, although this not the place to give even a brief proper review. Depending on the bare Newton coupling, the pure\footnote{In lattice QCD, using the pure gauge theory for computing hadron masses is called the `quenched' approximation (or `valence' approximation since it lacks dynamical fermion loops). The long-distance corrected Newton potential contains no massive scalar loops and our bound state calculations based on it are in this sense quenched approximations.} gravity model has two phases.
Deep in the weak-coupling phase the computer-generated simplicial configurations contain baby universes assembled in tree-like structures with `branched polymer' characteristics---hence the name `elongated phase'---very different from a four-sphere representing de Sitter space in imaginary time. Deep in the strong-coupling phase the configurations contain `singular structures', such as a vertex embedded in a  macroscopic volume within one lattice spacing---hence the name `crumpled phase'.
Only close to the transition between the phases the average spacetimes, as used in \cite{deBakker:1996qf}, have approximately properties of a four-sphere.
The transition was found to be of first-order \cite{Bialas:1996wu,deBakker:1996zx,Rindlisbacher:2015ewa} whereas many researchers were looking for a second- or higher-order critical point at which a continuum limit might be taken.
%(see e.g.\ the review \cite{Thorleifsson:1998jr}).
Primarily for these reasons
%(but also the quest for a suitable transfer matrix)
`causal dynamical triangulation' (CDT) was introduced, %\cite{Ambjorn:1998xu,Ambjorn:2001cv,Ambjorn:2012jv,Loll:2019rdj},
which has a phase showing a de Sitter-type spacetime, with fluctuations enabling a determination of a renormalized Newton coupling, and furthermore a distant-dependent spectral dimension showing dimensional reduction at short distances \cite{Ambjorn:2012jv,Loll:2019rdj}.
%\cite{Ambjorn:1998xu}--\cite{Loll:2019rdj}.
%\footnote{An additional reason was the desire for a suitable transfer matrix in four dimensions, which CDT provides. }
Another continuation of dynamical triangulation research uses a `measure term', which, when written as an addition to the action involves a logarithmic dependence on the curvature
%parametrized by an addition coupling parameter
%\cite{Laiho:2011ya}--\cite{Catterall:2018dns}.
%\cite{Laiho:2011ya,Ambjorn:2013eha,Coumbe:2014nea,Laiho:2016nlp,Laiho:2017htj,Catterall:2018dns}.
\cite{Bruegmann:1992jk,Ambjorn:1999ix,Ambjorn:2013eha,Laiho:2016nlp}.
Evidence was obtained for a non-trivial fixed point scenario in which the above 1st-order critical point is closely passed  on the crumpled side by a trajectory in a {\em plane} of coupling constants towards the continuum limit (cf.\ \cite{Laiho:2016nlp} and references therein).\footnote{The strong coupling side of the phase transition was also judged as physics-favoured in another lattice formulation approach to quantum gravity, see e.g.\ the review \cite{Hamber:2017pli}.}
Scaling of the spectral dimension was instrumental determining relative lattice spacings \cite{Coumbe:2014noa} and evidence for the possibility of a continuum limit %in this version of EDT
was also found in the spectrum of K\"{a}hler-Dirac fermions  \cite{Catterall:2018dns}.

Returning to the original SDT, reference \cite{Smit:2013wua} gives a continuum interpretation of average SDT spacetimes in terms of an approximation by an agglomerate of 4-spheres making up a branched polymer
in the elongated phase, and a four-dimensional negatively-curved hyperbolic space in the crumpled phase.\footnote{By modeling the {\rm continuum} path integral using such approximate saddle points one also finds a {\em first-order} phase transition \cite{JStbp}.}
A scaling analysis in the crumpled phase, away from the transition, led to an average curvature radius reaching a finite limit of order of the lattice spacing as the total four-volume increased to infinity. Similar behavior is expected to hold for the average radius of the four-spheres in the elongated phase; they have small volumes (still containing thousands of 4-simplices) and their {\em number} increases with the total volume.
Hence also this continuum interpretation implies that the UV cutoff given by the lattice spacing cannot be removed in SDT. However, models with a UV cutoff may still be able to describe truly non-perturbative aspects of quantum Einstein gravity at scales below the cutoff.

Section \ref{secpot} introduces the one-loop corrected long-distance Newton potentials of models I and II. When naively extended to short distances these potentials become more singular than $1/r$ and with a short-distance cutoff we calculate  in section \ref{secvreg} perturbative corrections to the binding energy. Section \ref{secevol} starts the derivation of the evolution equation, with a discussion of its properties in the two models. The running potential is then used to calculate s-wave binding energies in sections \ref{secbeI} (model-I) and \ref{secbeII} (model-II), with variational methods and with matrix-diagonalization in a discrete Fourier basis in finite volume.
A pleasant by-product of the latter method is a spectrum of eigenvalues and eigenfunctions, which is used in section \ref{secbouncecollapse} in real-time calculations of the spherical bouncing and collapse of a wave packet let loose far from the Schwarzschild-scale region.
In section \ref{secSDT} we return to the binding computation in SDT with an extended discussion of the renormalized mass and binding energy results. Relating some of the binding energy data to the very small mass region by a simple phenomenological formula yields an estimate of the renormalized Newton coupling $G$. The mass renormalization data are used in independent estimates of $G$, which improve the analysis of the binding energy results.

Results are summarized in section \ref{secsummary} with a conclusion in section \ref{secconclusion}. Solutions of the evolution equations are given in appendix \ref{appGr}. Appendix \ref{appnormal} starts with a formal definition of model-I and describes some consequences of its Hamiltonian being symmetric but not Hermitian. Further details of analytical and numerical treatments are in the remainder of appendix \ref{appI} and in appendix \ref{appII}. Classical motion of the relativistic classical-evolution models is studied in appendix \ref{appbounceCEN}. Appendix \ref{appmm0} sketches the derivation of a relation between the renormalized mass and the bare mass of the particles using renormalized perturbation theory.

\section{Perturbatively corrected Newton potential}
\label{secpot}

The potential is defined by a Fourier transform of the scattering amplitude of two scalar particles, in Minkowski spacetime, calculated to one-loop order and after a non-relativistic reduction \cite{Iwasaki:1971vb}.
Its long-distance form is UV-finite and calculable in effective field theory \cite{Donoghue:1993eb,Donoghue:1994dn}. Graviton loops give non-analytic terms in the exchange momentum $q$ at $q=0$, which determine the long-distance corrections. Terms analytic in $q$ correspond to short-distance behavior. They involve UV-divergencies; after their subtraction, finite parts remain with unknown coefficients, which are set to zero in our models. One-loop effects of the massive particle belong to the analytic type and are omitted this way.
Including the long-distance corrections the potential has the form
\be
V=-\frac{G m_1 m_2}{r}\left[1 + d\,\frac{G(m_1 + m_2)}{r} + c\, \frac{G}{r^2} \right] + \mO(G^3)\, ,
\label{Vbc}
\ee
where $G$ is the Newton coupling.
Actually, the $d$ term is a classical contribution (independent of $\hbar$) coming from classical General Relativity \cite{Iwasaki:1971vb,Holstein:2004dn,BjerrumBohr:2002ks}); the $c$ term is a quantum correction of order $\hbar$. Calculations were performed in harmonic gauge.

Intuitively one may think that the potential corresponds to dressed one-particle exchange. This leads to the so-called the one-particle-{\em reducible} (1PR) potential \cite{Burgess:2003jk}. The 1PR scattering amplitude does not include all one-loop diagrams and it is not gauge invariant. Since we are primarily interested in models that provide examples of bound-state energies, we accept this lack of gauge invariance and study also models based on the 1PR potential. Including all diagrams one arrives at a `complete' potential which may lead to gauge invariant results when calculating gauge-invariant observables. The dimensionless ratio of the bound-state energy to the mass of the constituent particles may be such an observable. The potential is not gauge invariant, as discussed in \cite{BjerrumBohr:2002kt}.

The constants $c$ and $d$ are given by \cite{BjerrumBohr:2002kt}
\bea
d&=& 3 ,\qquad c=\frac{41}{10\pi}\simeq 1.3\, ,\;\;\;\;\qquad\qquad \mbox{model-I (complete)}
\label{complete}
\\
d&=&-1 ,\quad\, c=-\frac{167}{30\pi}\simeq -1.8\,. \qquad\qquad  \mbox{model-II (1PR)}
\label{1PR}
\eea

\section{Calculations with the 1-loop potential}
\label{secvreg}

We continue with equal masses, $m_1=m_2=m$ and turn to the computation of the binding energy of the positronium-like system in which $G m^2$ plays the role of the fine-structure constant.
In terms of $f(r) = r \ps(\vec{r})$, with $\ps(\vec{r})$ the wave function, the time-independent non-relativistic radial s-wave Schr\"{o}dinger equation is to be
\be
H f(r) = E\, f(r),
\quad
H = K + V_{\rm reg}(r), \quad K=-\frac{1}{m}\,\frac{\partial^2}{\partial r^2}\,,
\ee
where the potential $V_{\rm reg}$ is a regularized version of $V$ in order to deal with the singular behavior of the $c$ and $d$ terms at the origin. Note that $m$ is twice the reduced mass. The binding energy is defined as the negative of the minimum energy
\be
\Eb=-E_{\rm min}\,.
\label{Eb1}
\ee
In case the average squared-velocity $v^2 = \langle K/m\rangle \gtrsim 1$ we also study relativistic models with kinetic energy operator
\be
\Krel = 2\sqrt{m^2 -\partial_r^2},
\ee
with
\be
\Eb=-(\Emin-2m)\,,\quad
\vrel^2 = \langle -\partial_r^2/(m^2 -\partial_r^2)\rangle\,.
\ee
In the following we shall tacitly be dealing with the nonrelativistic model unless mentioned otherwise.

The zero-loop potential
\be
V_{0} = - \frac{G m^2}{r} \, ,
\ee
gives the bound-state energy spectrum of the Hydrogen atom with $\al\to G m^2$ and reduced mass $m/2$,
\be
E_n=- \quart\,G^2 m^5\, \frac{1}{n^2}\,, \quad n=1,2, \ldots\, , \quad \Emin=E_1\,.
\ee
The eigenfunctions are given by
\be
u_n(r,a) = \frac{2ar}{(a n)^{5/2}}\, L^1_{n-1}\left(\frac{2r}{a n}\right) \exp\left(\frac{-r}{a n}\right)\, ,
\label{unswave}
\ee
where $L^1_{n-1}$ is the associated Laguerre polynomial ($L^1_0=1$) and $a$ the Bohr radius
\be
\aB=\frac{2}{G m^3}
\, .
\ee
The non-relativistic binding energy $\Eb=-E_1= G^2 m^5/4$.  It becomes very large as $m$ increases beyond the Planck mass $G^{-1/2}$, and then also the average squared-velocity of the particles becomes much larger than the light velocity: $v^2 =\langle K/m\rangle = G^2 m^4/4$.
In the relativistic version of the model the binding energy can be estimated estimate by a variational calculation using $u_1(r,a)$ as trial wave function with variational parameter $a$:
\be
\Eb=2m-\mE(\amin)\,,\quad
\mE(a)= \int_0^\infty u_1(r,a)\, (\Krel+V_0)\, u_1(r,a)
= \langle \Krel\rangle - \frac{G m^2}{a}\,,
\label{Evarrel}
\ee
where $\amin$ is the value of $a$ where $\mE(a)$ is minimal. Starting from small masses, when $m$ increases $a$ decreases from large values near the Bohr radius towards $a=0$.\footnote{The non-relativistic variational result is exact, $\amin=\aB$ and $\mE(\amin)= E_1$. } The relativistic $\langle \Krel\rangle$ scales like $(\mbox{const.})/a$ as $a\to 0$, like the potential part of $\mE(a)$ but with a $(\mbox{const.})$ independent of $m$. Consequently there is a maximum mass $\mc$ at which $a$ has reached zero and beyond which there is no minimum anymore. Since the variational energy provides an upper bound to the exact energy, the relativistic Newton model has no ground state for $m>\mc$.
The calculation is described in appendix \ref{appHtrial}:
\be
\mc= \frac{4}{\sqrt{3\pi\,G}}\,,\quad \lim_{m\uparrow \mc} \mE =0\,, \quad \Eb= 2\mc\,,\qquad \mbox{Newton model}
\ee
and $\vrel^2 \to 1$ as $m\uparrow\mc$ since $\langle -\partial^2\rangle\to \infty$.

Writing $V=V_0 + V_1$ and treating the one-loop contribution $V_1$ (order $G^2$ in (\ref{Vbc})) as a perturbation, with a simple short-distance cutoff,
\bea
V_{1\,\rm reg}(r) &=& V_1(r),\quad r> \ell
\nonumber\\
 &=& V_1(\ell), \quad 0<r<\ell,
\label{vreg}
\eea
the perturbative change in the minimum energy is given by
\bea
\Ebone &=& \int_0^\infty dr\, u_1(r,a)^2\, V_{1\,\rm reg}(r)\,,\qquad a=\aB\,,
\\
\nonumber\\
&=&- G^2 m^2 \left[\frac{4 d m}{a^2} + \frac{4c}{a^3}\left(\frac{1}{3} -\ln\!\left(\frac{2\ell}{a}\right)-\gm\right)\right]
\left(1 + \mO(\ell/a)\right)\,,
%\quad a=\aB\,,
\label{Ebonesmall}
\eea
where $\gm$ is the Euler constant and we assumed $\ell/a\ll 1$.
(The $d$ term in (\ref{Ebonesmall}) is finite as $\ell\to 0$, its presence in $V_1$ did not need a UV cutoff for this calculation.)
Choosing $\ell$ equal to the Planck length, $\sqrt{G}$,
this gives for small masses $(m\sqrt{G})^3\ll 1$,
\be
\frac{\Ebone}{m} =
-d (m\sqrt{G})^8 - c\left(\frac{1}{6}- \frac{3}{2}\, \ln (m\sqrt{G}) -\half\, \gm\right)\,
(m\sqrt{G})^{10}  + \mO((m\sqrt{G})^{12})\,.
\label{dlebone}
\ee
For masses smaller than $\simeq 0.54/\sqrt{G}$ this asymptotic expression is accurate to better than 10\%.
The ratio of the c and d term in (\ref{dlebone}) is maximal for $m\sqrt{G}=0.56$\,.

The perturbative evaluation looses sense when $|\Ebone/\Ebzero|> 1$,
which happens for $m\sqrt{G}\gtrsim 0.54$
and $m\sqrt{G}\gtrsim 0.66$, respectively in model-I and model-II. At these values the ratio of the binding energy to the mass is still small, $|\Ebzero+\Ebone|/m = 0.041$ in model-I, $\approx$ {\em zero} in model-II (for which $\Ebone$ is positive), while the Bohr radii are still much larger than the short-distance cutoff:
$\aB\simeq 13 \sqrt{G}$, respectively $\simeq 9\sqrt{G}$.

There is no physics reason to go to larger masses and treat $V_1$ non-perturbatively, but it is interesting to see what happens.
%(the model with only the $c$-term is scale invariant \cite{Paik:2017zqv}).
A first estimate is obtained in a variational calculation using $u_1(r,a)$ as a trial wave function with $a$ as a variational parameter, as in (\ref{Evarrel}) with $\Krel\to K$, $\Eb\to -\mE$, $V_0\to V_{0\,{\rm reg}}+ V_{1\,{\rm reg}}$ (putting the same cutoff on $V_0$ as on $V_1$). This estimate can be improved somewhat by using the
$u_n(r,\amin)$, $n=1, \ldots, N$, to compute matrix elements $H_{mn}$ for conversion to an $N\times N$ matrix problem (keeping $\amin$ fixed by the variational problem at $N=1$).
For $N\ge 3$ basis functions the minimum eigenvalue of this Hamiltonian matrix appears to converge rapidly to a limiting value,
$\Eb^{(N)}-\Eb^{(\infty)}\propto N^{-2}$ and the difference $\Eb^{(1)}-\Eb^{(\infty)}$ is only a few percent outside a crossover regime between small and large masses. But this convergence is misleading: the exponential fall off of the $u_n(r,a)$ sets in at increasingly larger $r$ ($\propto n^2$), such that the region around $\amin$ where the ground-state wave function is large is not well sampled well at large $n$. Calculations with the Fourier-sine basis introduced later in (\ref{sines})
indicate that $\Eb^{(1)}$ is accurate to about 10\%, 20 \%, for model-I, model-II.
Here we shall only record that the large-mass results of the variational calculation are asymptotic to\footnote{This can be understood as follows:
For model-I the result is simply the absolute minimum of the (large-mass approximation of the) regularized potential, $V_{\rm reg}(\ell)\simeq -2d  m^3 G^2 /\ell^2= - 6 m^3 G^2$.
For model-II the opposite sign of $V_1$ causes it to act like a small-$a$ `barrier' in $\mE(a)$, similar to the kinetic energy `barrier' $\langle K\rangle=1/(ma^2)$. At small masses this effect is negligible, since the kinetic energy pushes $\amin$ towards the then large $\aB$.  When $m$ increases $\amin$ decreases but it does not fall below $\simeq 11 \sqrt{G}$, after which it increases due to the $d$ term in $V_1$. At large masses
the potential may be approximated by $V \simeq - G m^2/r + 2 m^3 G^2/r^2 $, which gives $\mE(a)\simeq-m^2 G /a+4m^3 G^2 /a^2$ and results in a relatively small variational $\mE(\amin) \simeq - m/16$, which is $\ell$ and $G$ {\em in}dependent.}

\bea
\frac{\Eb}{m} &\simeq& \frac{6 m^2 G^2}{\ell^2}\,, \qquad \mbox{model-I, non-running}
\label{ebnonrunningI}
\\
&\simeq& \frac{1}{16} \,. \qquad\qquad \mbox{model-II, non-running}
\label{ebnonrunningII}
\eea

\section{Running potential models I \& II}
\label{secevol}

A distance-dependent coupling $G^{(1)}_r$ can be identified by writing
\be
V=-\frac{G^{(1)}_r m^2}{r} \;,
\ee
and from this a dimensionless $\tG$:
\be
\tG \equiv \frac{G^{(1)}_r}{r^2} =
\frac{G}{r^2} + \frac{2dmG^2}{r^3} + c\frac{G^2}{r^4} + \mO(G^3)\, .
\label{tG1}
\ee
We identify a beta-function for $\tG$ to order $G^2$:
\bea
-r\frac{\partial \tG}{\partial r} &=&
2\frac{G}{r^2} +6dm \frac{G^2}{r^3} + 4 c\frac{G^2}{r^4} + \mO(G^3)
=  \bt(\tG,m\sqrt{G}) + \mO(G^3)\,,
\label{bt1}
\\&&\nonumber\\
\bt(\tG,m\sqrt{G})&=&
2\tG + 2dm\sqrt{G}\,\tG^{3/2} + 2 c\,\tG^2 \, .
\label{defbt}
\eea
Here (\ref{tG1}) is used to eliminate $r$ to order $G^2$ on the r.h.s.\ of (\ref{bt1}).\footnote{For instance, solving (\ref{tG1}) for $G/r^2$ by iteration:
$G/r^2 = \tG-2d m\sqrt{G}\,(G/r^2)^{3/2} - c\,(G/r^2)^2 +\mO(G^3)=
\tG-2dm\sqrt{G}\, \tG^{3/2} -c\, \tG^2 + \mO(G^3)$. Alternatively, we can divide the l.h.s. and r.h.s. of (\ref{tG1}) by $G$, use Mathematica to solve for $1/r$, insert in (\ref{bt1}) and expand in $G$.}
We now {\em redefine} the running coupling $\tG(r)$ to be the solution of
\be
-r\frac{\partial \tG}{\partial r} = \bt(\tG,m\sqrt{G})\, ,
\label{evolution}
\ee
with the boundary condition
\be
G_r\equiv\tG r^2 \to G , \quad r\to\infty\, .
\ee
The corresponding running potential is defined as
\be
V_{\rm r} = -\frac{G_r m^2}{r} = - \tG\, m^2\, r.
\label{Vrdef}
\ee
A model-II type beta-function without the $d$-term but with the same negative $c$ was mentioned in \cite{Codello:2008vh} as a simple example generating a flow with a UV-attractive fixed point.

\FIGURE[t]{
\includegraphics[width=7cm]{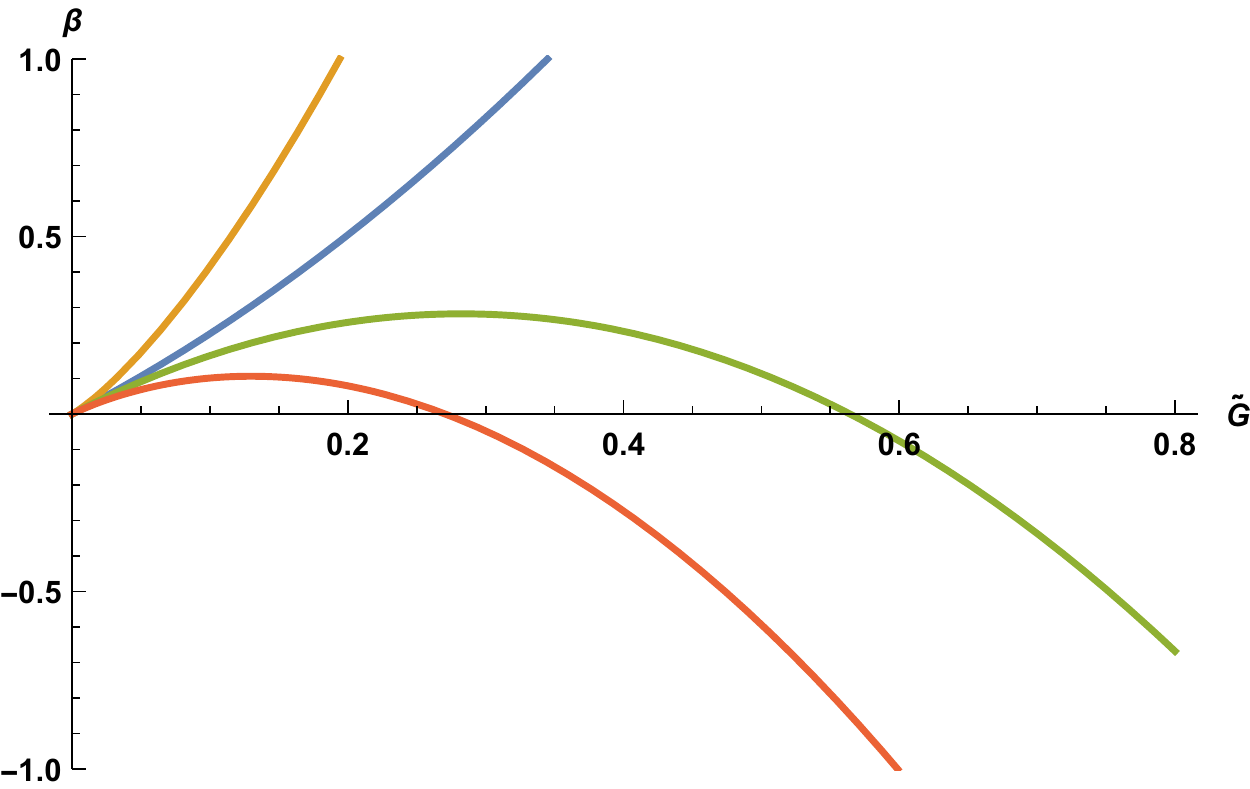}
\includegraphics[width=7cm]{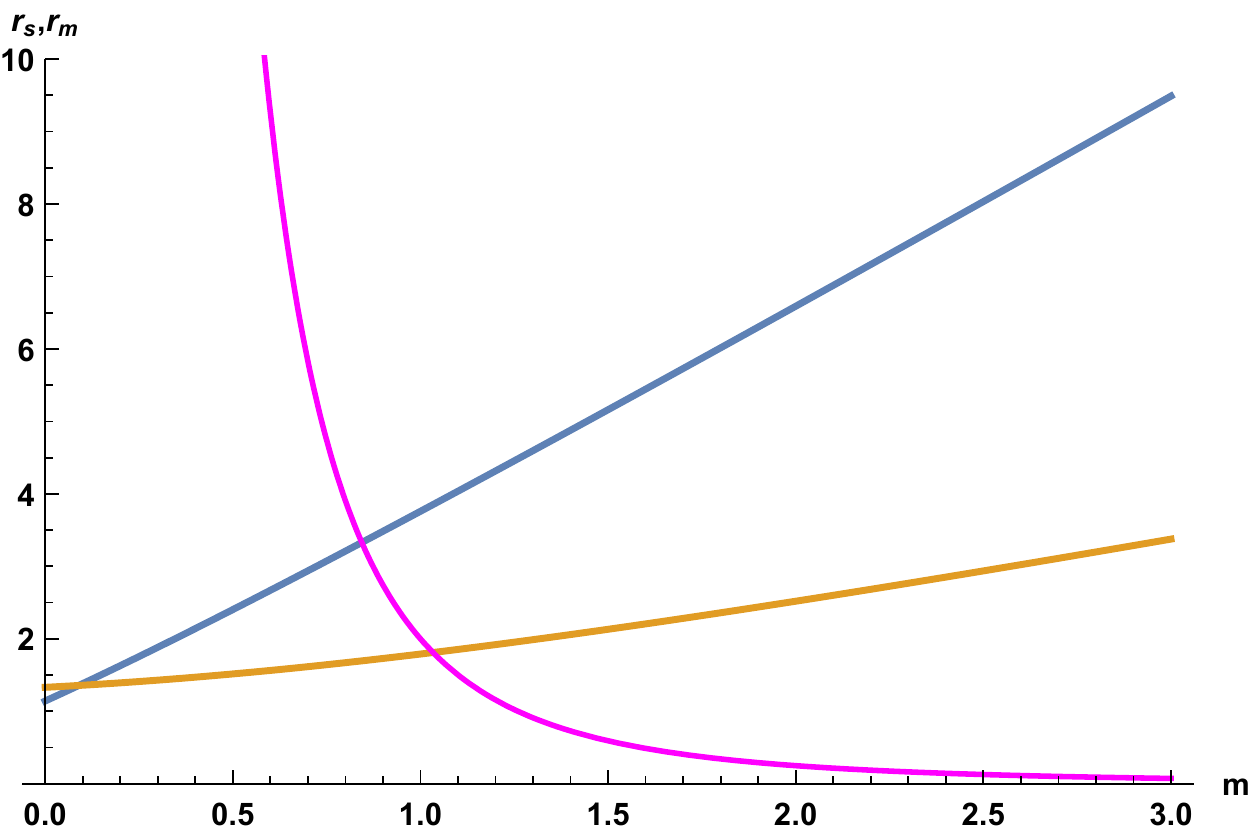} %from caseIvscaseII.nb
\caption{
Left: beta functions $\bt(\tG,m\sqrt{G})$; top to bottom:  $m=1$ (model-I, brown), $m=0$ (model-I, blue),  $m=0$
(model-II, green), $m=1$ (model-II, red).
Right: $\rs$ (model-I, blue) and $\rmin$ (model-II, brown) versus $m$.
Also shown is the Bohr radius $\aB$ (magenta).
Units $G=1$.
}
\label{figpbt12}
}
Figure \ref{figpbt12} shows a plot of the betas for two values of $m$. In model-I there is only the IR-attractive ($r\to\infty$) fixed point at $\tG=0$, for all $m$. In model-II there is in addition a UV-attractive ($r\to 0$) fixed point at positive $\tG$. It moves towards zero as $m$ increases:
\bea
\tG_* &=&\frac{1}{|c|}+ \frac{1}{2c^2}\left(G m^2-m\sqrt{G}\sqrt{|c| +G m^2/4}\right),
\qquad\mbox{model-II}
\label{NFPm}
\\
&=&\frac{1}{|c|} = 0.56 , \quad m=0\,,
\label{NFPmsmall}
\\
& = & \frac{1}{G m^2}- \frac{2|c|}{G^2 m^4} + \cdots, \quad m\to \infty\, .
\label{NFPmlarge}
\eea
Note that $\tG_*$ is not very large and it can even be close to zero for $m\sqrt{G}\gg 1$.
{\em For convenience, we use units $G=1$ in the following}.

The evolution equation (\ref{evolution})
is solved in appendix \ref{appGr}. For $d=0$, the solution simplifies to
\be
\tG(r) = \frac{1}{r^2 - c}
\, .
\label{Gtm0}
\ee
In model-II $c$ is negative and one recognizes the small-$m$ limit of the UV-fixed point as $r\to 0$.
In model-I $c$ is positive and when $r$ moves in from infinity towards zero, $\tG$ blows up at $r=\rL=\sqrt{c}$.
For non-zero masses $\rL$ moves to larger values (figure \ref{figpbt12}), which can be macroscopic,
%% very much larger than the Planck length:
\bea
\rL &=& 3 m  + \frac{c}{3m}[2\ln(3m)-\ln (c) -1 ]+\cdots , \quad m\to\infty \, ,
\qquad\qquad\mbox{model-I}
\label{lPmlarge}
\\
 &=& \sqrt{c} +  3m\,\frac{\pi}{4} +\cdots, \quad m\to 0\, .
\label{lPmsmall}
\eea
For general $m$, the running coupling has the expansion near $\rs$:
\be
\tG = \frac{\rL}{2c(r-\rL)} -\frac{3 \sqrt{2}\, m}{2 c^{3/2}}\,\frac{\sqrt{\rL}}{\sqrt{r-\rL}} + \mO(1)\,,
\qquad \mbox{ model-I}
\label{tGIas}
\ee
which shows an integrable square-root singularity in addition to a pole.
Hence, $\tG$ and the potential $V_{\rm r}$ are complex in $0<r<\rL$ and its Hamiltonian is not Hermitian.

Dropping the $c$-term in $\bt$ altogether at large $m$ gives a {\em classical} beta function (independent of $\hbar$) with a simple solution to its evolution equation,
\be
\bt = 2 \tG + 2dm\tG^{3/2} \Rightarrow \tG=\frac{1}{(r-d m)^2}\,,
\label{btclas}
\ee
and corresponding classical-evolution potentials,
\bea
V_{\CEI} &=& -\frac{m^2 r}{(r-3m)^2}\,,\qquad\qquad \mbox{CE-I model}
\label{CEImodel}
\\
V_{\CEII} &=&-\frac{m^2 r}{(r+m)^2}\,.\;\, \qquad\qquad \mbox{CE-II model}
\label{CEIImodel}
\eea
%The potential of the CE-I model has a double pole singularity at $r=3m$.

How to interpret the singularity in model-I? In usual terminology one might say that $\tG$ has a Landau pole at $\rL$ and a small-distance cutoff might have to be introduced to avoid it. However,
it seems odd to put a UV cutoff near $\rs$ when it is macroscopic. One option is to disallow macroscopic values of $m$---disallow huge values of the $d m$ term in the beta function and require $m\ll 1$; then $\rs=\mO(1)$ whereas the important distances are on the scale of the Bohr radius $2/m^3$ %$\mO(m^{-3})$
and a minimal distance $\ell$ of order of the Planck length would avoid problems. But then one would essentially be back to the previous section while
the region $m>1$ is interesting.

Encouraged by the following features we shall assume that the singularity represents black hole-like physics: At large $m$, $\rs\simeq 3m$ which is the right order of magnitude for the horizon when two heavy particles merge into a black hole.
In the relativistic classical-evolution model,
particles at rest released from a distance $r>\rs$ gain a relativistic velocity
approaching the light-velocity
as $r\downarrow \rs$ (cf.\ appendix \ref{appbounceCEN}). (The same happens with a test particle in the gravitational field of a heavy one,  when
%($m_1=M \ggg m_2=m$).)
$m=m_2\lll m_1$). Also for a Schwarzschild black hole the relativistic $|$velocity$|$ of a massive test particle approaches 1 at the horizon in finite (proper) time.
A non-Hermitian Hamiltonian also occurs with the Dirac equation in Schwarzschild spacetime when expressed in Hamiltonian form \cite{Lasenby:2002mc}.

We shall interpret the singularity in the potential as a
distribution. For the pole in (\ref{tGIas}) this is the Cauchy principal value.
One way to define the distributions is  \cite{Lighthill:1958},
\be
\frac{1}{(r-\rL)^n} = (-1)^{n-1}\frac{\partial^n}{\partial r^n}\, \ln|r-\rL|\,,
\quad n=1,2\,
\label{defdistr1}
\ee
($n=2$ refers to the CE-I model (\ref{CEImodel})). In terms of wave functions,
\be
\int_0^\infty dr\, \ph^*(r)\, \frac{1}{(r-\rL)^n}\, \ps(r)=
-\int_0^\infty dr\, \ln|r-\rL|\, \frac{\partial^n}{\partial r^n}\,\left[\ph^*(r) \ps(r)\right]\, ,
\quad n=1,2\,.
\label{defdistr2}
\ee
The wave functions are required to be smooth and to vanish sufficiently fast at the boundaries of the integration domain, such that the above partial integrations are valid as shown.
Eq.\ (\ref{defdistr2}) suggests that matrix elements of the potential are particularly sensitive to
{\em derivatives} of wave functions near $\rs$.

\FIGURE[t]{
\includegraphics[width=7cm]{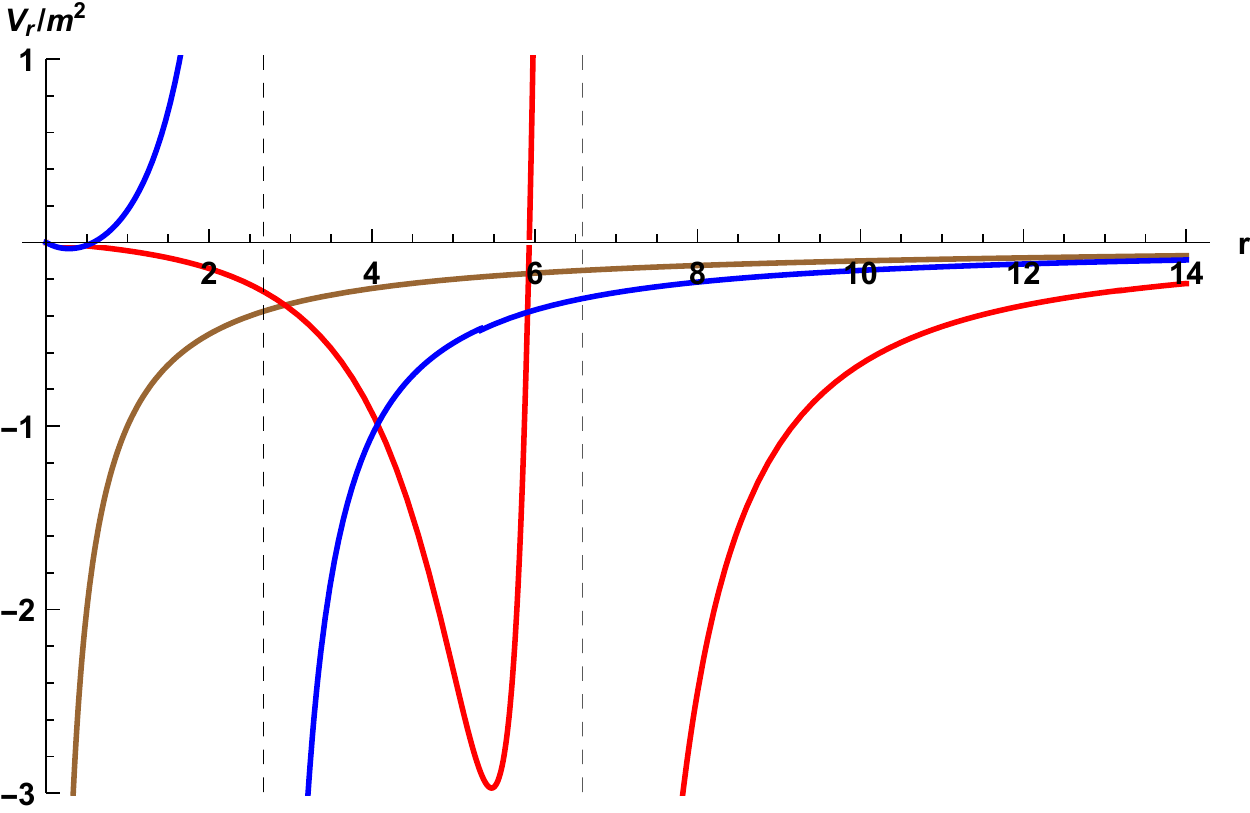} %see plots
\includegraphics[width=7cm]{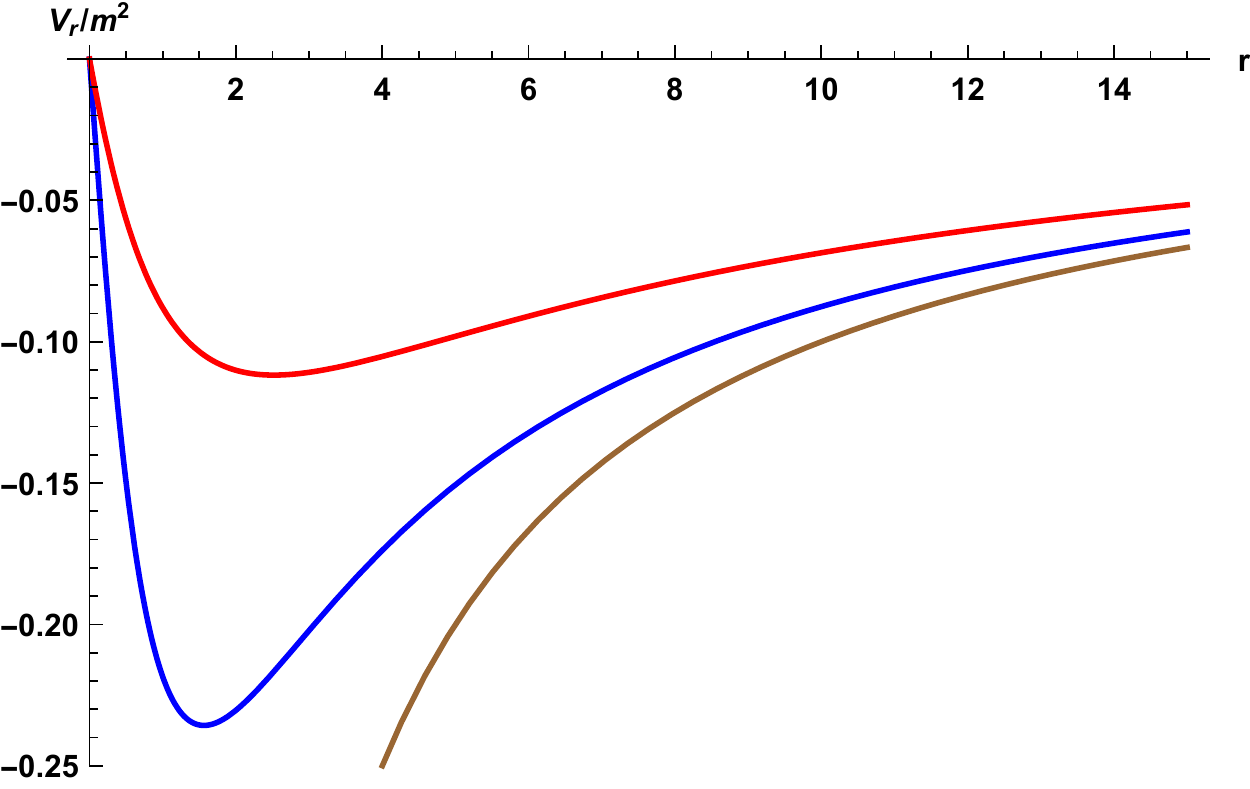} %see plots
\caption{
Running potentials $V_{\rm r}/m^2$ for $m=0.6$ (blue) and $m=2$ (red); also shown is the Newton form $-1/r$ (brown).
Left: model-I (real part); the dashed vertical lines indicate the position $\rs$ of the singularity. Right: model-II.
}
\label{figpVrm06m2}
}

Figure \ref{figpVrm06m2} shows the running potentials in models I and II for two masses (to facilitate visual comparison $V_{\rm r}$ was divided by $m^2$). They vanish at $r=0$ and at large distances they approach the Newton potential, which is also shown. In the left plot for model-I at $m=2$ one can imagine how the negative-definite classical double pole (\ref{btclas}) is ameliorated in the quantum model (\ref{tGIas}) into a single pole,
leaving a deep---still negative---minimum on its left flank and a steeper descent on its right flank. This minimum has nearly disappeared (just visible near the origin) for the smaller $m=0.6$.
In model-II, the potential is negative-definite and smooth with a minimum at $\rmin$:
\bea
\rmin &\to &\sqrt{|c|} ,\quad  \frac{V_{\rm min}}{m^2} \to -\frac{1}{2\sqrt{|c|}}\,,  \qquad m\to 0,  \qquad\mbox{model-II}
\\
 &\simeq& m, \qquad\; V_{\rm min} \simeq -\frac{m}{4}\,,\qquad\qquad\; m \gg 1.
\label{rminVrminas}
\eea
Remarkable here is the fact that $\rmin$ is for large $m$ also of order the Schwarzschild horizon scale, which suggests that this model might illustrate a `horizonless black hole'. The right plot in figure \ref{figpbt12} shows that  $\rs$ and $\rmin$ are rather featureless functions of $m$.

It is clear from figure \ref{figpVrm06m2} that when $m$ increases,
$\re[V_{{\rm r, I}}]$ can approach $V_{\CEI}$ only {\em non}-uniformly in $r$, since their singularity structures differ. One cannot expect simultaneous convergence of matrix elements. In case there is a UV cutoff on the wave functions that limits their first two derivatives one may expect uniform convergence of a finite number of matrix elements.  On the other hand, the approach of $V_{{\rm r, II}}$ to $V_{\CEII}$ {\em is} uniform in $r$ (as can be clearly illustrated by plotting their ratio). In this case also the approach of the $\bt$-functions is uniform in $\tG$, since the latter is restricted to $\tG<\tG_*$ and $\tG_*\to 0$.

\section{Binding energy in model-I}
\label{secbeI}

Appendix \ref{appI} describes details of the numerical treatment of the singularity; special aspects of non-hermitian but symmetric Hamiltonian are the subject of appendix \ref{appnormal}.

We start with variational calculations using $N$ s-wave bound state eigenfunctions $u_n(r,a)$ of the hydrogen atom,  (\ref{unswave}). Let $\Emin(N,a)$ be the eigenvalue with minimal real part of the hamiltonian matrix
\be
H_{mn} = \int_0^\infty dr\, u_m(r,a)\,(K+V_{\rm r})\, u_n(r,a),\quad (m,n) = 1,\,2,\,\ldots, N\,.
\ee
In section \ref{secvreg} we mentioned that keeping $a$ fixed by the variational method with $N=1$ leads to a (probably misleading) fast convergence when increasing $N$. Here we allow $a$ to depend to depend also on $N$.
The eigenvectors $f_{jn}(a)$ of the Hamiltonian matrix determine eigenfunctions
$f_j(r,a) =\sum_n f_{jn}(a) u_n(r,a)$.
Using the eigenfunction corresponding to $\Emin(N,a)$ as a trial function in the energy functional $\mE$ (cf.\ appendix \ref{appnormal}) and its real part for minimization the variational method becomes
\bea
\re[\mE] &=& \re[E_{{\rm min}}(N,a)]\equiv
F_N(a) \,,
\\
\frac{\partial}{\partial a} F_N(a)|_{a =\amin}&=& 0,
\quad \Emin = E_{{\rm min}}(N,\amin)\,,
\eea
where $\amin$ corresponds to the deepest local minimum of $F_N(a)$.

\FIGURE[t]{
\includegraphics[width=7cm]{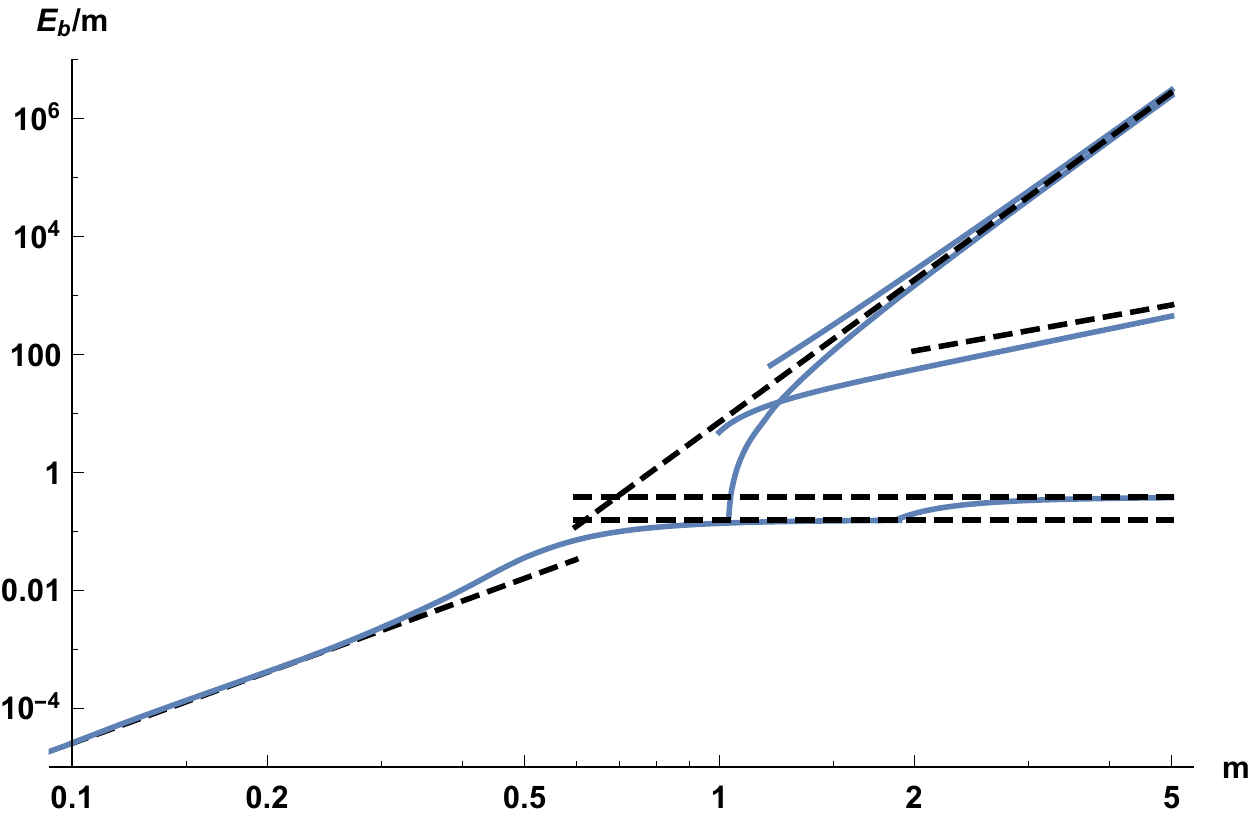} %plots
\includegraphics[width=7cm]{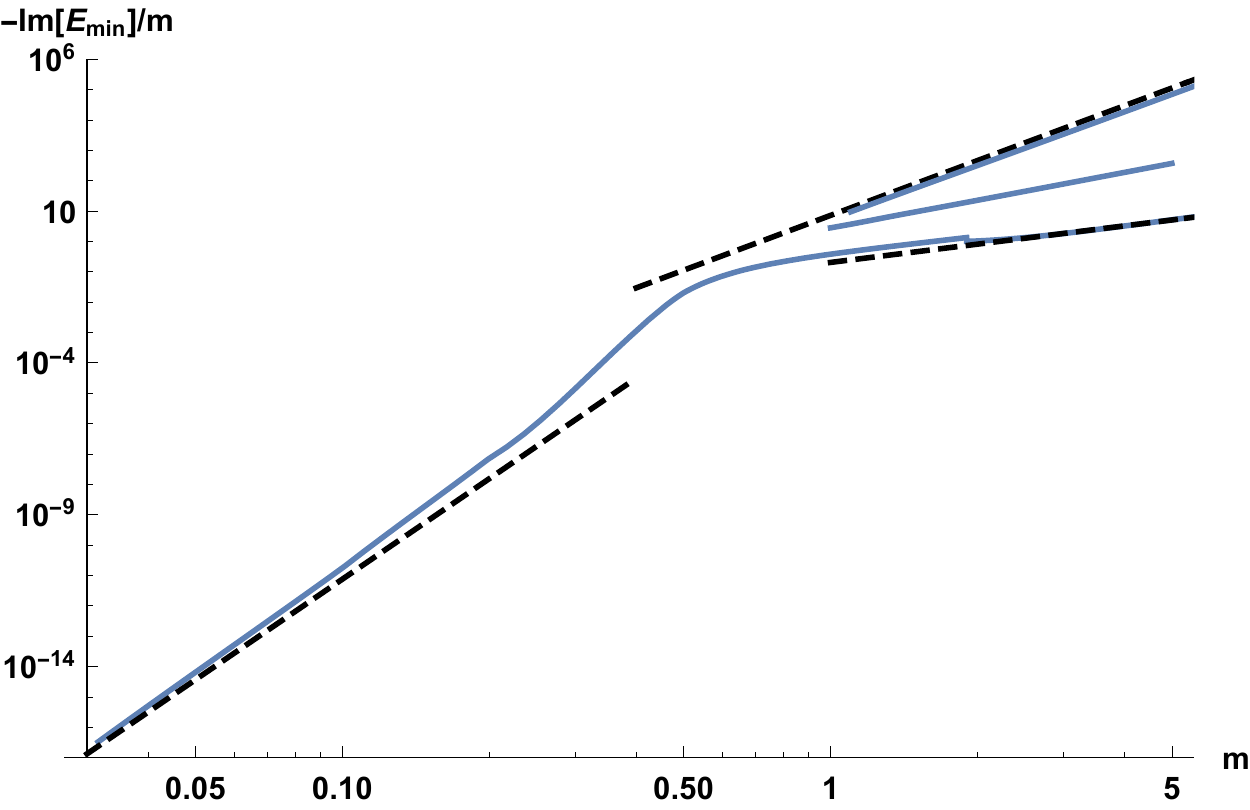} %plots
\caption{Left: Variational estimates of $-\re[\Emin]/m$.  The lowest blue curve is obtained with $F_1(a)$ and $u_1(r,a)$, with asymptotes into the small and large mass regions (black, dashed). Next in height in $m> 2$ is an estimate using a Gaussian wave function $f_{\rm G}(r,a,s)$ at fixed variance $s=0.18$ with asymptote (dashed) provided by the CE-I model,  (\ref{ECEIPlanckas}). Also shown are 'variational bounds'
obtained with Gaussian and Breit-Wigner functions (appendix \ref{appG}), $\mE_{\rm GP}(a,s)$ and
$\re[\mE_{\rm BWPSR}(a,s)]$ (highest and next highest  blue curves  in  $m>2$) with enclosed GP-asymptote (black, dashed).
Right: imaginary parts. The small-mass asymptote (black-dashed) represents (\ref{Gm}). The large-mass asymptote to the variational $-\im[\mE_{\rm min}]/m$  (lowest black-dashed line in $m>1)$ is a fit $0.20\, m^2$ to the numerical data. The highest asymptote and curve represent the BWPSR result. (Absent is the CE-I model which has a real potential.)
}
\label{figpevarI}
}

In the simplest approximation, $N=1$,
\be
\Emin(1,a)=H_{11}(a)\,.
\label{F1a}
\ee
For small masses we find again a single minimum $\amin\simeq \aB$ with binding energy $\Eb= -\re[\Emin] \simeq m^5/4$ shown in the left plot of figure \ref{figpevarI}.\footnote{Figure \ref{figpevarI} shows many other results for the binding energy which will be explained in due course.}
New in model-I is the imaginary part of $\Emin$, shown in the right plot.
Its asymptotic form for small $m$ is approximately given by
\be
\Gm_{\rm b} \equiv -2\,\im[\Emin]\approx \frac{32\sqrt{2}}{105}\,d \sqrt{c}\, m^{12} = 1.48\, m^{12}
\label{Gm}
\ee
(cf.\ appendix \ref{appHtrial}).

When raising $m$ beyond $0.9$ a second local minimum appears in $F_1(a)$ at much smaller $a$ (figure \ref{figpF1a}). This second minimum becomes the lowest one when $m$ increases between 1.8 and 1.9 -- the new global minimum for determining $\Emin$. The resulting extension of $\Eb/m$ into the large mass region is shown in the left plot of figure \ref{figpevarI}, wherein the dashed horizontal asymptotes come from the classical-evolution potential for $m\to\infty$ (cf.\ (\ref{FCE-I})). The right plot in figure \ref{figpevarI} shows the imaginary part.
%For large masses $-\im[\Emin]\simeq 0.20\, m^3$.

Increasing $N$, for masses $m\ge 2$,  each step $\Delta N=1$ introduces a new local minimum at still smaller $a$, while the previous minima change somewhat and then stabilize.\footnote{This does not happen in the small mass region where the kinetic energy contribution $1/(m a^2)$  to the variational function allows only one H-like minimum near $\aB$. In the Newton model (potential $-m^2/r$) new local minima do not appear when raising $N$.}
For example, for $N=6$ and $m=2$, $F_6(a)$ has seven local minima  (right plot in figure \ref{figpF1a}); the sixth is the lowest, with $\Emin/m \simeq -4.2 - 2.4\, i$ and an
$|{\rm eigenfunction}|^2$ consisting of two Gaussian-like peaks to the left and right of $\rs$. The energy of the first minimum is much higher and has changed little, its corresponding eigenfunction has the qualitative shape of the first hydrogen s-wave -- it is still H-like.
Increasing $N$ further leads to even lower values of $\re[\Emin]$ and this line of investigation rapidly becomes numerically and humanly challenging. We could not decide this way whether $\re[\Emin]$, at fixed $m\geq 2$, reaches a finite limit or goes to minus infinity as $N\to\infty$.

\FIGURE[t]{
\includegraphics[width=7cm]{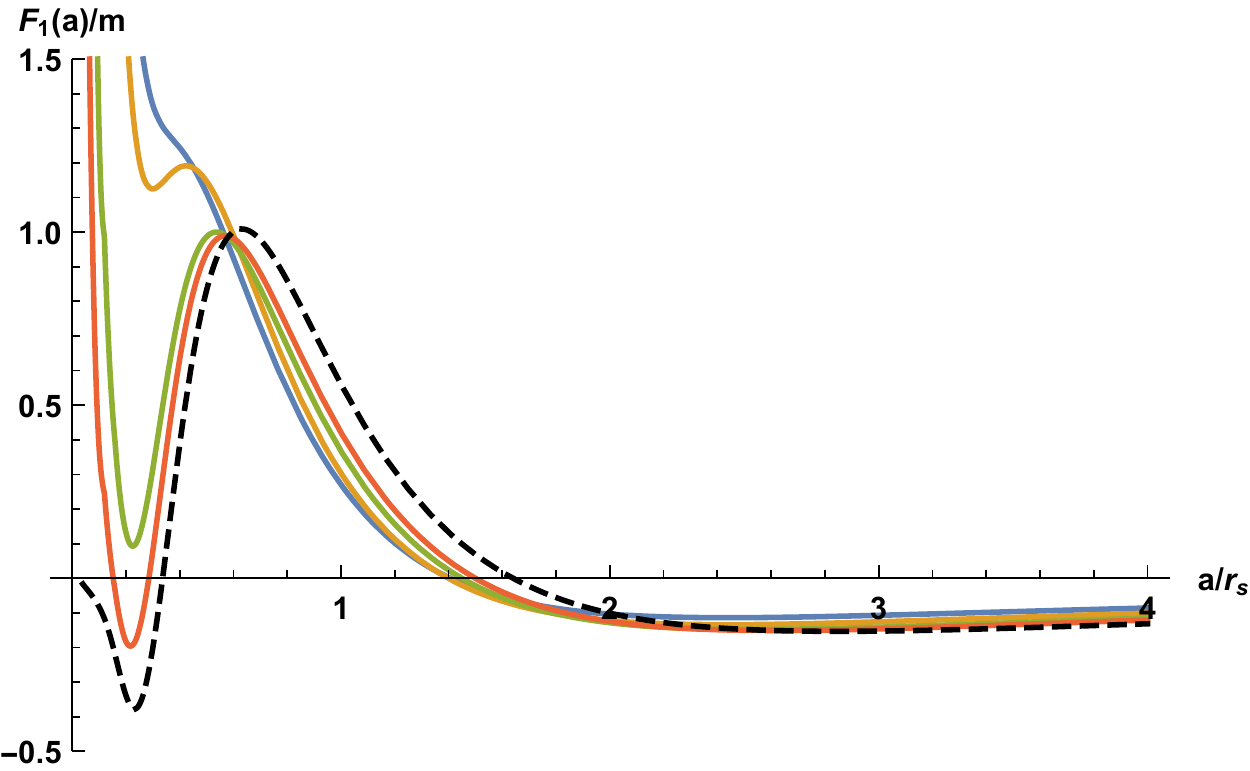} %plots
\includegraphics[width=7cm]{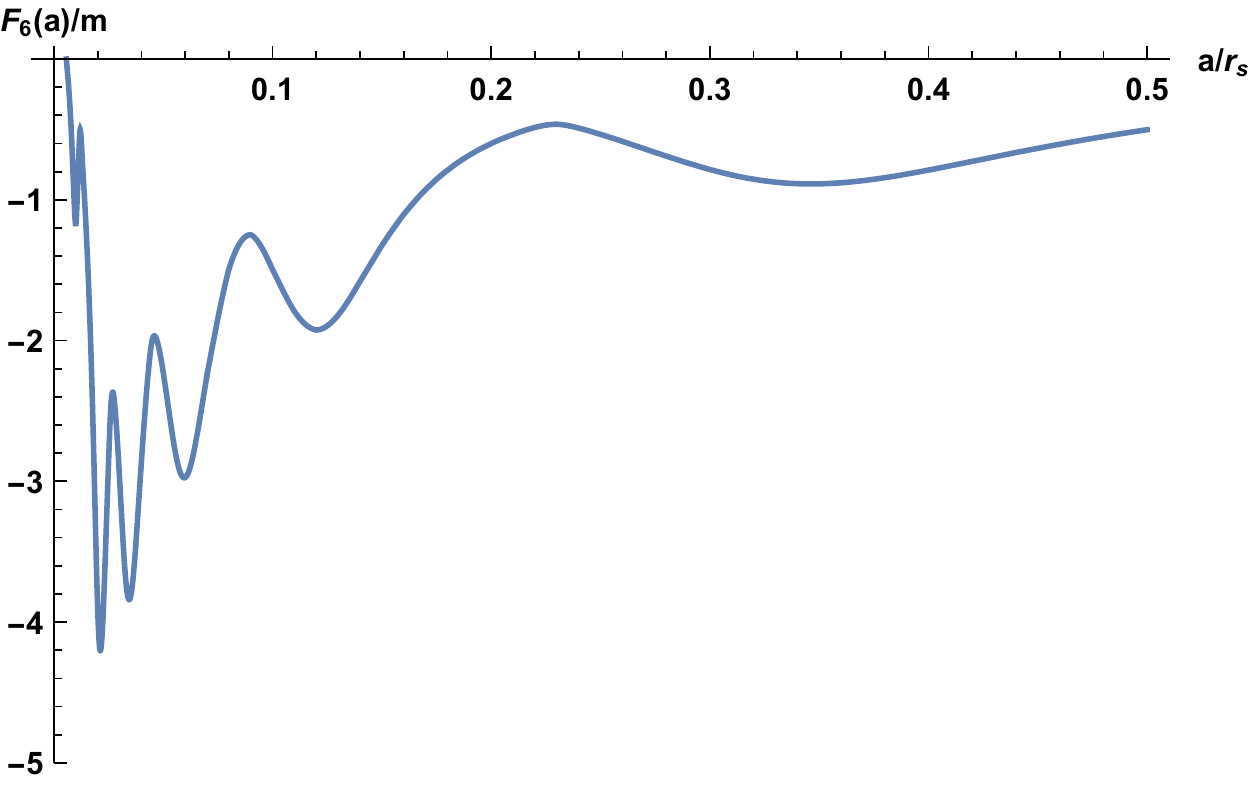} %plots
\caption{Left: Variational function $F_1(a)/m$ versus $a/\rs$ for $m=0.9$, 1, 1.5, 2 (top to bottom near $a/\rs = 0.2$), and the classical-evolution limit function (\ref{FCE-I}) for $m\to\infty$ (dashed).
The first minimum is the shallow one in the region $2< a/\rs <3$.
Right: $F_6(a)/m$ for $m=2$; the 2nd to 7th minima are shown, the first minimum is outside of the plot.
}
\label{figpF1a}
}

The s-wave Hydrogen eigenfunctions have nice asymptotic behavior for $r\to\infty$, but they are not well suited to investigate the evidently important region around the singularity at $\rs$. For this region much better sampling is obtained with the Fourier-sine modes
\be
b_n(r,L)=\sqrt{\frac{2}{L}}\, \sin\left(\frac{n\pi r}{L}\right) \theta(L-r),
%\quad 0\le r\le L,
\quad n=1,\,\ldots,\, N\, ,
\label{sines}
\ee
where $\theta$ is the unit-step function.
The modes are chosen to vanish at $L$ which is large relative to the region where the wave function under investigation is substantial; $L$ controls finite-size effects.
The sampling density is controlled by the
minimum half-wavelength $\lmmin/2= L/N$; the equivalent maximum momentum
$p_{\rm max}=\pi N/L$ serves as a UV cutoff on derivatives of the basis functions.

\FIGURE[t]{
\includegraphics[width=7cm]{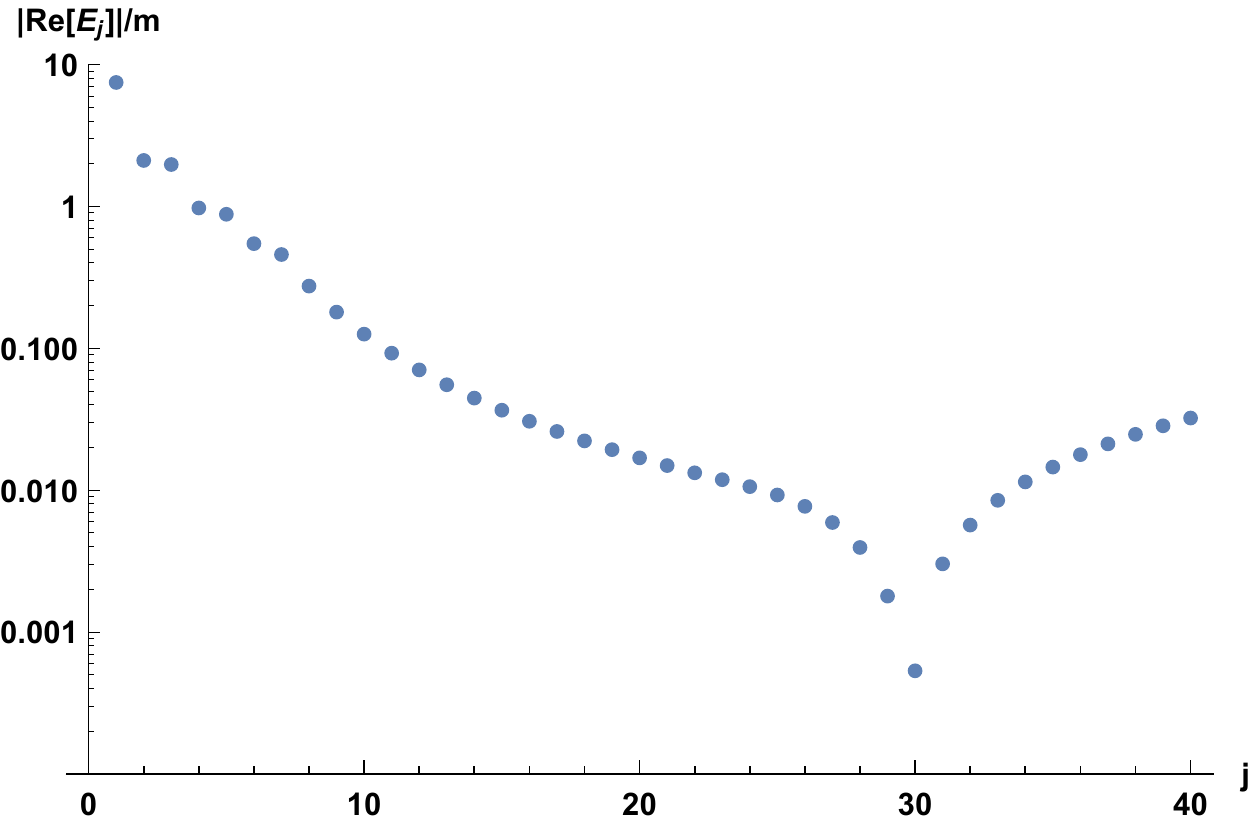} %plots
\includegraphics[width=7cm]{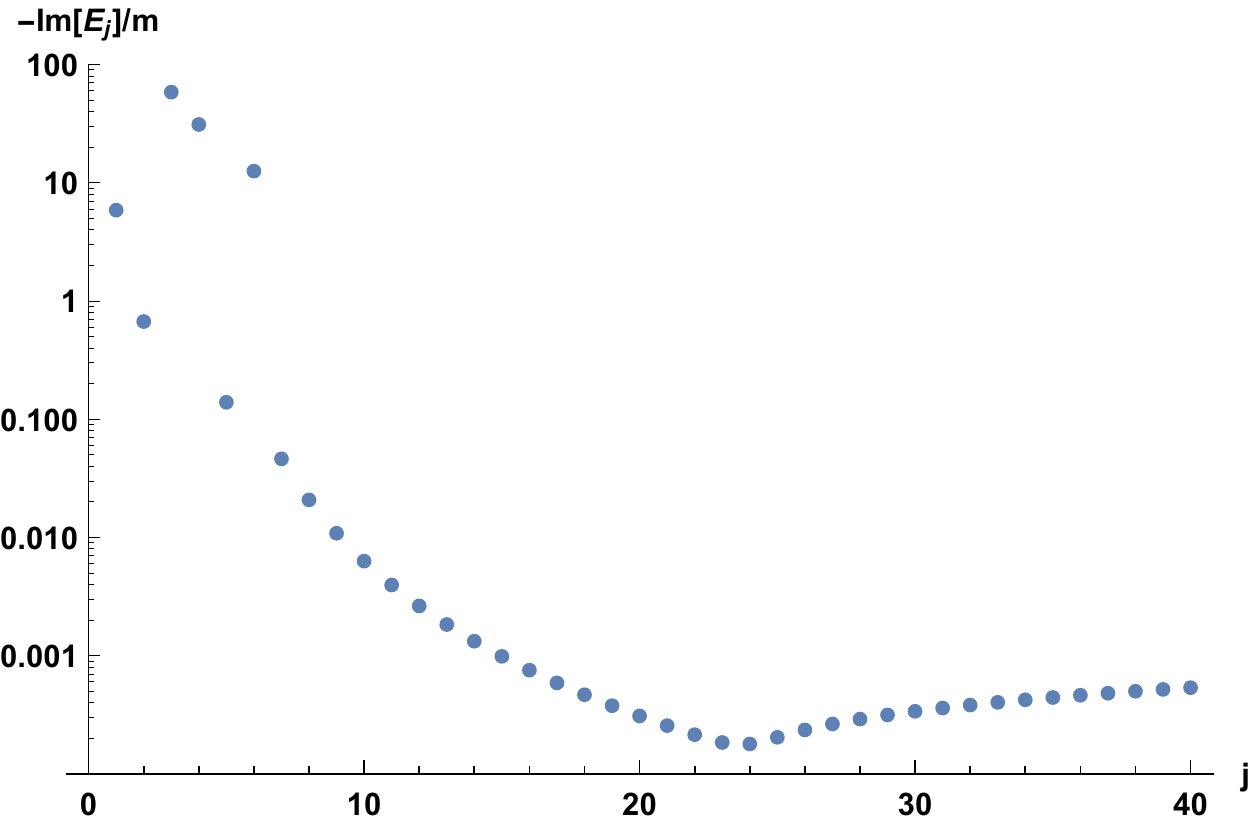} %plots
\caption{Left:
Absolute value of $\re[E_j]/m$ of the first 40 eigenvalues for $m=2$, $L=32\, \rL$, $N=128$ ($E_j$ is negative for $j<30$).
Right: Corresponding $-\im[E_j]/m$. % ($\im[E_j]<0)$.
}
\label{figpREjm20L32N128}
}

Some results follow now first for the case $m=2$ in the large-mass region for which $\rL= 6.6$ and $\aB=1/4$. Figure \ref{figpREjm20L32N128} shows part of the eigenvalue spectrum for $L=32\, \rL$, $N=128$, ordered by increasing $\re[E_j]$. The real parts of the eigenvalues start negative and change sign near mode number $j=30$, beyond which they increase roughly quadratically with $j$ (linearly when using $\Krel$) where they correspond to the unbound modes. The mode number where the eigen-energy changes sign increases with $L$ at fixed $\lmmin$.
The binding energy $|\re[E_1]|$ is large compared to $m$ and first few $\re[E_j]$ look a bit irregular; their imaginary part is very large at $j=3$, 4 and 6.
Exceptional is the $j=N$ eigenvalue: the last three eigenvalues are $E_{126} = 1.73- 0.00088\, i$, $E_{127}= 1.77 - 0.00070\, i$, $E_{128} = 40.8 - 201\, i$.

\FIGURE[t]{
\includegraphics[width=7cm]{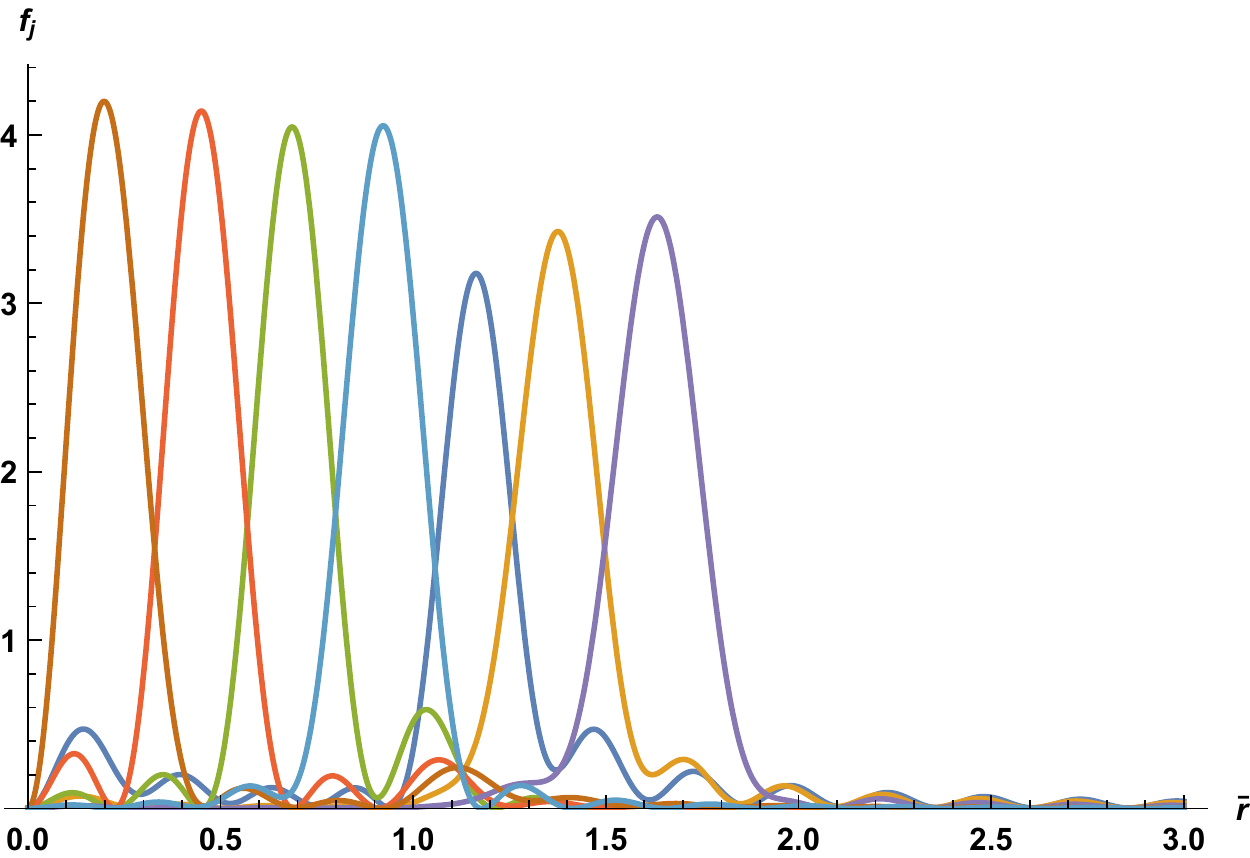} %plots
\includegraphics[width=7cm]{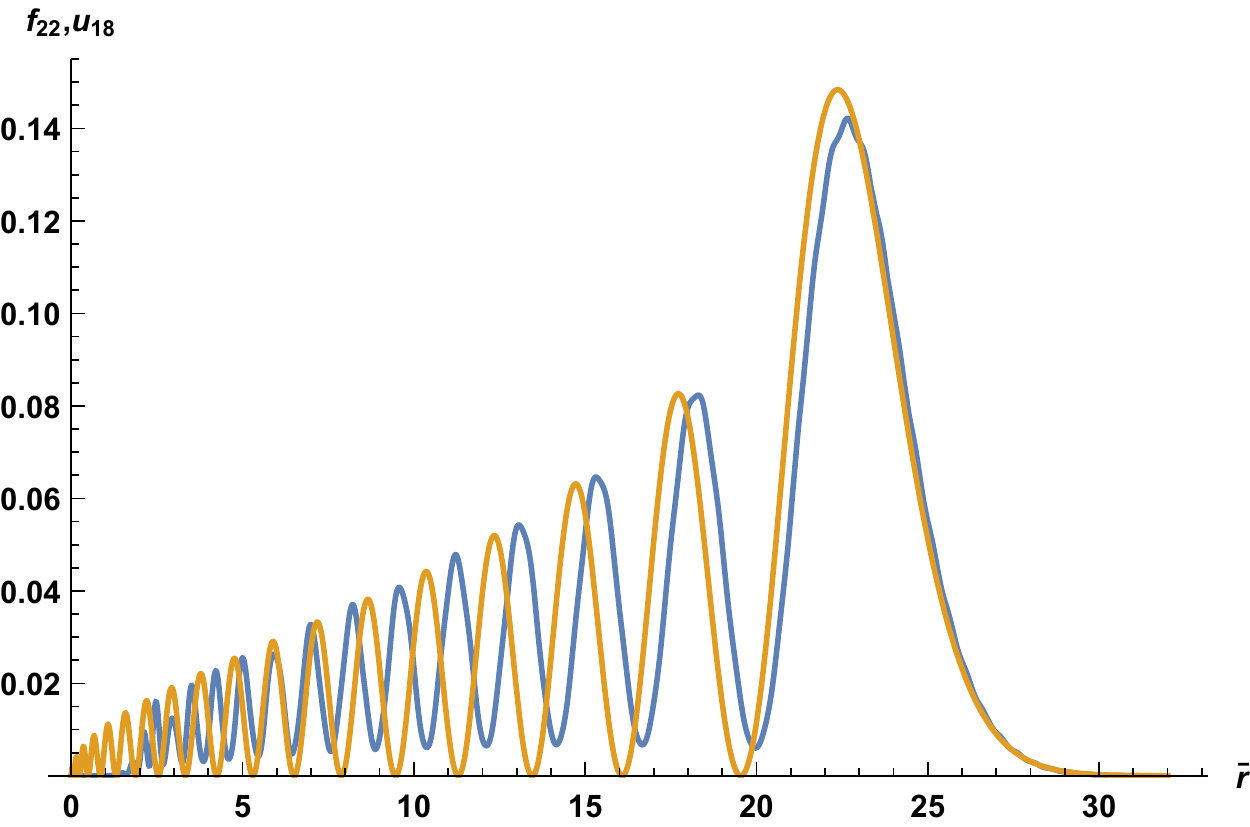} %plots
\caption{Left: first six eigenfunctions $\rs |f_j(r)|^2$ and $\rs |f_{128}(r)|^2$, as a function of $\rb=r/\rL$, for $m=2$, $L/\rs=32$, $N=128$. From left to right: $j=6,\,4,\,3,\,128,\,1,\,2,\,5$. Right: $\rs |f_{22}(r)|^2$ (blue, smallest peak at $\rb\approx 23$) and $\rL u_{18}(r,\aB)^2$ (brown).
}
\label{figpfcnm20L32N128}
}

The left plot in figure \ref{figpfcnm20L32N128} shows the first six eigenfunctions $\rs |f_j(r)|^2$ versus $\rb=r/\rL$, normalized under Hermitian conjugation (the factor $\rs$ stems from the Jacobian in $d r = \rs\, d\rb$).
Also added is the last one, i.e.\ $f_{128}(r)$;
it straddles $\rs$ and reaches into $r<\rs$ where $\im[V_{\rm r}]$ is large. The eigenfunctions $f_j(r)$, $j=1$, 2, 5, tunnel a little through the pole barrier into the region $r<\rs$ and become small for
$r\lesssim 0.95\, \rs$.

Eigenfunctions for which $|\im[E_j]| > |\im[E_1]|$ peak in the region $0<r < \rs$ where the potential is complex. At smaller $L$ such very large $-\im[E_j]$ `outliers' also occur in the unbound part of the spectrum, whereas their $\re[E_j]$ appear mildly affected relative to neighboring $j$. The mode numbers of the outliers vary wildly when varying $L$ or $\lmmin$, but their number appears to be roughly given by the sampling density times $\rs$: $(N/L)\rs$. In figures \ref{figpREjm20L32N128} and \ref{figpfcnm20L32N128}, $N\rs/L=128/32=4$ and including $j=N$ there are four outliers.

With increasing $j>6$ the negative-energy eigenfunctions slowly become H-like, but without support in the region $0<\rb\lesssim 1$ and with relatively small imaginary parts, $\im[E]/\re[E]\ll 1$. The right plot in figure \ref{figpfcnm20L32N128} shows an example in which $f_{22}(r)$ is compared with $u_{n}(r,\aB)$, $n=18$, chosen to give a rough match at the largest peak.

Finite-size effects appear under control when the wave function fits comfortably in $0<r<L$, which is true in the left plot of figure \ref{figpfcnm20L32N128}, and reasonable well also in the right plot. Beyond $j=23$ the wave functions get squeezed in the limited volume and finite-size effects become large.
The large $j$ eigenfunctions ({\em except} $f_{N}(r)$) look a bit like the sine functions of a free particle in the region $\rs <r<L$.
A domain size of the  minimal-energy eigenfunction can be defined by the distance $r_{90}$ containing 90\% of the probability, which is for the current example given by
\be
\int_0^{r_{90}} dr\, |f_1(r)|^2=0.9\,, \qquad r_{90} \simeq 1.95\,\rs \simeq 13,
\qquad\qquad \mbox{($m=2$, model-I)}
\label{r90I}
\ee
much smaller indeed than $L= 32\,\rs\simeq 211$. But the {\em peak} of $|f_1(r)|^2$ is just outside $\rs$ (figure \ref{figpfcnm20L32N128}). For larger $N/L$ most of this $r_{90}$ consists of $\rs$ since then the width of the peak is {\em much} smaller than $\rs$ (appendix \ref{appbounds}), and the same holds in general for larger masses (appendix \ref{appG}).

\FIGURE[t]{
\includegraphics[width=10cm]{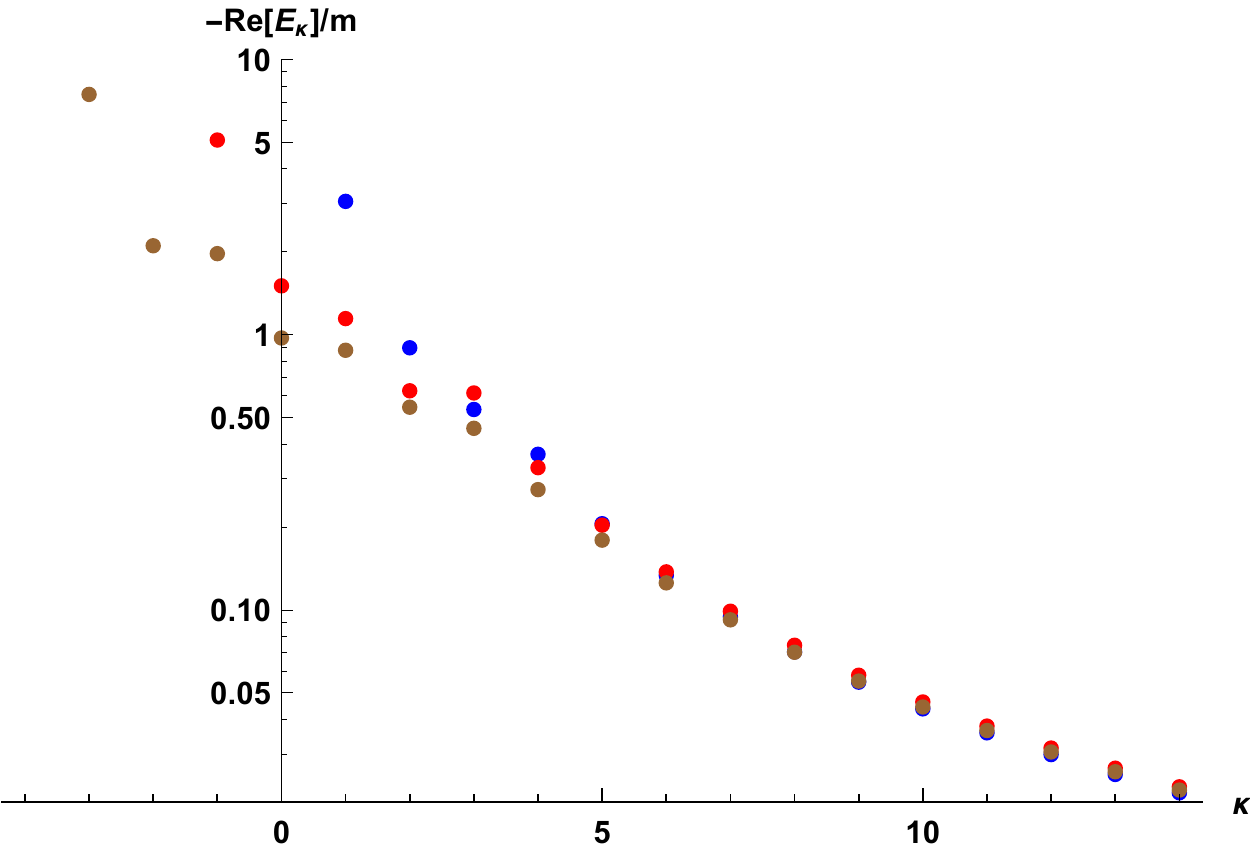} %binding1loopISine-m20L8N128A.nb
\caption{Shifted spectra for $m=2$, $L/\rs=32$, $N=64$, 96, 128, equivalently $\lmmin/\rs=1$, 2/3, 1/2. The shifts $\sg$ are respectively 0, 2, 4
(blue, red, brown dots or upper, middle, lower dots at $\kp=1$).
}
\label{figpspeckappa}
}

The minimal energy $\re[E_1]$ is quite sensitive to $N/L$  because matrix elements $V_{mn}$ are sensitive to the derivatives of the basis functions at the singularity. Comparing different $N/L$ we can shift the sequence $j$ by an amount $\sg$
and label $E$ by $\kp=j-\sg$ with $\kp$ `anchored' at some value where the energy and eigenfunction are H-like: $f_j(r)\approx u_\kp(r)$ and $\re[E]\approx - m^5/(4\kp^2)$ .
For example, $f_{22}$ was compared to $u_{18}$  in figure \ref{figpfcnm20L32N128} and thus $\kp=18$ and $\sg=4$.
Figure \ref{figpspeckappa} shows shifted spectra for three values of $N/L$.
The sequence reaching to $\kpmin=-3$ ($\sg=4$) corresponds to case shown in figures  \ref{figpREjm20L32N128} and \ref{figpfcnm20L32N128}. The dots match visually at
$\kp=6$, 7, \ldots, where UV-cutoff effects are reasonably small.

The question whether the binding energy is bounded is investigated further in appendix \ref{appbounds} where we come to the conclusion that it is finite in the non-relativistic model. But it is huge for large masses, $\Eb/m\approx 7 m^8$,  and the squared average velocity $v^2=\langle K \rangle/m $ is of the same order of magnitude.\footnote{In figure \ref{figpREjm20L32N128}, $\Eb/m$ is already very large for $m=2$ but $v^2=0.51$ is still moderate. Repeating the computation for the relativistic model gave $v^2=0.52$, $\vrel^2=0.30$, whereas the other results changed little compared to the non-relativistic model.} The number of eigenfunctions with dominant support in the region $r\lesssim\rs$ is expected to stay finite but large in the limit $N/L\to\infty$, with a finite large negative minimal $\kp\equiv\kpmin$, in the shifted labeling.

In the relativistic model-I (with the kinetic energy operator $\Krel$) we find that there is {\em no} lower bound on the energy spectrum in the large mass region (in the small mass region $m\lesssim 0.61$
the binding energy with $\Krel$ is finite and approaches that with $K$ as $m\to 0$). When $N/L \to \infty$, all energies $\re[E_j]$ near the ground state move to $-\infty$; in the shifted labeling $\kpmin$ moves to $-\infty$.

\FIGURE[t]{
\includegraphics[width=10cm]{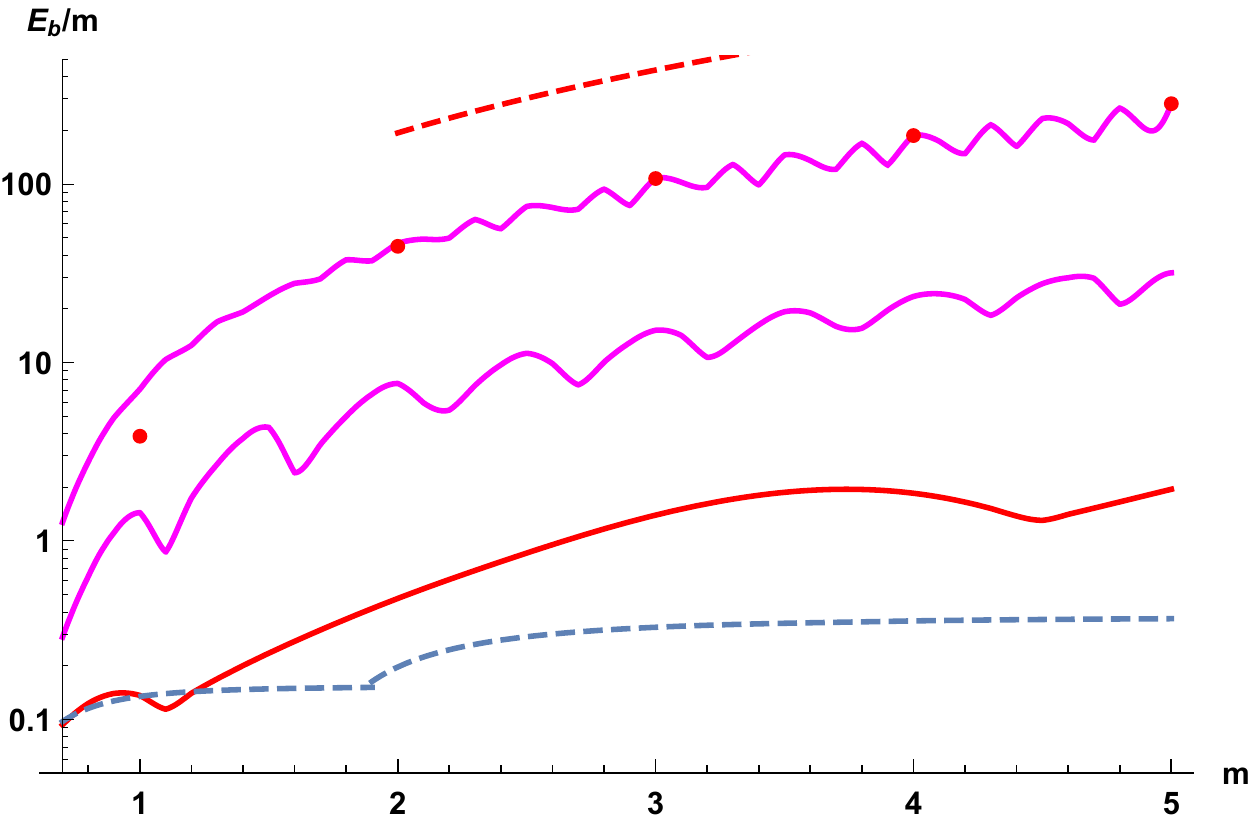} %...Planck2rel.nb
\caption{
Examples of the mass dependence of $\Eb/m$ at various fixed minimal wavelengths $\lmmin$,
and for comparison also the variational result obtained with $u_1(r,a)$ shown earlier in figure \ref{figpevarI} (lowest blue dashed curve); higher at $m=2$, in succession: $\lmmin=19.8$ with nonrelativistic $K$ (red); $\lmmin=3.29$ with the relativistic $\Krel$ (magenta); $\lmmin=1$ with $\Krel$ (magenta), $K$ (red dots) and asymptote (\ref{EblmminCEI}) from the CE-I model (red, dashed).
}
\label{figfixedlmmin}
}
\FIGURE[h]{
\includegraphics[width=7cm]{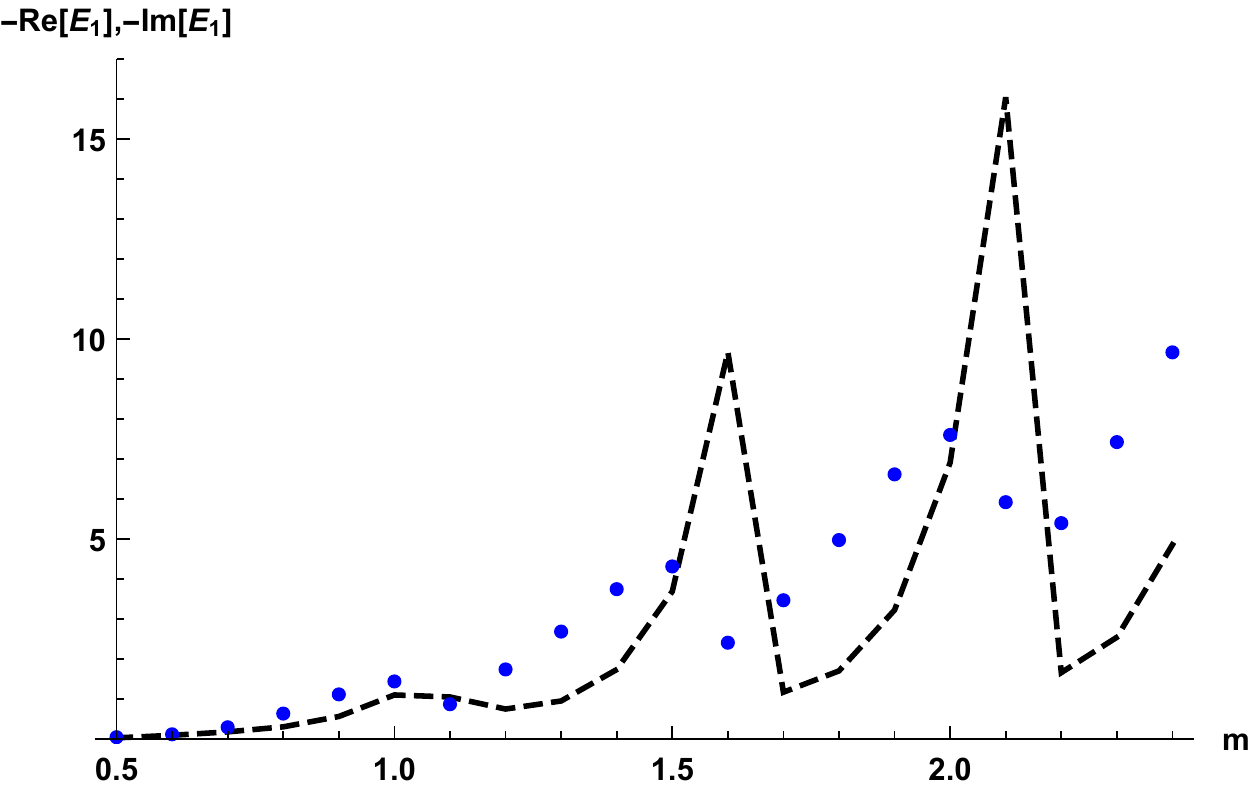} %...Planck4rel.nb
\includegraphics[width=7cm]{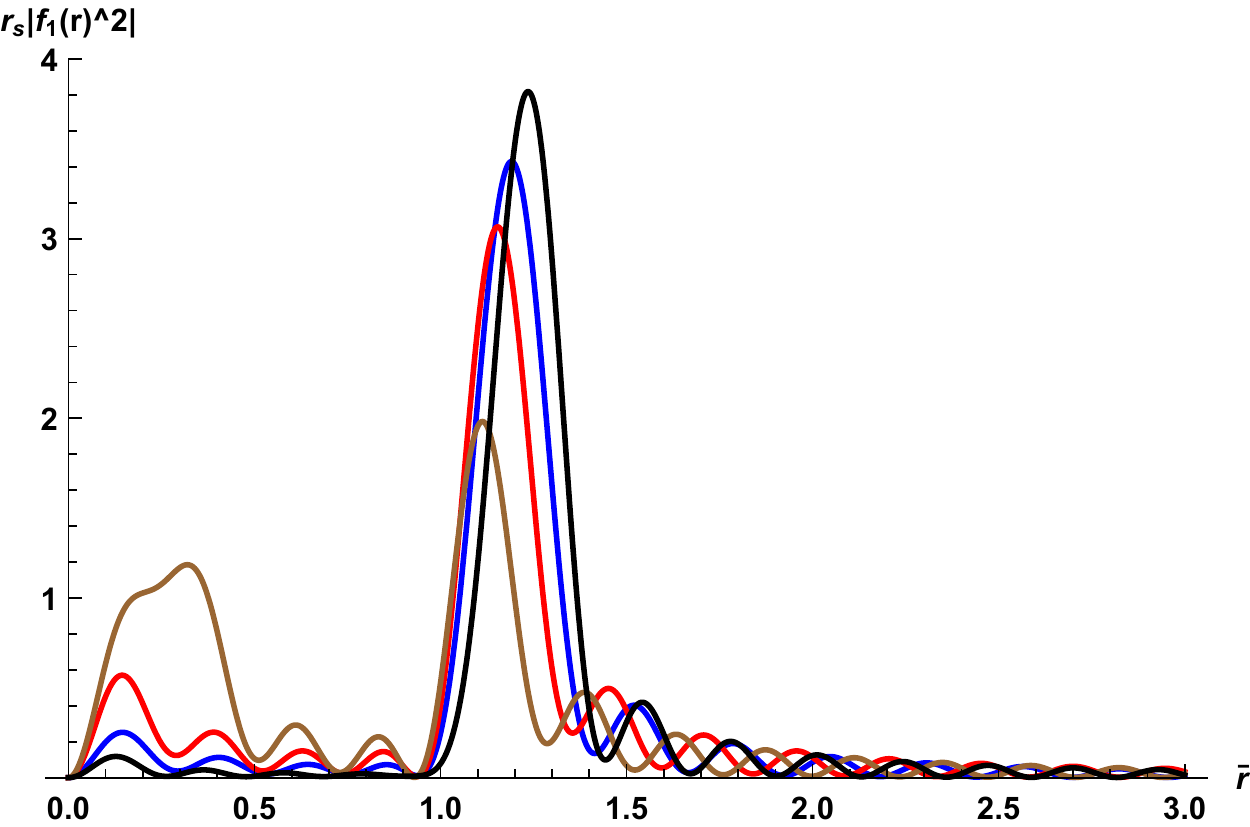} %...Planck4rel.nb
\caption{Left: real and imaginary parts of $-E_1$ vs.\ $m$ , for $\lmmin=3.29$ (respectively blue dots and black dashed straight line segments connecting data points).
Right: $\rs |f_1(r)|^2$ vs.\ $\rb=r/\rs$ ; large to small peak-heights around $\rb=1.2$: $m=2.2$ (black), 1.9 (blue), 2 (red), 2.1 (brown).
}
\label{figPlanck}
}

But one may question wether it makes sense to allow arbitrarily large derivatives in non-relativistic eigenfunctions when the binding energy is so sensitive to this. Let us put a cutoff on the Fourier momenta, $p_{\rm max} = N \pi/L$, equivalently, require a minimum
$\lm_{\rm min}$. Examples are  shown in figure \ref{figfixedlmmin}. For comparison, also shown is the earlier variational result obtained with $u_1(r,a)$ (same as in figure \ref{figpevarI}), and the large mass result obtained in the CE-I model with $\lmmin=1$ (cf.\ appendix \ref{appCEIsine}),
\be
\Eb/m \simeq 48\, m^2\, \qquad\qquad\qquad\qquad \mbox{CE-I model }\,.
\label{EblmminCEI}
\ee
This large mass result seems quite far off; comparison with results using Gaussian variational trial functions with a fixed width support it (cf.\ end of appendix \ref{appG}; a quadratic dependence $\Eb/m\propto m^2$ was found earlier in (\ref{ebnonrunningI})).
The surprising dips in the mass dependence are accompanied by large variations in the imaginary part of $E_1$, as shown in the close-up in figure \ref{figPlanck}. Large $|\im[E_1]|$ imply eigenfunctions that are substantial in $r<\rs$ (cf.\ the right plot), which diminishes the contribution to $\re[E_1]$ from the right flank of the singularity.
The occurrence of substantial contributions to $f_1(r)$ in $r<\rs$ is perhaps an effect of rendering it orthogonal (under transposition) to all other eigenfunctions, a property involving also the imaginary part of the Hamiltonian and its eigenfunctions.
%However, the large imaginary part in the potential occurs on the left flank of the singularity, it diverges as $r\uparrow \rs$ where the $|f_1(r)|^2$ in figure \ref{figPlanck} appear not particularly large.
In the CE-I model the potential is real;  the potential and the ground-state wave function are nearly symmetrical around $\rs$ and we found no dips in the binding energy as a function of $m$.

For large masses the variational trial function $u_1(r,a)$ is evidently wrong in its estimate of a small and constant $\Eb/m \simeq 0.23$. Its only parameter $a$ cannot simultaneously monitor two properties of the wave function: a large derivative, near the singularity.

\section{Binding energy in model-II}
\label{secbeII}

\FIGURE[h]{
\includegraphics[width=7cm]{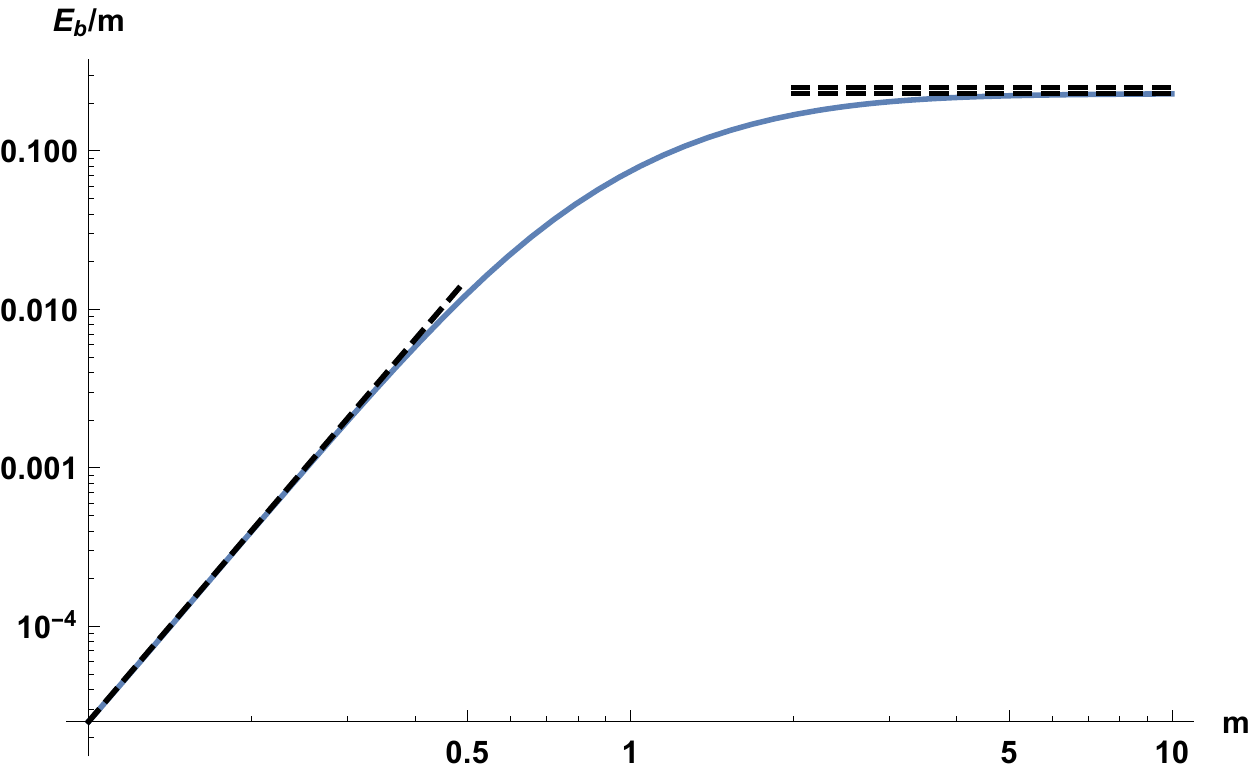} %from plots.nb
\includegraphics[width=7cm]{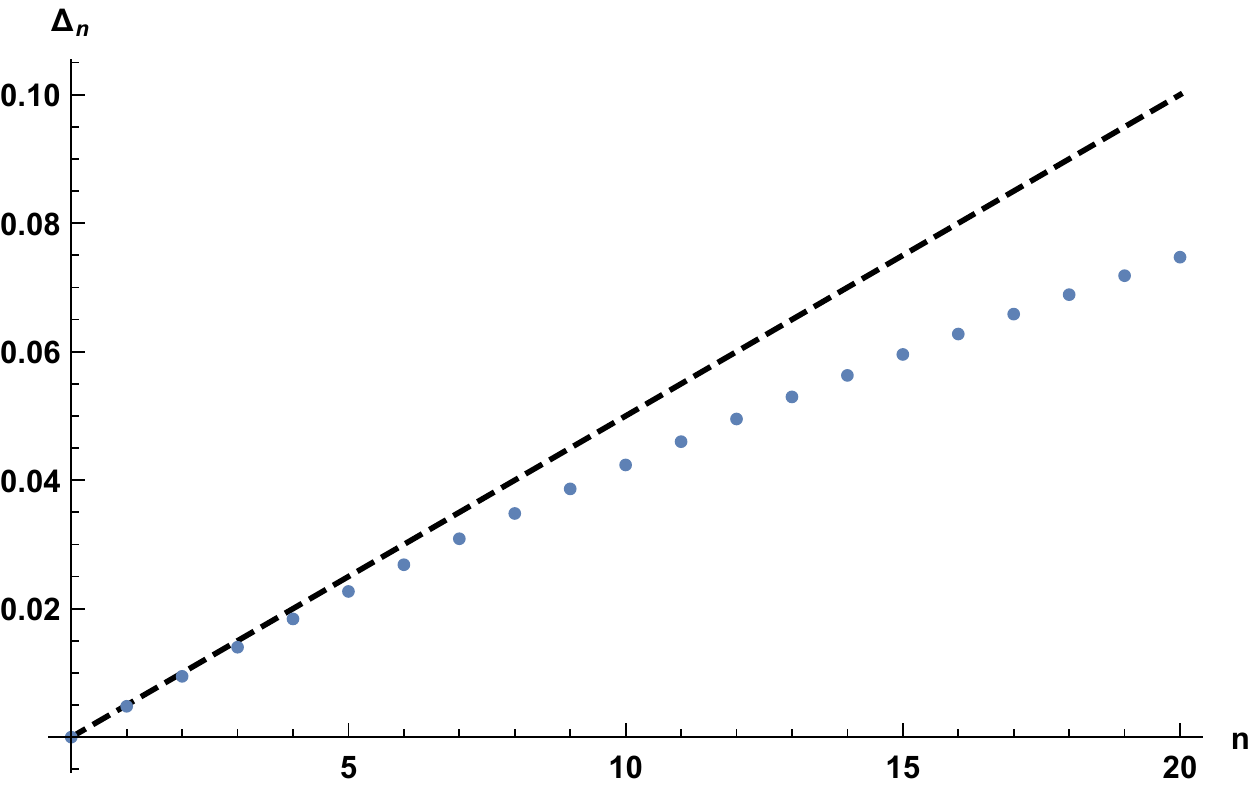} % from binding-1loopIISine-m100L64N64
\caption{Left: Variational binding energy of model-II, with its large-$m$ asymptote. The slightly higher black-dashed line represents $1/4$.
(At $m=2$ the Fourier-sine basis with $N=128$ and $L=64$ gives a 4 \% larger $\Eb/m$ than the variational estimate; the relativistic value is another 28 \% higher.) Right: excitation spectrum $\Delta_n =(E_{n+1}-E_1)/m$ near the ground state for $m=10$, $L=64$, $N=64$. The dashed line shows $n\om/m=n/(2m^2)$.  }
 \label{figpevarII}
}

Figure \ref{figpevarII} shows the variational estimate of the binding energy with the s-wave trial function $u_1(r,a)$. For small masses $\Eb/m$ is again close to the perturbative values in section \ref{secvreg}.
At large $m$ it becomes constant as in model-I where this behavior was misleading. However, here the mismatch of the variational value ($\simeq 0.23$) with the ideal value ($1/4$) is moderate because the running potential approaches uniformly that of the classical-evolution model CE-II (section \ref{secevol}, (\ref{CEIImodel})), for which $E_{\rm b}/m$ becomes constant at large $m$. The spectrum near the ground state is approximately that of a harmonic oscillator (HO), which can be understood from the expansion of the CE-II potential near its minimum at $r=m$,
\bea
V_{\CEII}(r)&=&-\frac{m^2 r}{(r+m)^2}
=  - \frac{m}{4} +\frac{(r-m)^2}{16 m} -\frac{(r-m)^3}{16 m^2} + \cdots
\nonumber\\
&=&- \frac{m}{4}
%+\frac{(r-m)^2}{16 m} -\frac{(r-m)^3}{16 m^2} + \mO((r-m)^4/m^3)\,,
+\frac{m_{\rm red}}{2}\,\om^2 (r-m)^2 + \dots\;,\qquad
\om =\frac{1}{2m}\,
 \label{HO}
\eea
(recall $m_{\rm red}=m/2$).
Hence, we expect the large-$m$ spectrum near the ground state to be approximately given by
\be
\frac{E_{n+1}}{m} \simeq -\quart + \left(n +\half\right) \frac{1}{2 m^2},\quad n=0,\,1,\,2,\, \ldots\,
\label{EbHO}
\ee
($j=n+1$), with corrections primarily of order $\mO(m^{-4})$ from the terms omitted in (\ref{HO}). These can be substantial because the potential is quite asymmetrical (figure \ref{figpVrm06m2}) with its $1/r$ tail at large $r$ where the true eigenfunctions fall off slower than a Gaussian.
There are also exponentially small corrections due to the fact that the eigenfunctions of this anharmonic oscillator have to vanish at the origin.
The right plot in figure \ref{figpevarII} compares the excitation spectrum near the ground state with (\ref{EbHO}), for $m=10$ ($\rmin=10.1$), using the basis of sine functions. The ground state energy $E_1$ differs only 0.5\% from the $-1/4$ in (\ref{EbHO})  (which may be compared with the $-1/6$ in (\ref{ebnonrunningII})).  The first few eigenfunctions are closely HO-like;
for large $n$ they should become H-like, $\approx u_n(r,\aB)$, but it would require much larger $L$ and $N$ to verify this.

At substantially smaller masses the spectrum near the ground state is neither closely HO-like nor H-like. For $m=2$ ($\rmin = 2.52$),  part of the spectrum is shown in the left plot of figure \ref{figpspecIIm20};
in this case even the first few eigenfunctions are still H-like (right plot).
The $r_{90}$ domain size of the ground state for $m=2$:
\be
\int_0^{r_{90}} dr\, |f_1(r)|^2=0.9\,, \qquad r_{90} \simeq 7\,,
\qquad\qquad \mbox{($m=2$, model-II)}
\label{r90II}
\ee
is somewhat smaller than the 13 in (\ref{r90I}) for model-I; it approaches $1/\om =2m=2\rmin$ for larger masses.

\FIGURE[t]{
 \includegraphics[width=7cm]{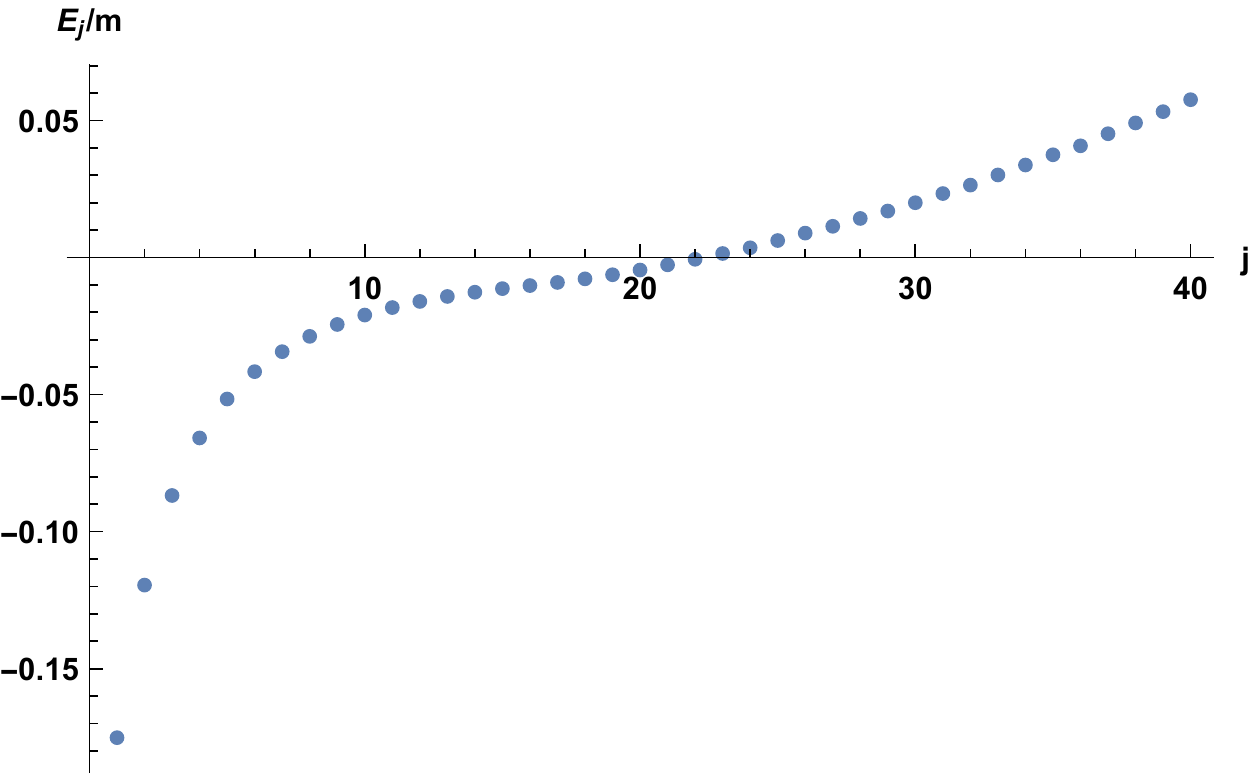}%from binding1loopIISine-m20L210N128b
 \includegraphics[width=7cm]{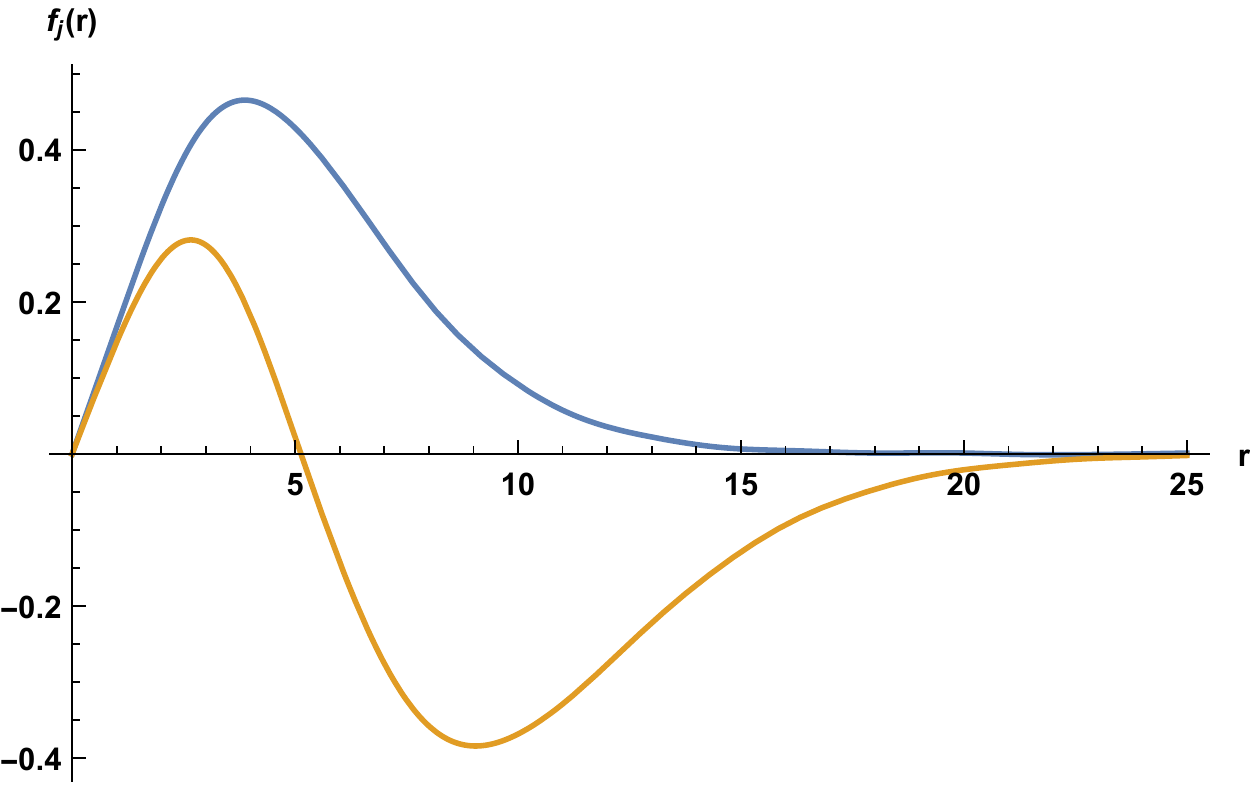}%from binding1loopIISine-m20L210N128b
 \caption{Left: model-II spectrum for $m=2$, $L=211$, $N=128$  ($\lmmin=3.3$); the remaining positive energies increase approximately quadratically. Right: first two eigenfunctions, $j=1,2$. }
 \label{figpspecIIm20}
}

\section{Spherical bounce and collapse}
\label{secbouncecollapse}

Using the spectrum and eigenfunctions obtained with the Fourier-sine basis we study here the time development of a spherically symmetric two-particle state.  Consider a Gaussian wave packet at a distance $r_0$ from the origin, at time $t=0$,
\be
\ps(r,0) = \mu^{-1/2} \exp\left[-\frac{(r-r_0)^2}{4 s_0^2}\right],
\qquad \int_0^{\infty}dr\, \ps(r,0)^2 = 1\,.
\ee
Assuming $r_0$ sufficiently far from the origin and $s_0/r_0$ sufficiently small, $\ps(0,0)$ is negligible such that it qualifies for a radial wave function, and extending the normalization integral to minus infinity
$\mu = 2\pi s_0$. The Fourier-sine basis at finite $L$ and $N$ is accurate (visibly) provided that $L/r_0$ is large enough and the wave packet not too narrow. We can then replace $\ps(r,0)$ by its approximation in terms of the models' eigenfunctions (for model-I these are here normalized under transposition). Using the notation of appendix \ref{appnormal}, let
\bea
\ps_n&=&
\int_0^L dr\, b_n(r)\,\ps(r,0)\,, \\
\ps_j &=&
\int_0^L dr\,f_j(r)\,\ps(r,0)=\sum_{n=1}^N f_{jn} \ps_n\,.
\eea
We now {\em redefine} $\ps(r,0)$,
\be
\ps(r,0) = \mu^{-1/2} \sum_{n=1}^N \ps_n\, b_n(r) =
\mu^{-1/2} \sum_{j=1}^N \ps_j\, f_j(r)\,,
\label{redefineps}
\ee
with $\mu$ such that $\ps(r,0)$ is normalized again,
\be
\sum_{n=1}^N\ps_n^2 = \sum_{j=1}^N \ps_j^2 = 1\,.
\ee
The coefficients $\ps_j$ are real in model-II and complex in model-I (in the latter
$\sum_j \im[\ps_j^2]=0$).
This initial wave function satisfies the boundary conditions at $r=\{0,L\}$ and it
should be an accurate approximation to the original Gaussian.
The time-dependent wave function is given by
\be
\ps(r,t) = \sum_{j=1}^N \ps_j\, f_j(r)\, e^{-i E_j t}\,.
\ee

The case with the pure Newton potential is informative for interpreting the results, as is also the `free-particle' case $V=0$ in $0<r<L$.
In addition to looking at detailed shapes the packet may take in the course of time, quantitative observables are useful: the squared norm $\nu$, average distance $d$ and its root-mean-square deviation $s$ that we shall call {\em spread}:
\bea
\nu(t) &=&||\ps||^2= \int_0^L dr\, |\ps(r,t)|^2\,,
\\
d(t) &=&\langle r \rangle = \nu(t)^{-1} \int_0^L dr\, |\ps(r,t)|^2\, r\,,
\\
s(t) &=& \sqrt{\langle r^2-\langle r\rangle^2 \rangle} =
 \left\{\nu(t)^{-1}\int_0^L dr\, |\ps(r,t)|^2\,\left[r^2-d(t)^2\right]\right\}^{1/2}\,.
\eea
Following the norm is only interesting for model-I with its non-Hermitian Hamiltonian; for the other models (II, Newton, free) it stays put at $\nu=1$.

The free pseudo-particle with reduced mass $m/2$ is not entirely free because of the boundaries at $r=0$ and $L$. As time progresses the wave packet broadens. When it reaches the origin its composing waves scatter back and the average $\langle r \rangle$ increases. Similar scattering starts when the packet reaches $L$. After some `equilibration' time $|\ps(r,t)|^2$ becomes roughly uniform with fluctuations, and
$\langle r \rangle\approx L/2$, $\langle r^2-\langle r \rangle^2\rangle\approx L^2/12$.

Examples in model I and II now follow for mass $m=2$, which implies $\aB=1/4$, $\rs=6.59$, $\rmin=2.52$, with $N=128$, $L=32\, \rs=211$, which implies $\lmmin=2 L/N = 3.29$. These values of $m$, $L$ and $N$  are also used in figures \ref{figpfcnm20L32N128} and \ref{figpspecIIm20}. Let us start with model-II in which the Hamiltonian is Hermitian.

\FIGURE[t]{
\includegraphics[width=7cm]{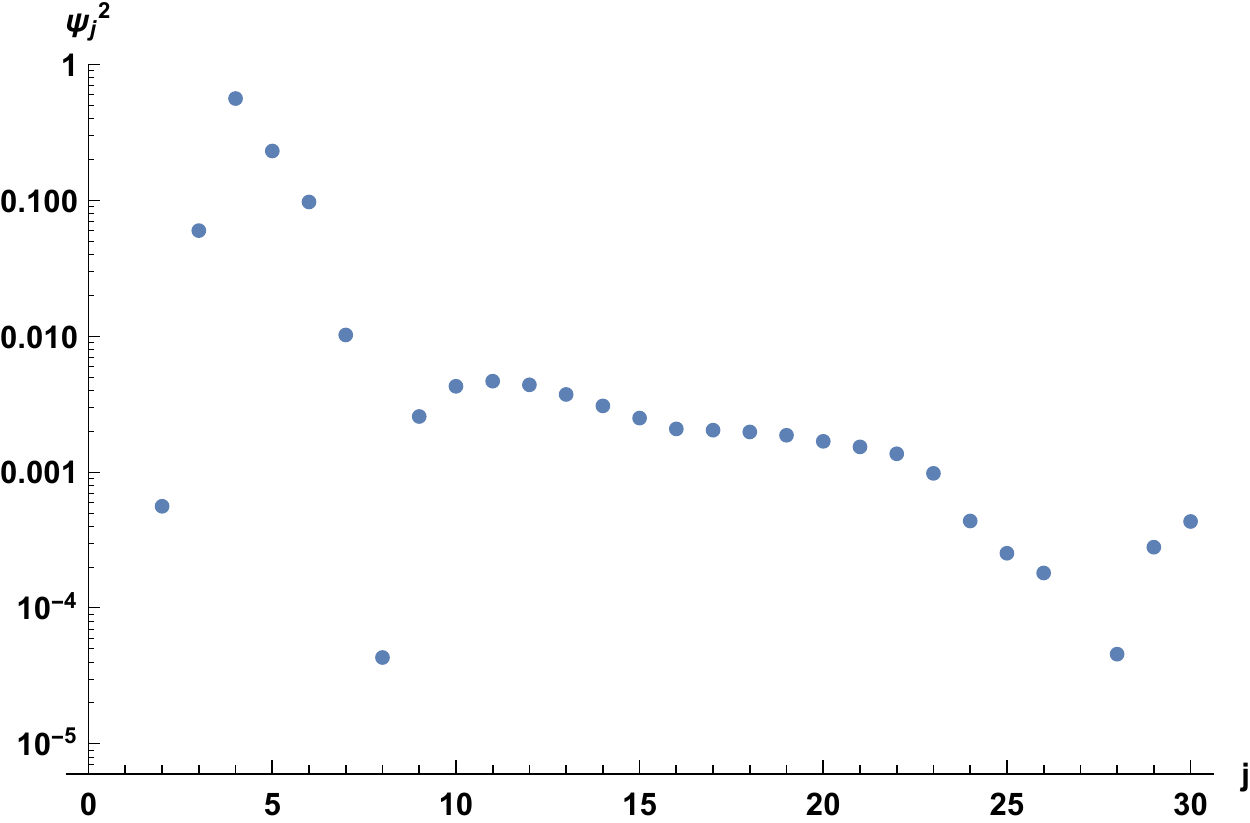}%from timeII-m20L210N128b
\includegraphics[width=7cm]{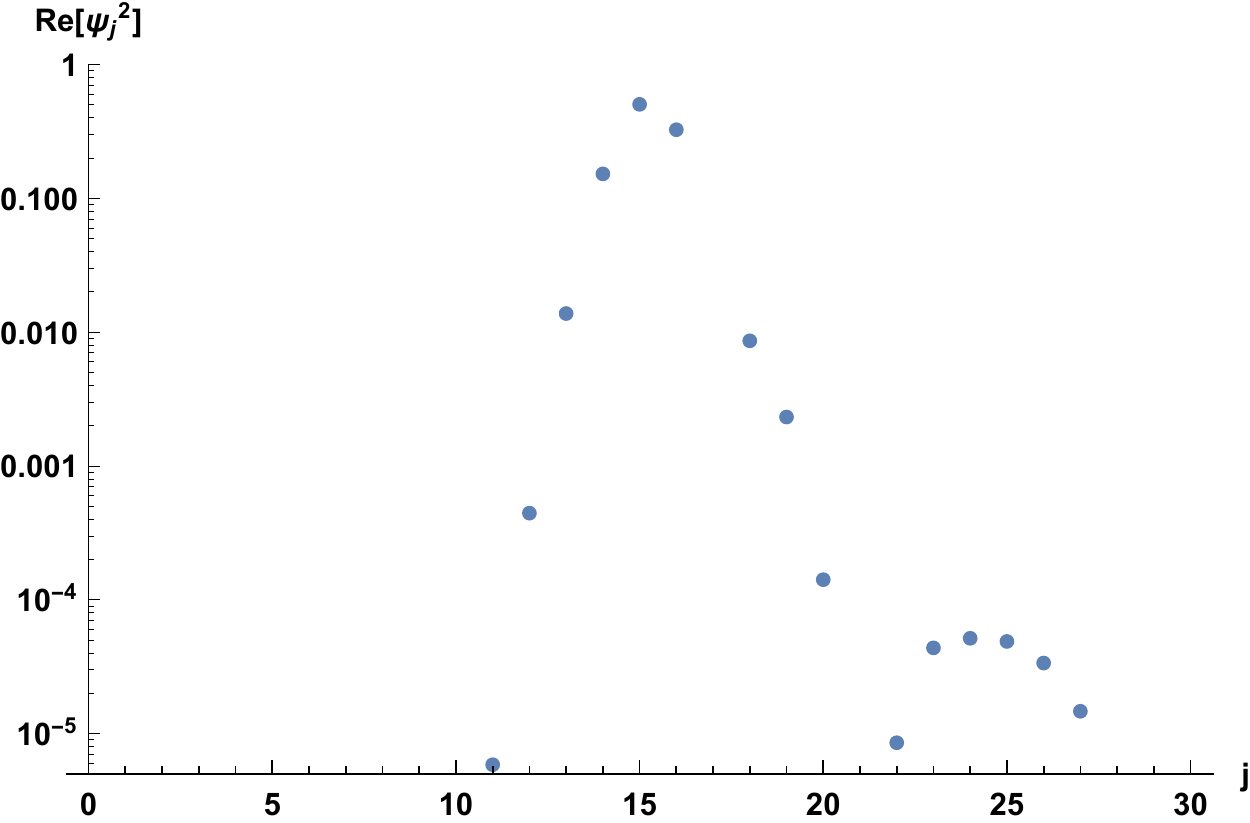}%from time-m20L32N128bb
 \caption{ Left: model-II coefficients $\ps_j^2$ ($\ps_1^2 = 5\times 10^{-7}$). Right: model-I coefficients $\re[\ps_j^2]$, for the first 10 modes these vary from $\mO(10^{-20})$ to $\mO(10^{-8})$.
 }
 \label{figpsfjs}
 }

\subsection{Bouncing with model-II}
\label{secbounceII}

The parameters
of $\ps(r,0)$ are $r_0=10\, \rmin= 25.2$ and $s_0=\rmin = 2.52\,$. With these the initial Gaussian is negligible at the origin and $\lmmin=1.3\, s_0$ turns out to be sufficiently small to enable a reasonably accurate approximation in the basis of sine functions or eigenfunctions $f_j(r)$. Figure \ref{figpsfjs} (left plot) shows coefficients $\ps_j^2$; the dominant modes are $j=4$ and 5. The ratio $r_0/\aB$ is large (100.8) and in the Newton case 39 bound-state $u_n(r,\aB)$ enable a good representation of $\ps(r,0)$. The energy $\langle H \rangle=-0.59\, m$.

\FIGURE[t]{
\includegraphics[width=7cm]{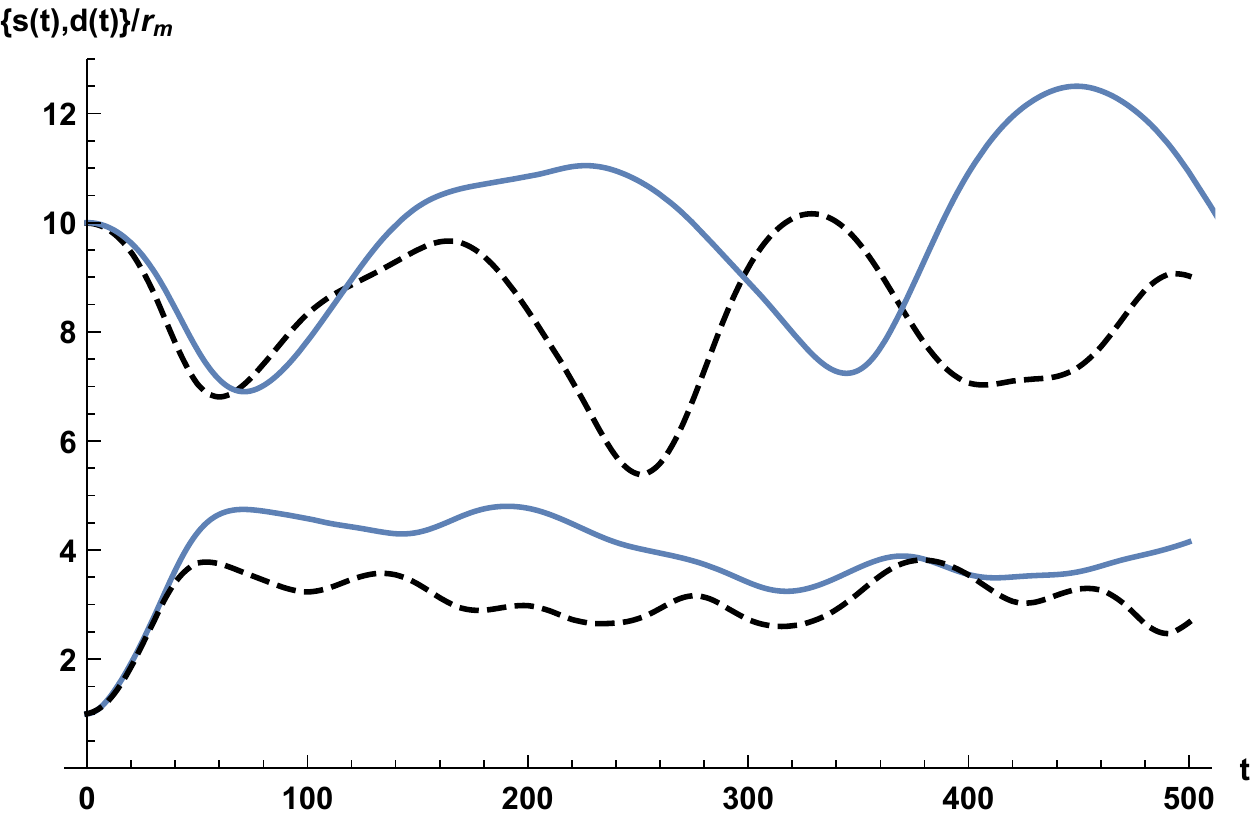}%from timeII-m20L210N128b
\includegraphics[width=7cm]{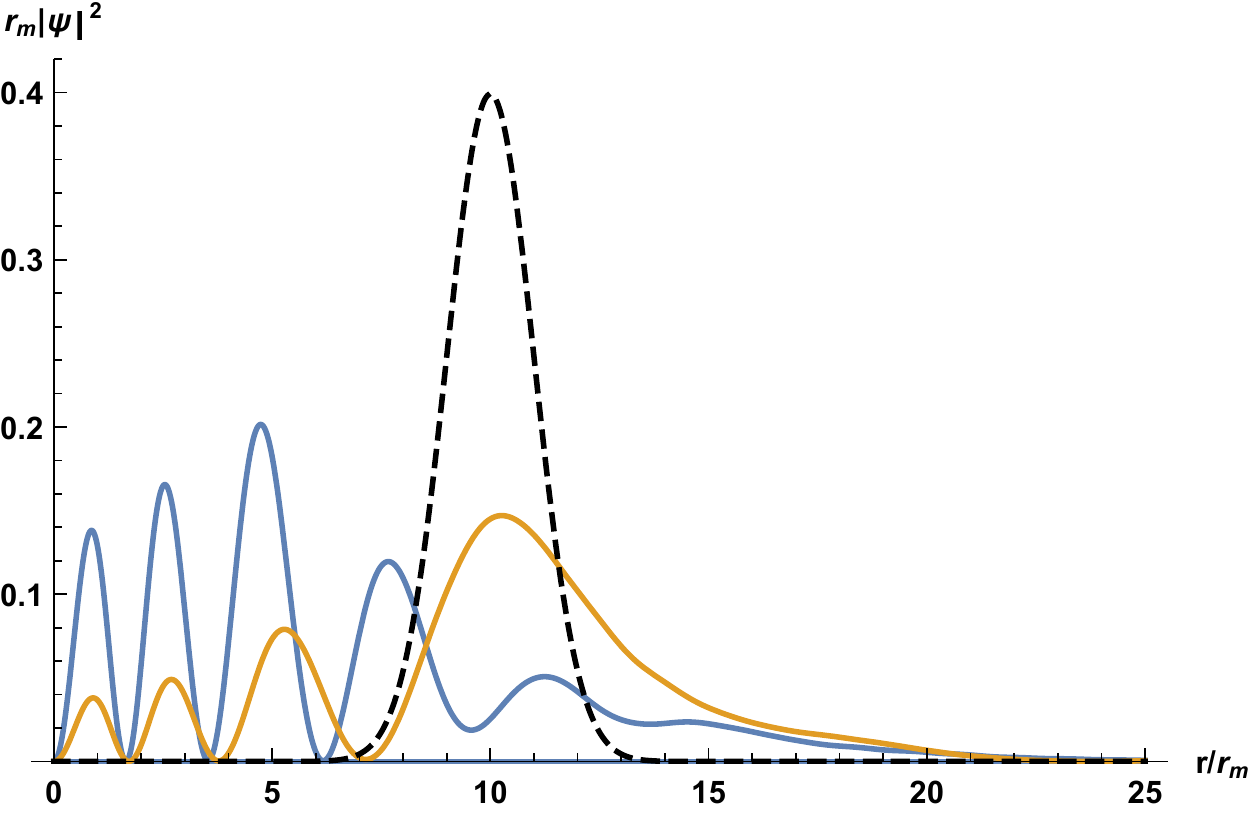}%from timeII-m20L210N128b
 \caption{ Left: $d(t)/\rmin$ (upper curves) and $s(t)/\rmin$ (lower curves) of model-II (fully drawn) and Newton (dashed); $m=2$, $\rmin=2.52$. Right:
 $\rmin |\ps(r,t)|^2$ vs.\ $r/\rmin$ at the time of the first bounce (blue, $t_{{\rm b},1}=73$) and at the time of the first fall-back (brown, $t_{{\rm f},1}= 220$). The initial $|\ps|^2$ is also shown (black, dashed). % $L=64$, $N=128$.
 }
 \label{figpIIbf}
 }

Initially the packet spreads and moves towards the origin with roughly the classical acceleration, then it decelerates and bounces back to a distance near the starting point, after which the process repeats. The left plot in figure \ref{figpIIbf} shows the oscillation of $d(t)$ (upper curves). The initial acceleration $d^{\prime\prime}(t)$ in model-II is smaller than that of Newton which' force is stronger (figure \ref{figpVrm06m2}). The spread $s(t)$ (lower curves) has similar oscilations, its maximum values are much smaller than the free-particle value $\sqrt{L^2/12}=61$ and the scattered wave from the boundary at $L$ is negligible.
The right plot in figure \ref{figpIIbf} shows the packet at the time of the first bounce (minimum of $d(t)$) and at the time of the subsequent fall-back (maximum $d(t)$): $\{t_{{\rm b},1}, t_{{\rm f},1}\}=\{73, 220\}$.
The number 4 to 5 of large maxima
may reflect that $j=4,5$ dominate in the expansion (\ref{redefineps}).

These plots will not change much in the limit $\lmmin\to 0$ or in the infinite volume limit
$L\to\infty$.

\subsection{Bouncing collapse with model-I}
\label{seccollapseI}

The parameters here are that of model-II with $\rmin\to\rs$:  $r_0=10\, \rs = 65.9$ and $s_0=\rs = 6.6$\, (here $\lmmin/s_0=1/2$ and $r_0/\aB=L\simeq 211$). With $r_0$ here larger than in model-II the dominant $\ps_j$ are around $j=15$ (figure \ref{figpsfjs}, right plot); the
energy, $\langle H \rangle = -0.036\, m$ is smaller in magnitude and the time scale on which things change is larger. But the major difference is the imaginary part in the eigenvalues $E_j$, which leads to a rapid decay of all eigenfunctions with a sizable imaginary part, typically those with support in $r\lesssim \rs$ (figures \ref{figpREjm20L32N128} and \ref{figpfcnm20L32N128}).

\FIGURE[h]{
\includegraphics[width=7cm]{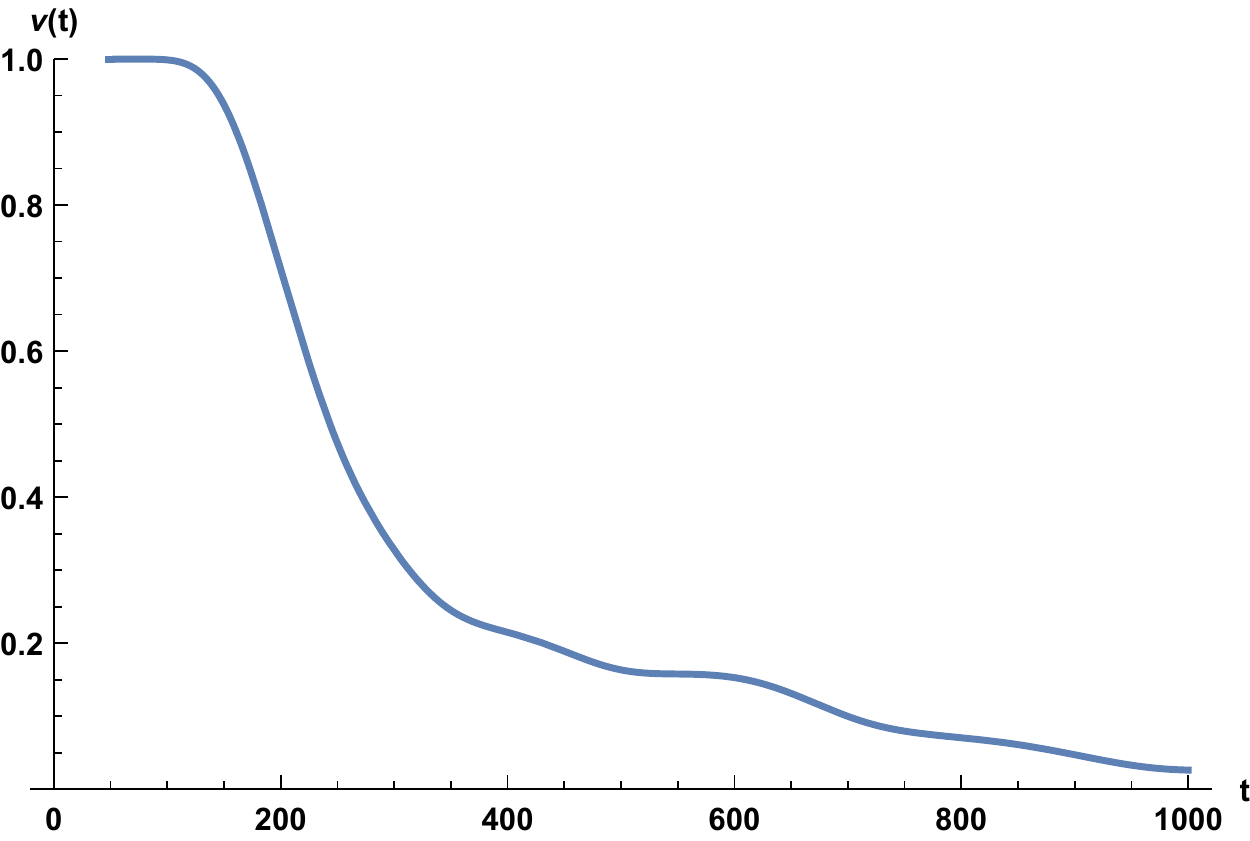}%from time-m20L32N128bb
 \includegraphics[width=7cm]{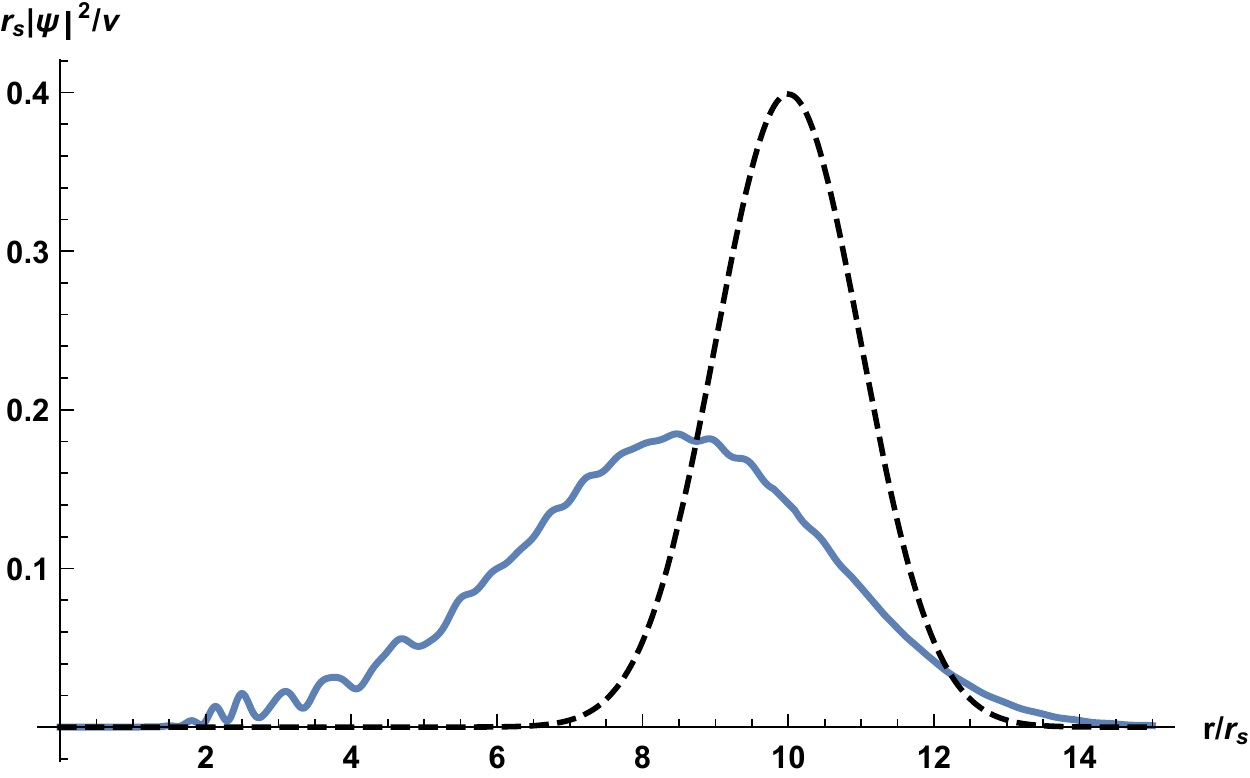}%from time-m20L32N128bb
 \caption{Left: Time-dependence of the squared-norm in model-I. Right: $\rs|\ps(r,t)|^2$ at $t=0$ (dashed), and at $t=124.2$, when $\dot\nu(t)=-0.001$ and $\nu(t)=0.987$. }
 \label{figpInorm}
}

Figure \ref{figpInorm} shows the squared norm $\nu(t)$ (left plot). Up to times of about 100 it hardly changes, the wave packet has not reached the region $r\approx \rs $ yet.
Beyond that the norm starts diving down. The `norm-velocity' $\dot\nu(t)\equiv d\,\nu(t)/dt$ is maximal at $t=201$, $\dot\nu(201) =-0.0053$.  At the earlier $t=124$ this velocity is already -0.001 and although the norm has changed little, $|\ps(r,t)|^2$  has changed quite a lot
as can be seen in the right plot of figure \ref{figpInorm}.

\FIGURE[t]{
\includegraphics[width=7cm]{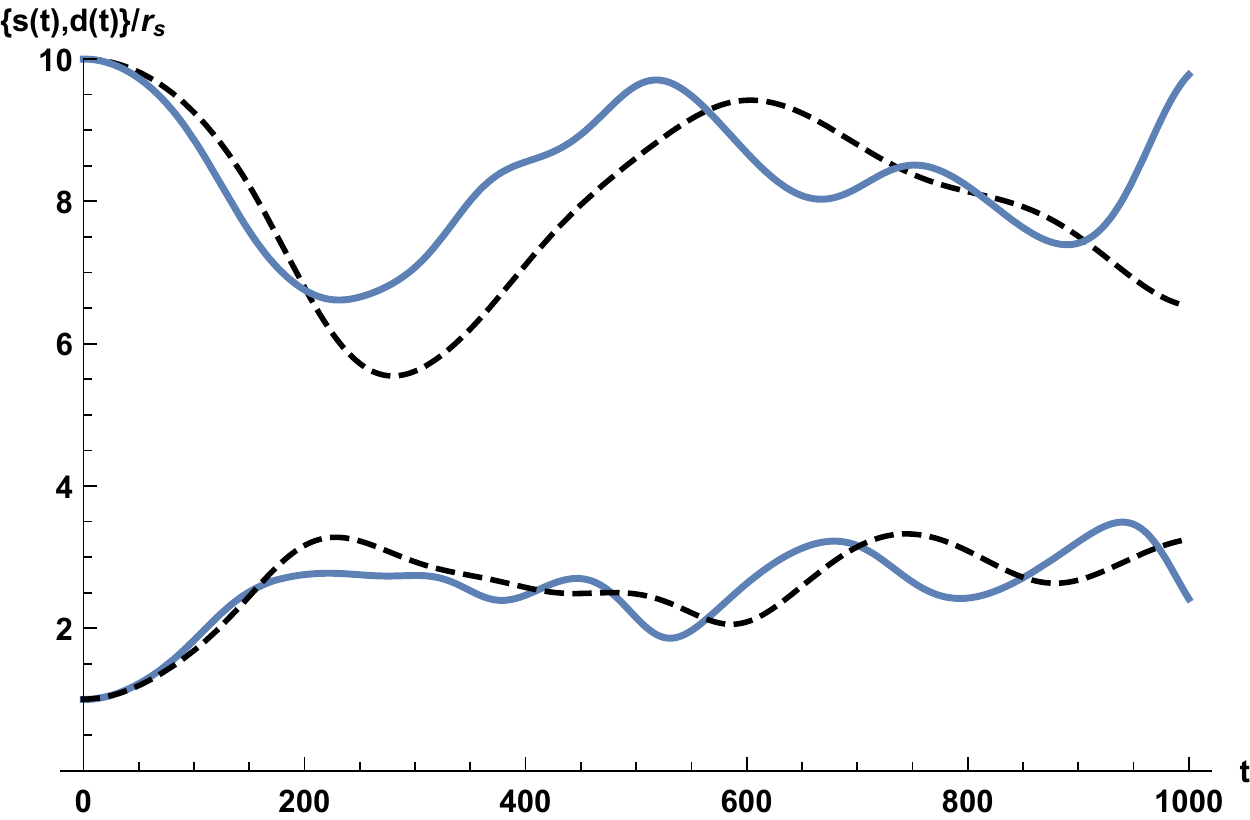}%from timeI-m20L32N128bb
\includegraphics[width=7cm]{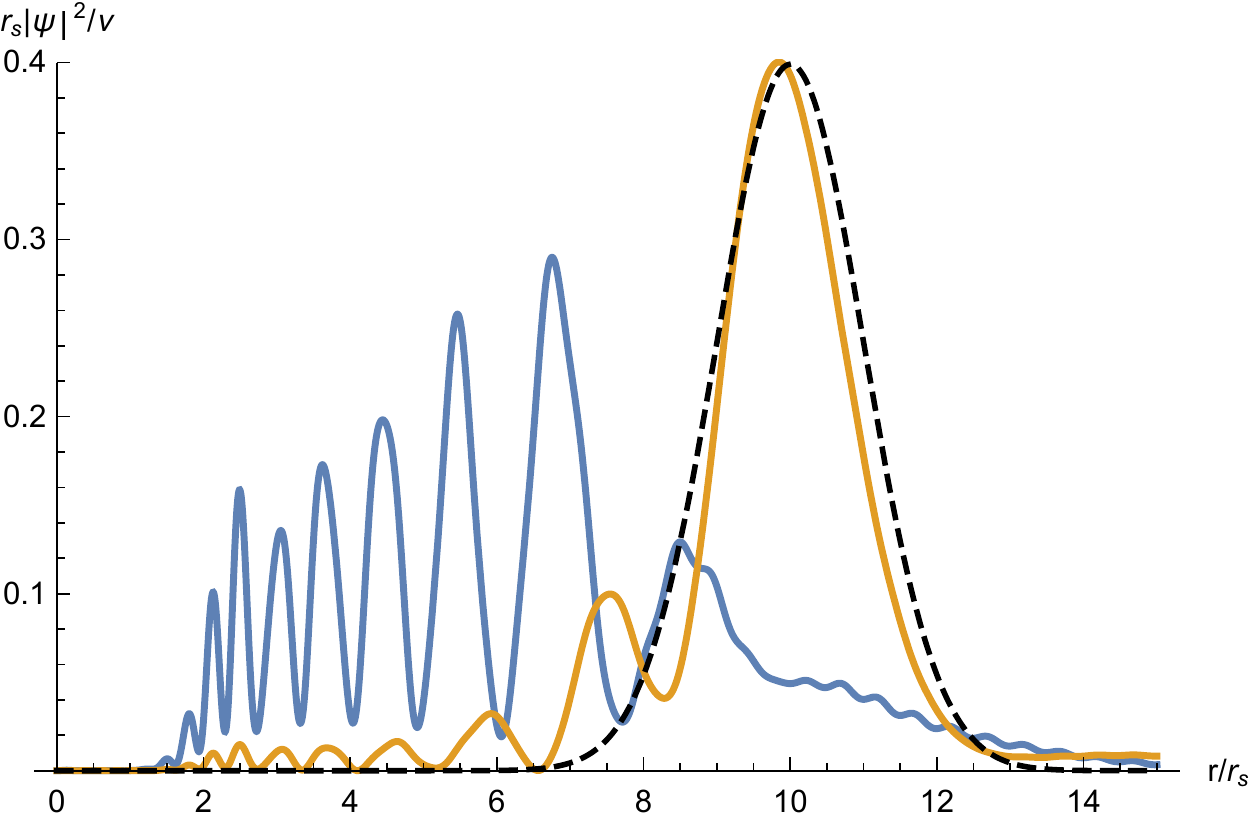}%from timeI-m20L32N128bb
 \caption{As in figure \ref{figpIIbf}, here for model-I; for $m=2$ ($\rs =6.6$); the first bounce and fall-back time are $t_{{\rm b},1}= 231$, $t_{{\rm f},1}=518$. The right plot shows $|\ps(r,t)|^2/\nu(t)$ at $t_{{\rm b},1}$ (blue) and $t_{{\rm f},1}$ (brown). }
 \label{figpIbf}
}

The distance and spread shown in the left plot of figure \ref{figpIbf} display similar bouncing and falling back as for model-II in figure \ref{figpIIbf}. The Newton force is in this case the smaller one. Wave functions at the first bounce and fall-back times are shown in the right plot. A gap in the region $0<r\lesssim \rs$ is clearly visible. Also remarkable is the approximate recovery of the initial shape of $|\ps|^2$ at the fall-back time ($|\ps(r,t)|^2$ in the right plot is `renormalized' by $\nu(t)$, the remaining total probability at the fall-back time is $\nu(518)=0.16$).

Also here in model-I the infinite volume limit $L\to \infty$ at fixed $\lmmin$ will have little effect on $|\ps(r,t)|^2$.
With $\lmmin\to 0$ it is useful to revert to the shifted labeling $\kp=j-\sg$, as in figure \ref{figpspeckappa}. With the anchoring of that plot we expect
the important contributing modes to stay put around the $\kp$ value corresponding to $j=15$ in the right plot of figure \ref{figpsfjs}. For example, in figure \ref{figpspeckappa}, this $\kp=j-\sg=15-4=11$ for the sequence with the same $N$ and $L$ as here (brown dots); the modes with $j\leq 10$, $\kp\leq 6$ are negligible. In the relativistic model $\kpmin=-\infty$ and the contribution of the modes
$\kp=6,\,5,\,4,\, \ldots,\, -\infty$, is also expected to be negligible. In particular, huge negative imaginary parts of energy eigenvalues make all such modes irrelevant after times small compared to the Planck time $\sqrt{G}$.

\section{Revisiting SDT results}
\label{secSDT}

A few lattice details:
configurations contributing to the imaginary-time path integral regulated by the simplicial lattice were generated by numerical simulation, with lattice action $S=-\kp_2 N_2$; $N_2$ is the number of triangles contained in a total number $N_4$ of equal-lateral four-simplices. The bare Newton coupling $G_0$ is related to $\kp_2$ by
\be G_0 = \frac{4 v_2}{\kp_2},
\quad v_2 = \frac{\sqrt{3}\,a^2}{4}\,,
\ee
where $v_2$ is the area of a triangle and $a$ is the lattice spacing (called $\ell$ in \cite{deBakker:1996qf}).
The scalar field was put on the {\em dual lattice} formed by the centers of the four-simplices; the dual lattice spacing $\at = a/\sqrt{10}$. The inverse propagator of the scalar field depends on a bare mass parameter $m_0$.
%\footnote{A more appropriate definition of the bare mass would be $m_{0{\rm c}} =\sqrt{8/5}\, m_0$, as follows by embedding a dual lattice point and its five neighbors in flat Euclidean spacetime.
The `renormalized' mass $m$ was `measured' from the (nearly) exponential decay of the propagator at large distance. The lattice-geodesic-distance between two centers is defined as the minimal number of dual-lattice links connecting the centers, times $\at$.
Not too far away from the phase transition point the propagators  on the dual lattice do not seem to be affected by singular structures or fractal branched polymers.

The numerical simulations were carried out with $N_4=32000$
and two values of $\kp_2$ on either side of, but close to, the phase transition at
$\kp_2^{\rm c}\simeq1.258$:
$\kp_2=1.255$ ($G_0=0.863\,\at^2$) in the crumpled phase and $\kp_2=1.259$ ($G_0=0.860\,\at^2$) in the elongated phase. A way of envisioning the generated spacetimes was suggested by their similarity to a four-sphere (de Sitter space in imaginary time), stemming from a comparison of an averaged volume-distance relation with that of a $D$-sphere of radius $r_0$, up to an intermediate distance, which gave $\{D,r_0\}=\{4.2, 13.4\, \at\}$ and $\{3.7,\,14.2\,\at\}$ respectively at $\kp_2=1.255$ and $\kp_2=1.259$\,.
More local analyses, strictly in $D=4$ dimensions, of such volume-distance relations led to comparisons with four-spheres in the elongated phase and 4D hyperbolic spaces in the crumpled phase \cite{Smit:2013wua}. A factor $\lm$ was proposed that converts the zigzag-hopping lattice-geodesic distance $d_\ell$ to an effective continuum geodesic-distance $d_{\rm c}$ through the interior of the lattice:\footnote{The value of $\lm$ depends somewhat on $\kp_2$ and the lattice size, but much more on its application: the so-called A-fit \cite{Smit:2013wua} is appropriate here for comparison with the exponential decay of the propagators in \cite{deBakker:1996qf}.}
\be
d_{\rm c} = \lm\, d_\ell, \qquad \lm \simeq 0.45\,.
%\qquad (N_4 = 32000)\,.
\label{lambda}
\ee
In the following we use in this section {\em cutoff units} $\at=1$.

\FIGURE[t]{
\includegraphics[width=9cm]{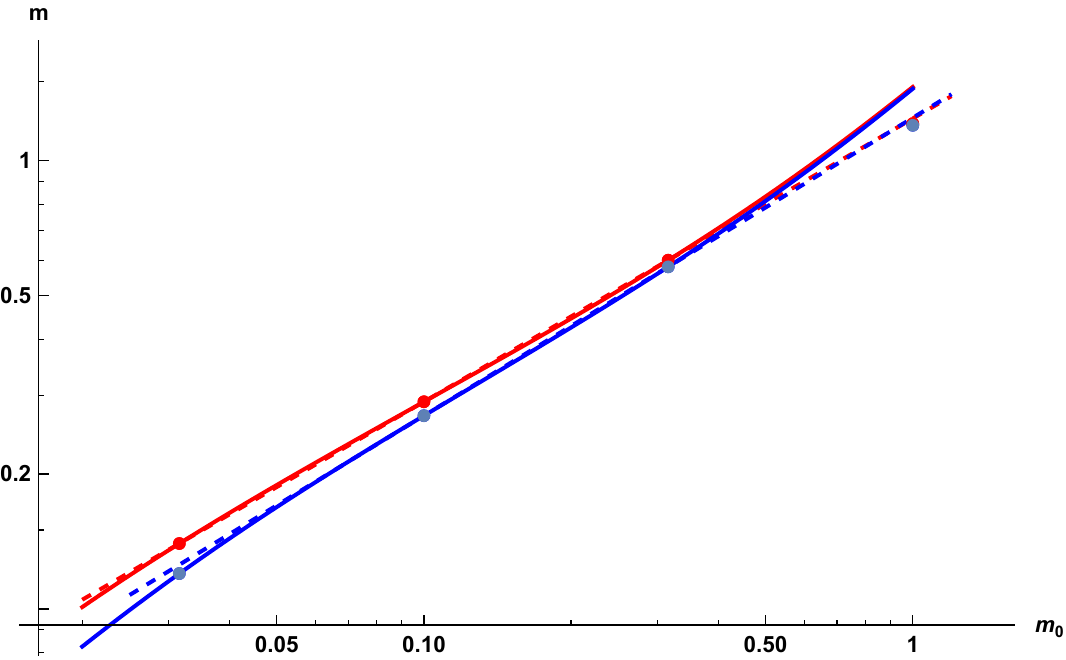}%from bindingNew
 \caption{Renormalized mass $m$ vs.\ bare mass $m_0$. The dashed straight lines are fits of $m=x\, m_0^y$ to only the data points at $m_0=0.1$ and 0.316\,, with $x(1.255)=1.24$, $y(1.255)=0.63$ (upper data, red) and $x(1.259)=1.25$, $y(1.259)=0.66$ (lower data, blue)\,. The curves represent (\ref{mm0rat}) with (\ref{fresults}).
% and (\ref{lPresults}).
Units $\at=1$.}
 \label{figpmm0}
}

From Tables 1 and 2 in \cite{deBakker:1996qf} we find the binding energies and masses:
\be
\begin{array}{lcccllccc}
\kp_2=1.255&m_0&m&\Eb&&\kp_2=1.259&m_0&m&\Eb\\
&0.0316&0.14&0.035(2)&&            &0.0316&0.12&0.019(2)\\
&0.1&0.29&0.064(2)&&               &0.1&0.27&0.038(2)\\
&0.316&0.60&0.078(2)&&             &0.316&0.58&0.053(1)\\
&1&1.21&0.054(1)&&                  &1&1.20&0.045(1)
\end{array}
\label{bindpars}
\ee
It is interesting to focus first on the renormalized mass, which represents in perturbation theory a binding of a `cloud of gravitons' to a bare particle. Based on the shift symmetry of the scalar field action it was argued in \cite{Agishtein:1992xx} that the mass-renormalization should be multiplicative and not additive. A power-like relation compatible with this was noted in \cite{deBakker:1996qf}: $m\propto m_0^y$, with $y=\ln(2.1)/\ln(\sqrt{10})= 0.64$ (the values of $m_0^2$ used in the computation differed by factors of 10). A check on this is in the log-log-plot figure \ref{figpmm0}, where the dashed straight lines are fits to only the intermediate data points at $m_0=0.1$ and $\sqrt{10}=0.316$\,; the lines miss the other data points only by a few percent or less.
Similar fits to all four data points support also remarkably precise power behavior.

However, if the power $y$ stays constant in the limit $m_0\to 0$, this is only compatible with absence of additive renormalization, multiplicative renormalization suggests that $y$ should approach 1 in the zero mass limit. Numerical evidence for this was presented in \cite{Jha:2018xjh} using so-called degenerate triangulations in which finite-size effects are reduced compared to SDT. Estimating by eye, the plots in this work appear compatible for small masses $m_0\leq 0.1$ with a multiplicative relation $m=f\, m_0$, $f\approx 1.8$.

To see whether the numerical results can be interpreted by comparing with `renormalized perturbation theory'
we have calculated the bare $m_0^2$ as a function of the renormalized $m^2$ to 1-loop order in the renormalized $G$ using dimensional regularization in the continuum (cf.\ appendix \ref{appmm0}). Surprisingly, the result comes out UV- and IR-finite:
\be
m_0^2 = m^2 + \frac{5}{2\pi}\, G m^4\,. \qquad \qquad\qquad\qquad\qquad \mbox{(continuum)}
\ee
Transferring this relation to the SDT lattice while keeping the unambiguous nature of its right hand side, only the coefficient of the bare mass $m_0^2$ may be affected by the lattice regularization differing very much from dimensional regularization in the continuum, suggesting that
\bea
f^2 m_0^2 &=&m^2\left( 1 + \frac{5}{2\pi}\,G\, m^2\right)  \qquad\qquad\qquad\qquad\qquad \mbox{(lattice)}
\label{m0sqmsq}
\eea
where $f$ depends on $G_0/a^2$  (or equivalently $\kp_2$) but not on $m_0$ or $m$ (in the currently quenched approximation).\footnote{In SDT, the permutations of the labels assigned to vertices form a remnant of the diffeomorphism gauge-group, which is effectively summed-over in the numerical computations. The renormalized $m$ and $G$ are defined in terms of gauge-invariant observables. In the quenched approximation we can alternatively think of $G$ to be defined by the terms of order $m^2$ and $m^4$ in an expansion of $m_0^2$ vs.\ $m^2$, as in (\ref{m0sqmsq}), and compare with the binding-energy definition.}

Using the renormalized Planck lengths $\ellP=\sqrt{G}$ in (\ref{lPresults}) obtained from the binding energy, a fit of $f$ to the renormalized mass at the smallest bare mass $m_0=0.0316$ gives
%\be
$f(1.255)= 5.25$, $f(1.259) = 4.19\,$.
%\label{fresults}
%\ee
With these $f$ and $G$ the formula (\ref{m0sqmsq}) turns out to describe surprisingly well also the other three masses (within a few percent for the next two larger masses and within 20\% for the largest).
Fitting the data at more masses it is possible to estimate also $\ellP$.
A fit of (\ref{m0sqmsq}) to the renormalized masses at the {\em two} smaller bare masses ($m_0=0.0316$ and 0.1) yields similar values for $f$ and the Planck lengths come out as $\ellP(1.255) = 6.6$, $\ellP(1.259)=5.3\,$, and again (\ref{m0sqmsq}) fares quite well for the two other  masses.
However, at the two fitted masses the factor $1+5 G m^2/(2\pi)$ comes out much larger than 1: $\{1.7,\, 3.9\}$ and $\{1.3,\,2.6\}$, respectively for $\kp_2=1.255$ and $1.259$; for the not fitted masses this factor is even very much larger. The perturbative formula appears to work too well, as if it is nearly exact, which is of course hard to believe.
We avoided this problem by using a rational function representation of the renormalization ratio
\be
\frac{m_0^2}{m^2} = \frac{1}{f^2}\, R(m^2),\qquad
R(m^2) = \frac{1+ p\, m^2}{1+q\, m^2}\,,
\label{mm0rat}
\ee
and identified $G$ from the expansion
\be
R(m^2)= 1+(p-q)m^2 + (q^2-p q)m^4 + \cdots\,, \qquad
G=(2\pi/5)(p-q)\,,
\label{GfromR}
\ee
in which we can think of the $\mO(G m^2)$ term as applying to very small masses. Fitting $R(m^2)/f^2$ to the renormalization ratio of the {\em three} smaller masses gives
\bea
\{f,\,p,\,q\} &=& \{6.2,\,55.7,\,2.6\},\quad \ellP=8.1\,, \qquad \kp_2=1.255\,,
\nonumber\\
\{ f,\,p,\,q\} &=& \{4.5,\,32.9,\,3.0\},\quad \ellP=6.1\,, \qquad \kp_2=1.259\,.
\label{fresults}
\eea
%At $\kp_2=1.255$  the Planck length is rather large and suggests strong finite-size effects may be at play.
The fit is shown in figure \ref{figpmm0} as $m$ versus $m_0$ (which can be obtained easily in exact form from (\ref{mm0rat})). In Planck units the resulting renormalized masses $m\ellP$ are given by
\be
\begin{array}{llcllccc}
\kp_2=1.255&m_0&m \ellP&&\kp_2=1.259&m_0&m\ellP&\mbox{(from mass renormalization)}\\
&0.0316&1.13&&               &0.0316&0.736&\\
&0.1&2.35&&               &0.1&1.66&\\
&0.316&4.86&&             &0.316&3.56&\\
&1&9.80&&               &1&7.36&
\end{array}
\label{mGren}
\ee
They are in the intermediate to large mass regions of models I and II.

\FIGURE[t]{
\includegraphics[width=7cm]{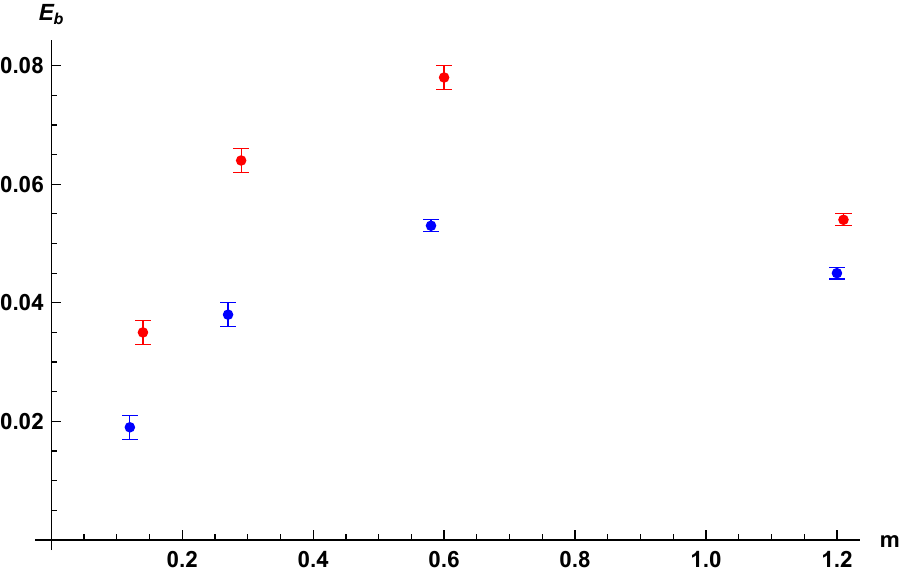}%from bindingNew
\includegraphics[width=7cm]{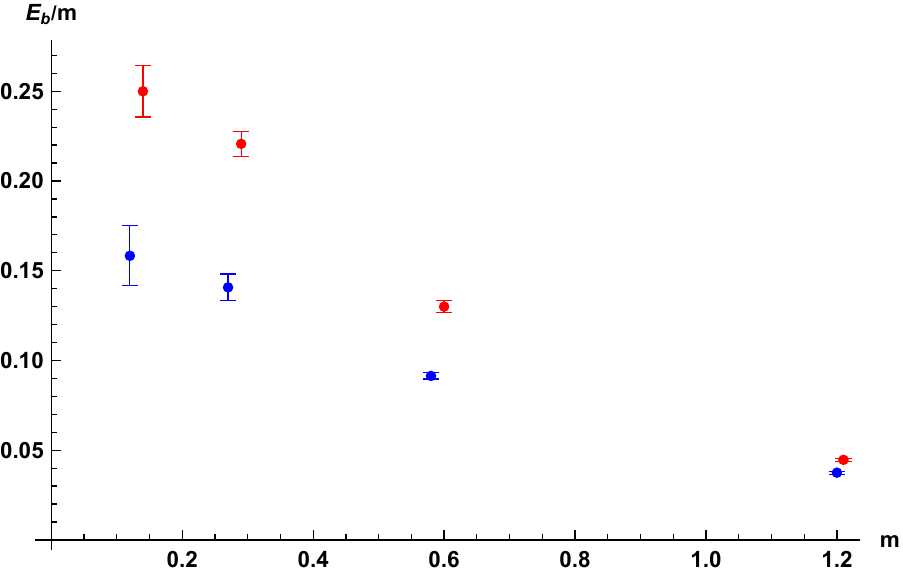}%from bindingNew
 \caption{Left: Binding energy vs.\ $m$\,. Right: Binding-energy ratio  $\Eb/m$  vs.\ $m$\,. Upper data (red): $\kp_2=1.255$\,, lower data (blue): $\kp_2=1.259$\,.
 %double-logarithmic plots of $\Eb/m$ vs.\ $m\ellP$ obtained with the  `minimal Ansatz fit' using only data at $m_0=0.1$ and 0.316\,. Upper data and curves (red): $\kp_2=1.255$\,, lower (blue): $\kp_2=1.259$\,. Downward shifted smallest mass data are indicated by blank spots on the curves.
 }
 \label{figpebmm}
}

For the mass renormalization, the data at the smallest bare mass appears to still make sense when neglecting finite-size effects.
Yet, there are good reasons to distrust the {\em binding energy} data at $m_0=0.0316$, and at $m_0=1$: the renormalized mass of the first is too small for a reasonable determination of binding energies on the distance scale of the simulations,\footnote{In figure 4 of \cite{deBakker:1996qf}, the propagators show exponential falling at large distances for all masses, but the effective $\Eb(r)$ of the smallest mass in figure 5 lacks a stationary region as for the other masses ($\kp_2$ is the same in both figures.)}
and the renormalized mass of the second is so large that strong lattice artefacts are to be expected. There is no reason to suspect the data at the other two mass values.
The left plot in figure \ref{figpebmm} shows the binding energy versus the renormalized masses.
At the largest mass they have dropped, which seems odd. The ratio $\Eb/m$ in the right plot of figure \ref{figpebmm} shows an almost linear behavior. But a linear extrapolation of the first (left) three points towards $m=0$ would give silly physics, since one expects that $\Eb/m$ vanishes rapidly as $m$ goes to zero.  These plots strengthen our suspicion of the binding energy data at the smallest (and largest) mass.

Assuming that the SDT data for the two intermediate masses ($m_0=0.1$ and 0.316) can be connected with the Newtonian behavior $\Eb/m\to G^2 m^4/4$ as $m\to 0$, consider fitting them by functions of the form
\be
\frac{\Eb}{m}= F_n(\mlp) =
\frac{(m\lP)^4}{4 P_n(m\ellP)}\,,\quad
P_n(x) =1+\sum_{k=1}^n c_k\,  x^k\,,
\label{fitSDT}
\ee
in which $\lP$ is the renormalized Planck length in lattice units. With $n=5$ the polynomial in the denominator can implement the trend of $\Eb/m$ falling with increasing $m$. Without further input the minimal Asatz
$P_5(m\ellP)=1 + c_5 (m\ellP)^5$ leads to the fit:
\bea
\{\ellP, c_5\} &=&\{5.10, 0.625\}, \qquad \kp_2=1.255\,,
\nonumber\\
&=&\{4.37, 1.07\}, \qquad \kp_2=1.259\,.
 \label{lPresults}
\eea
The ordering in magnitude,
$\ellP(1.255) > \ellP(1.259)$ follows that of the bare Planck lengths
$\ell_{\rm P0}(1.255)  = 0.9287$, $\ell_{\rm P0}(1.259)  = 0.9273$\,(this applies also to (\ref{fresults})).
The fits are shown in figure \ref{figpebmmfitA}.
Implicit in the form of the fit function is the assumption that sufficiently to the left of its maximum it represents continuum behavior of $\Eb/m$ on huge lattices with negligible finite-size effects.

\FIGURE[t]{
\includegraphics[width=9cm]{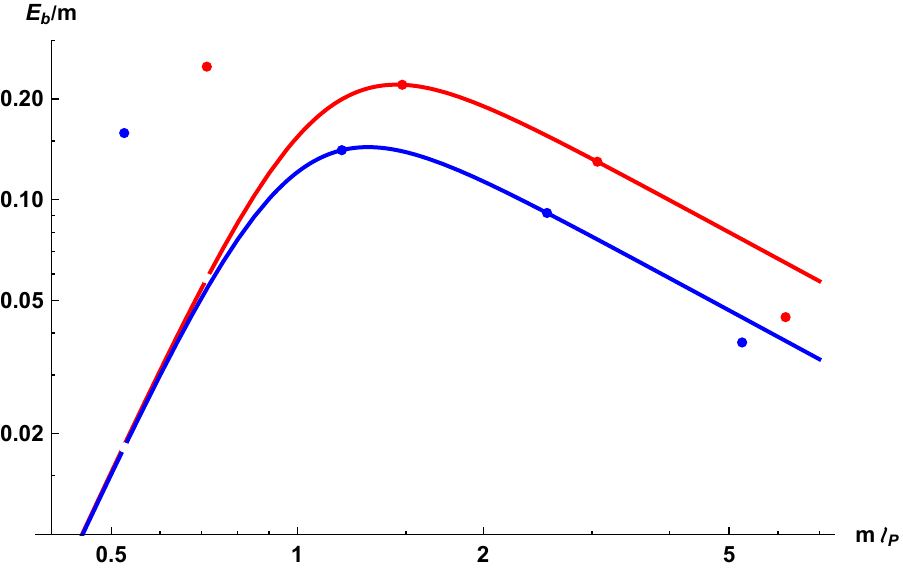}%from bindingNewHelp
 \caption{
Double-logarithmic plot of $\Eb/m$ vs.\ $m\ellP$ obtained with the minimal-Ansatz fit (\ref{lPresults}) which uses only data at $m_0=0.1$ and 0.316\,. Upper data and curves (red): $\kp_2=1.255$\,, lower (blue): $\kp_2=1.259$\,. Downward shifted smallest mass data are indicated by blank spots on the curves.
 }
 \label{figpebmmfitA}
}

In the `minimal Ansaz' fit (\ref{lPresults}) the renormalized masses come out in Planck units as
\be
\begin{array}{llcllccc}
\kp_2=1.255&m_0&m \ellP&&\kp_2=1.259&m_0&m\ellP&\mbox{(from $\Eb/m$ minimal Ansatz)}\\
&0.0316&0.71&&               &0.0316&0.52&\\
&0.1&1.5&&               &0.1&1.2&\\
&0.316&3.1&&             &0.316&2.5&\\
&1&6.2&&               &1&5.2&
\end{array}
\label{mG}
\ee

The smallest renormalized masses (left out of the fits, $m_0=0.0316$) are not particularly small in Planck units---more like in the intermediate mass region of models I and II. Their $\Eb/m$ ratios are also shown in figure \ref{figpebmmfitA}.
The blank spots on the curves indicate values they should have on huge lattices, assuming the curves are right. These shifted values seem rather small, too different from the actual (although distrusted) numerical data. One would like to take into account also the smallest mass $\Eb/m$ data, somehow. Including it in a chi-squared fit does not work: the resulting curves turn out to go practically through the {\em smallest} mass point while missing the other two by several standard deviations. This puts the minimal Ansatz into question.
Using a modified $P_5(m\ellP) = 1+c_4 (m\ellP)^4 + c_5 (m\ellP)^5$ in a three parameter fit ($c_1$, $c_2$ and $\ellP$) leads to satisfactory looking fit curves, with  $\ellP(1.255) =9.9$, $\ellP(1.259)\simeq 11$. However, these are rather large Planck lengths, which moreover violate the ordering $\ellP(1.255) > \ellP(1.259)$. As good compromise is found to be: fix $\ellP$ by the mass renormalization results (\ref{fresults}) in a two-parameter ($c_1$ and $c_2$) chi-squared fit to the three smaller mass data; then
\bea
\{\ellP, c_4, c_5\} &=&\{8.1, 0.179, 0.368\}, \qquad \kp_2=1.255\,,
\qquad\qquad \mbox{(mixed fit)}
\nonumber\\
&=&\{6.1, 0.589, 0.603 \}, \qquad \kp_2=1.259\,.
 \label{c12fixelP}
\eea
The results are shown and compared with the minimal Ansatz fits in figures \ref{figAren255} and \ref{figAren259}. The left plots show $\Eb/m$ versus the bare masses. We see that for $\kp_2=1.255$ the downward shift of the minimal mass data to the curves has reduced to an acceptable extent; for $\kp_2=1.259$ it is still substantial.

\FIGURE[t]{
\includegraphics[width=7cm]{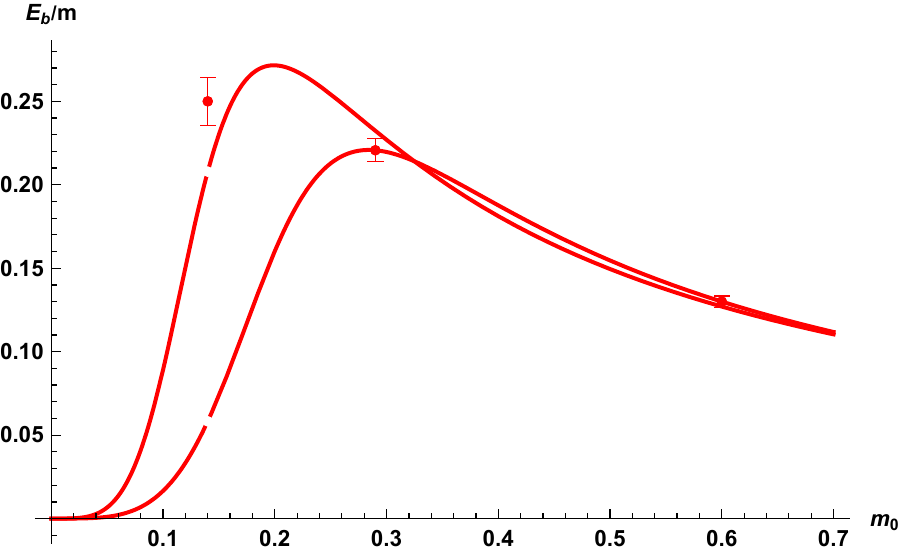}%from bindingNewHelp
\includegraphics[width=8cm]{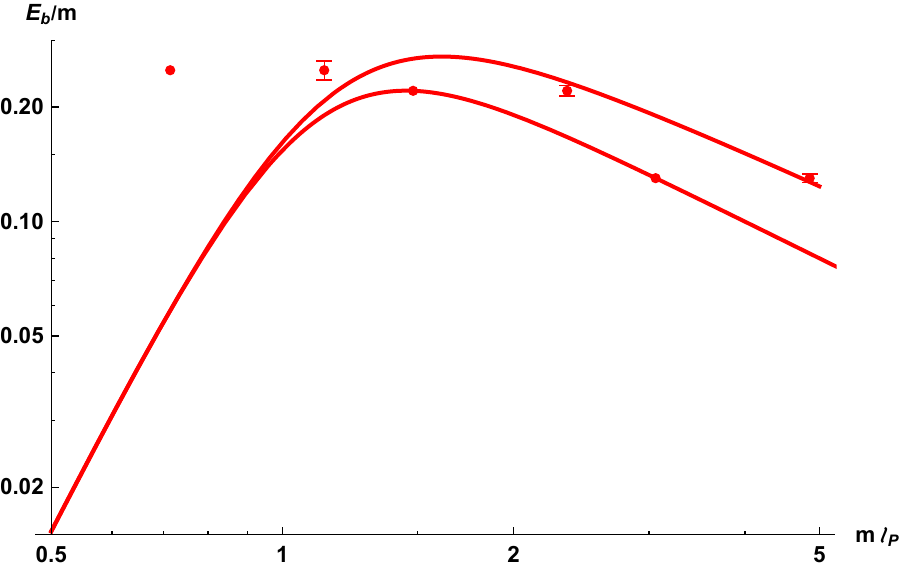}%from bindingNewHelp
 \caption{Plots showing results of the mixed fit (\ref{c12fixelP}) and the minimal-Ansatz fit (\ref{lPresults});  $\kp_2=1.255$. Left: $\Eb/m$ vs.\ the {\em bare mass} $m_0$  (upper curve left of the maxima: mixed fit). Downward shifted smallest mass data are again indicated by blank spots on the curves. Right: double-logarithmic plot. Error bars have been left out in the minimal-Ansatz case (lower curve right of the maxima) for easier recognition of corresponding data points.
 }
 \label{figAren255}
}

\FIGURE[t]{
\includegraphics[width=7cm]{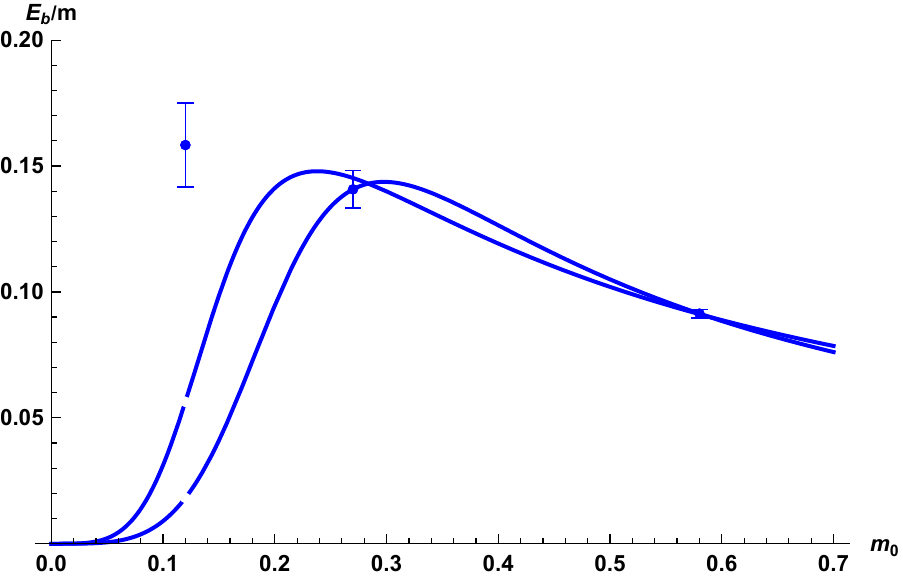}%from bindingNewHelp
\includegraphics[width=8cm]{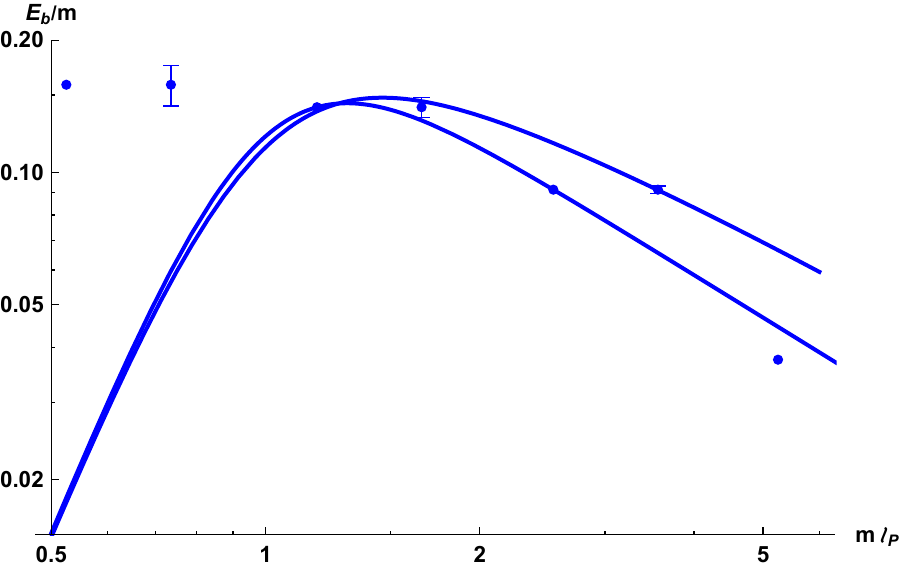}%from bindingNewHelp
 \caption{As in figure \ref{figAren255}, here for $\kp_2=1.259$.
 }
 \label{figAren259}
}

Applying the conversion factor (\ref{lambda}) to the Planck lengths,
their continuum version is $\ell_{\rm P,c}=\lm\, \ellP$\,:
\be
\ell_{\rm P,c}(1.255)= 3.6  =  1.15\,a\,,\quad
\ell_{\rm P,c}((1.259)= 2.8 =  0.87\,a\,.
\label{ellPc}
\ee

The two intermediate masses in the minimal Ansatz fits are already in the large-mass region of models I and II. Inspired by these models for interpreting the data here, realizing fully well that the jump from the continuum to SDT is a big one,
we recall the size of the bound states, $13\,\ellP$ and $7\,\ellP$ respectively in model-I ($\lmmin=3.3\,\ellP$) and model-II, for
$m \ellP=2$ (cf.\ (\ref{r90I}) and (\ref{r90II})).
These bound-state sizes are similar to half the circumference of the above-mentioned four-spheres approximating the average SDT spacetimes, e.g.\  for $\kp_2=1.259$, $r_0\pi \approx 45\, \approx 10\, \lP$\,. Hence, the SDT bound-state could well be squeezed---suffering from a finite-size effect---which raises the energy and {\em lowers} $\Eb$.
In models I and II the size of the bound states {\em increases} with {\em increasing} $m$ (since $\rs$ and $\rmin$ are roughly proportional to $m$) and such squeezing may explain the curious lowering of $\Eb/m$ with increasing $m$, here in SDT.

Since models I and II illustrate such different possibilities as, respectively, a potential singular at $\rs>0$, and a slowly varying potential with a minimum at $\rmin>0$ reflecting a running coupling with a UV fixed point,
%both at a Schwarzschildian distances,
it is interesting to compare with SDT some more of their qualitative features:

\begin{enumerate}
\item At small $m\ellP$ in both models,  the position $\rmax$ of the maximum of the ground-state wave function's magnitude $|f_1(r)|^2$ is near the Bohr radius $\aB=2\ellP/(m\ellP)^3$; for $m\ellP=0.1$ this is about $2000\, \ellP$ (!). As $m$ increases $\rmax$ first goes to a minimal value after which it increases asymptotically $\propto m$. In model-I the scale of $\rmax$ is then set by $\rs=3m\ellP^2$, in model-II by $\rmin = m\ellP^2$.

\item In the small mass region $\mlp\lesssim 0.4$ the ratio $\Eb/m$ in model-I increases faster with $m$ than the Newtonian $(m\ellP)^4/4$; in model-II this increase is slower.

\item
    %In model-I with a UV cutoff  $\Eb/m$ tends to grow rapidly already at intermediate masses $m\ellP \gtrsim 0.5$.
    Different implementations of a UV cutoff in the relativistic model-I imply different versions of the model. Binding energies for minimum wavelength cutoffs $\lmmin/\ellP=1$, 3.3 and 19.8 were shown in figure \ref{figfixedlmmin}.
    %in the region $m\ellP \gtrsim 0.5$.
    Effects of the singularity come to the fore when $\lmmin \ll \rs$, i.e.\ $\lmmin/\ellP \ll 3 m\ellP$. Typically, $\Eb/m$ rises rapidly above 1 when $m\ellP$ increases beyond 0.6 and at large masses it increases $\propto m^2$.

\item Model-II's intermediate mass region is rather broad (figure \ref{figpevarII}) and the ratio $\Eb/m$ rises slowly to a limiting value 0.25, a value {\em much} smaller than typical in model-I.
\end{enumerate}

%Point 1 is illustrated in figure \ref{figamin} in appendix \ref{appIIISDT}.
Above we came to the major conjecture in this paper is that the  decrease of $\Eb/m$ with increasing mass in our SDT results is due to finite size effects getting larger as a reflection of the Schwarzschild scale. This incorporates large mass aspects in point 1.
For point 2: expansion of the fit function, $F_5(m\ellP) = (1/4) (m\ellP)^4 (1-c_4 (m\ellP)^4 - c_5 (m\ellP)^5 + \cdots)$, shows that model-II is favoured since $c_{4,5}>0$.
Point 3:
With the Planck lengths $\ellP \approx 6$ and 8 of the mass-renormalization fit (\ref{fresults}), $m\ellP > 1.5$ for the two intermediate masses in (\ref{mGren}). On the dual lattice the minimal wavelength is $2\at=2$; hence $(\lmmin/\ellP)_{\rm lat}\lesssim 1/3$.  Then $m\ellP$ is large enough for the intermediate masses to satisfy $\lmmin/\ellP \ll 3 m\ellP$. When comparing with model-I features the case $\lmmin=19.8$ is not relevant. For $\lmmin/\ellP=3.3$ the ratio $\Eb/m$ shoots up rapidly when $m\ellP\gtrsim 4$, way beyond the values here in SDT study. Smaller $\lmmin$ gave binding-energy ratios which are already in the intermediate mass region orders of magnitude larger than found in SDT. There is no indication of model-I behavior in the SDT data.
%Model-I behavior may have slipped through the SDT data points but one would still expect the typical magnitude of $\Eb/m$ to be larger than 1.
Point 4: the magnitude of the numerical $\Eb/m$ is smaller than $0.25$,   as in model-II.

All in all the explorative SDT data are compatible with model-II behavior,  not with model-I's.

\section{Summary}
\label{secsummary}

%(Planck length $\ellP=G^{1/2}$ and Planck mass $\mPl=G^{-1/2}$ will be used were convenient).

In the previous sections, binding energies in model I and II were found to depend very much on whether $m$ is in a small-mass region or a large-mass region. At very small masses it approaches the Newtonian form $\Eb=G^2 m^5/4$.
Requiring that perturbative one-loop corrections are smaller than the zero-loop value
gives $m/\mPl\lesssim 0.54$ ($0.66$) in case I (II).\footnote{The numbers
depend logarithmically on a UV-cutoff length $\ell$ in the potential, which was chosen equal to the Planck length (cf.\ (\ref{Ebonesmall})).} This gives an idea of where the small mass regions end.
%Exceptionally, the non-relativistic binding energy $E_{b \rm N}= G^2 m^5/4$ applies to {\em all} masses with the Newton potential, but the average squared-velocity $v^2 \propto m^4\gg 1$.
The relativistic Newton model turns out to have no ground state, $\Emin = -\infty$, for
$m> \mc \simeq 1.3\, \mPl$, and the average squared-velocity
%$\vrel^2\to 1$
approaches one when $m\uparrow \mc$.

Model-I 's energy eigenvalues have a non-vanishing imaginary part, a probability decay rate $\Gm = -2\,\im[E]$\,. In the small-mass region, the decay rate of the ground state
%$\Gm_{\rm b}\simeq 1.5\, G^{11/2}\, m^{12}$.
$\Gm_{\rm b}\propto G^{11/2}\, m^{12}$  (cf.\ (\ref{Gm})).
%It depends on the presence of both classical and quantum terms in the beta function, without either of them it vanishes.
In the large-mass region the binding energy $\Eb\equiv -\re[\Emin]$ is huge in the non-relativistic model,
$\Eb \propto G^4 m^9$, $\Gm_{\rm b} \propto G^3 m^7$ (figure \ref{figpevarI} and appendix \ref{appG}). The {\em relativistic} model-I lacks a ground state for $m\gtrsim 0.61\,\mPl$\,,
and this number certainly represents the end of its  small-mass region. With a UV cutoff on the derivative of the wave function $\Eb$ is finite.
With a minimum wavelength $\lmmin$, $\Eb$ and $\Gm_{\rm b}$ approach infinity as $\lmmin\to 0$.
For $\lmmin= \ellP $ and $m\gtrsim 2\, \mPl$, the binding energy is even close to the non-relativistic one (figure \ref{figfixedlmmin}). Peculiar undulations occur in the mass dependence of $\re[\Emin]$, which are accompanied by a wildly varying $\im[\Emin]$ (figure \ref{figPlanck}). At fixed $\lmmin$, $\Eb\propto m^3$ for large masses.

Model-I's ground state $|\mbox{eigenfunction}|^2$ peaks near $\rs\simeq 3 G m$  in the large-mass region. Eigenfunctions with a large decay rate have their domain in the inside region,
$r\lesssim \rs$, and they are not excited when an initial wave packet does not penetrate this region. In the study of collapse (figure \ref{figpInorm}) during bouncing  (figure \ref{figpIbf}) the time scales stem from the excited modes\footnote{Mode numbers $j$ around 15 in the right plots of figures \ref{figpsfjs} and  \ref{figpREjm20L32N128}.} which have decay rates $\Gm\lll \mPl$.
After dividing out the absorbtion effect on the norm of the wave function, the bouncing is for $m=2\,\mPl$ qualitatively similar to the Newton case.

Model-II has a singularity-free potential with a minimum at a finite distance $\rmin$ that increases with $m$; at large masses $\rmin\simeq G m$ and $V_{\rm min} \simeq -m/4$. Its non-relativistic and relativistic versions differ only substantially in an intermediate mass region ($\lesssim 30$\% for $\Eb/m$). The ground state $|\mbox{eigen function}|^2$ is large near $\rmin$ and the hydrogen-like spectrum in the small-mass region changes slowly to that of an anharmonic oscillator at large masses, where $\Eb/m\to 1/4$, a value much smaller than typically in model-I.
For $m=2\,\mPl$ its bouncing behavior of an in-falling wave packet appears to deviate somewhat more from the Newton case than model-I (figure \ref{figpIIbf}).

Model-I shares the absorbtion effect with black holes.
%The interpretation of the model as representing a sort of black hole is also supported by
The {\em classical} motion in the classical-evolution (CE) models (in which the quantum term in the beta function is neglected) can be extended through the singularity into one of perennial bouncing and falling back (appendix \ref{appbounceCEN}). In the CE-I case, the relativistic velocity of particles falling-in from a distance $r_0>\rs$  reaches that of light at $\rs$. (In the relativistic Newton model the particles also reach the light  velocity, but only strictly at the origin where they may pass each other---the model has no inside region.)
In case II both properties are absent (no absorbtion and $\vrel <0.46$ even when falling in from infinity). The model still shares the interesting possibility of quantum physics at macroscopic distances $\propto m$ where the bound-state wave function is maximal. Since the potential in both models is regular at the origing they show features similar to `regular black holes' \cite{Frolov:2016pav,Bonanno:2020fgp}.

In reanalyzing the SDT results, the data at the largest renormalized mass was not used since one expects its value
%($m>1$ in units of the inverse dual lattice spacing)
to cause large lattice artefacts. The remaining mass-renormalization results were compared to a formula derived from renormalized perturbation theory to order $G$ and adapted to the lattice. The formula described the results surprisingly well, too good to be believed and it was therefore re-interpreted as the $\mO(G)$ term in the expansion of a phenomenological function fitted to the data. This led to an estimate of the renormalized Newton coupling from mass renormalization.

The binding-energy results at the smallest mass were treated with caution since their determination in \cite{deBakker:1996qf} is not convincing. Discarding them initially, phenomenological fits with the Newtonian constraint $\Eb/m\to G^2 m^4/4$ as $m\to 0$ led to estimates of $\sqrt{G}$ somewhat smaller than the ones from mass-renormalization. Treating the latter as fiducial values in improved fits which included also the smallest-mass data finally led to a reasonable understanding of the binding-energy results. The values $m\ellP$ of the trusted masses for the binding energy came out as lying clearly in the large-mass region of models I and II. This offered the explanation of the puzzling mass dependence in the $\Eb/m$ data as a large-mass finite-size effect. Further comparison with characteristic features of models I and II, in particular the magnitude of $\Eb/m$, then led to the conclusion that the explorative SDT results are compatible with model-II behavior, and not with that of model-I.

\section{Conclusion}
\label{secconclusion}

Models I and II are interesting in their own right. Model-I, with its pole and inverse square-root singularities at $\rs$, required considerable numerical effort. The imaginary part of its potential depends on the presence of both classical and quantum corrections in the beta function. It occurs in the region $r<\rs$ and is maximal near $\rs$, which is a finite distance from the origin for all mass values.\footnote{This is different from the Dirac Hamiltonian in a Schwarzschild geometry in which the non-hermitian part is concentrated at the origin \cite{Lasenby:2002mc}.} For small masses the ground state decays slowly\footnote{$\Gm_{\rm b}$ is $\mO(G^{1/2}\, m)$ smaller than the two-graviton decay rate of equal-mass `gravitational atoms',
%bound by the zero-loop Newton potential,
$\Gm_{\rm atom} = (41/(128\,\pi^2))\, G^5\, m^{11}$,
%= 0.032\, G^5 m^{11}$,
which depends primarly on the wave function at the origin \cite{Nielsen:2019izz}.}
at a rate $\Gm_{\rm b} \approx 1.5\, G^{11/2}\, m^{12}$.
For large masses the relativistic model lacks a ground state.
Yet, a spherical wave packet state falling in from a distance $r\gg\rs$ is primarily composed of exited states with small decay times and the packet still exhibits bouncing and falling back during its slow decay.
It is desirable to extend the model by including decay channels into gravitons.

The non-trivial UV fixed point in model-II leads to a regular potential at all $r$.
The increase of its minimum at $\rmin$ with $m$ suggests the possibility of a macroscopic an-harmonic oscillator when $\rmin$ becomes of order of the Schwarzschild scale. In-falling spherical states keep their norm while bouncing. Some of the local probability should diminish eventually by the familiar `spreading of the wave packet'.

Using the SDT results in \cite{deBakker:1996qf}, Planck lengths obtained with perturbative mass renormalization or with matching binding energies to the Newtonian region were similar; the first were actually employed to improve the analysis of the latter.\footnote{The renormalized `continuum Planck lengths' in lattice units, $\ell_{\rm Pc}/a\approx 1.15$ to $0.9$ happen to be larger than the value 0.48 found in CDT based, on a different method (\cite{Ambjorn:2012jv} section 11).} The magnitude of the binding energy is roughly compatible with values found in model-II.
The growing of $\rs$ and $\rmin$ in models I and II suggested a reasonable interpretation of the binding energy data. The relevance of the Schwarzschild scale in this interpretation came as a surprise.

Simulations on larger lattices are necessary to see whether these conclusions hold up to further scrutiny. This should be possible with current computational resources when carried out in a large-mass region, and may tell us something non-perturbative about black holes in the quantum theory.\footnote{As the volume increases $\Eb/m$ vs.\ $m$ should stop decreasing; it might flatten as in model-II or even increase as in model-I. In a plot like figure 5 of \cite{deBakker:1996qf} one might see an oscillation in the effective $\Eb(r)$ beyond $r=6$ indicating a complex energy (and its conjugate), something like
$\exp(-\re[E] \ta)\cos(\im[E] \ta)$ with $\ta=r + {\rm const}$.}
Simulations at small masses, $m\lP\ll 1$, aiming at observing binding energies of Newtonian magnitude seem very difficult because of the rapid increase of the equal-mass Bohr radius $2\ellP/(m\ellP)^3$.

\section*{Note added}
Shortly after the previous version of this article a new EDT computation of the quenched binding energy of two scalar particles appeared in \cite{Dai:2021fqb}. The authors used the `measure term' and an extended class of `degenerate' triangulations as in \cite{Laiho:2016nlp,Jha:2018xjh}. Their analysis included short distances in which dimensional reduction was expected to influence the results. This was taken into account by assuming a corresponding mass dependence of the binding energy,
$\Eb=G^2 m^\al/4$.
%$\Eb\propto m^\al$.
Subsequently an infinite-volume extrapolation and a continuum extrapolation led to the Newtonian $\al=5$ in four dimensions and a renormalized Newton coupling $G$ with relatively small statistical errors.
The computation used very small masses and is in this sense complementary to \cite{deBakker:1996qf} in which (as concluded here) binding energies were computed in a large-mass region. A follow-up article \cite{Bassler:2021pzt} addressed the relation of the Newton coupling to the lattice spacing more closely and described also the computation of a differently defined $G$, which agreed quite well with
\cite{Dai:2021fqb}.

\section*{Acknowledgements}

Many thanks to the Institute of Theoretical Physics of the University of Amsterdam for its hospitality and the use of its facilities contributing to this work. I thank Raghav Govind Jha for drawing my attention to the results in \cite{Jha:2018xjh}.

\appendix
\section{Evolution equation}
\label{appGr}

The equation $-r\partial \tG/\partial r =\bt(\tG)$ simplifies when $\bt$ in (\ref{defbt}) is expressed in terms of $\sqrt{\tG}\equiv z$ (units $G=1$, a notation $b=d m/2$ is introduced for convenience):
\bea
-\frac{r \partial z}{\partial r} &\equiv&\bt_z = z+ 2b z^2 + c z^3,
\quad z=\sqrt{\tG}, \quad b= d m/2,
\\
&=& cz (z-z_1)(z-z_2), \quad z_1 = \frac{-b-\sqrt{b^2-c}}{c},
\quad z_2 = \frac{-b + \sqrt{b^2 -c}}{c}
\eea
We note that $c$, and $b$ are positive in model-I and negative in model-II ((\ref{complete}), (\ref{1PR})).
The critical coupling in model-II is
\be
z_* = z_1 \, ,
\ee
and (\ref{NFPm})--(\ref{NFPmlarge}) in the main text follow.
Separating variables, integrating and imposing the boundary condition
$z\to 1/r$ for $r\to\infty$, the solution can be obtained in the form
\bea
\ln(r) &=& -\ln(z) - f(z) + f(0),
\label{intbtz}
\\
f(z) &=&\frac{\ln(z-z_1)}{cz_1(z_1-z_2)} + \frac{\ln(z-z_2)}{cz_2(z_2-z_1)}
\qquad\qquad \mbox{(model-I)}
\\
&=&\frac{\ln(z_1-z)}{cz_1(z_1-z_2)} + \frac{\ln(z-z_2)}{cz_2(z_2-z_1)}
\qquad\qquad \mbox{(model-II)}\, .
\label{rtozII}
\eea
The second form is chosen for model-II to avoid $f(0)$ being complex, since
$z < z_1$ in this case.
For $b=0$ ($m=0$), $f(z)$ simplifies to $-(1/2)\ln(z^2 + 1/c)$, resulting in $z^2=1/(r^2 - c)$, as used in (\ref{Gtm0}).
(This follows more easily directly from $\bt(\tG,0)$).

\section{More on model-I}
\label{appI}

Since $f(z) \to -\ln(z)+\mO(1/z^2)$ for $z\to\infty$, the position of the singularity is given by
\be
\rL=\exp[f(0)]\, .
\ee
Expanding $r$ as a function of $z$ for $z\to \infty$ gives
\be
\rb \equiv \frac{r}{\rL} = \frac{1}{z}\, e^{-f(z)} = 1 +\frac{1}{2 c z^2} - \frac{2b}{3 c^2 z^3 } + \mO(z^{-4}),
\ee
with the inversion
\bea
z &=&\frac{1}{\sqrt{2c}\,\sqrt{\rb-1}} - \frac{2b}{3c} -\frac{8 b^2-3c}{12
\sqrt{2}\,c^{3/2}}\,\sqrt{\rb-1} + \mO(\rb-1)\, ,
\\
z^2 &=& \frac{1}{2c(\rb-1)} -\frac{2\sqrt{2}\,b}{3 c^{3/2}}\,\frac{1}{\sqrt{\rb-1}} -\frac{40b^2-9c}{36 c^2} + \mO(\sqrt{\rb-1})\, .
\label{zsqrbas}
\eea
Coefficients of even (odd) powers $(\sqrt{\rb-1})^k$ in the expansion (\ref{zsqrbas}) happen to be even (odd) polynomials in $b$ of order $k+2$. For later use we note that keeping only the terms linear in $b$ gives a series that converges in $\rb \in (0,1)$, where $z^2$ is imaginary.
%$\sqrt{\rb-1}=-i \sqrt{1-\rb}$ (the negative sign leads to $\im[V_{\rm r}]<0$).

Keeping in  $V_{\rm r}=-m^2 r z^2$  only the first term of the expansion (\ref{zsqrbas}), or the first two terms,  gives models which can be used to study the effect of the singularity on the binding energy:  the pole model, respectively the pole+square-root model:
\bea
V_{\rm P} &=& -m^2 r\,\frac{\rs}{2c(r-\rs)}\, , \qquad \qquad \qquad \qquad \qquad  \qquad \qquad \qquad\mbox{(P model)}
\label{Pmodel}\\
V_{\rm PSR} &=& -m^2 r \left[\frac{\rs}{2c(r-\rs)} -
%\frac{2\sqrt{2}\,b}{3 c^{3/2}}\,
\frac{\sqrt{2}\,m}{c^{3/2}}\,
\sqrt{\frac{\rs}{r-\rs}}\right]\,
 \qquad \qquad \qquad \; \mbox{(PSR model)}
 \label{PSRmodel}
\eea
(we used $b=3m/2$) .
These potentials do not vanish as $r\to\infty$ and are intended to be used only in matrix elements that focus on a neighborhood of the singularity. For the PSR potential at large $m$ the square-root contribution should not overwhelm that of the pole, because if it would, then all terms left out in the expansion (\ref{zsqrbas}) would contribute substantially and we are back to model-I.
%\P We recall also the limiting classical evolution potential (\ref{btclas}) of model-I:
%\be
%V_{\CEI}= -m^2 r\, \frac{1}{(r-3m)^2}\,. \qquad \qquad \qquad \qquad \qquad  \qquad \qquad \qquad\mbox{(CE-I model)}
%\label{CEImodel}
%\ee

For values of $r$ not close to $\rL$ the dependence of $z$ on $r$ was determined by solving (\ref{intbtz}) numerically for the real and imaginary parts of $z$ as a function of $r$ in the region $0<r<2\rs$ ($z$ is real for $\rs<r<2\rs$). There are two solutions with opposite signs of $\im[z]$. The one with $\im[z]<0$ is chosen to get a decaying time dependence of the eigenfunctions of the Hamiltonian. For remaining integral $\int_{2\rs}^\infty$ we used the inverse of a small $z$ expansion of $r$, or changed variables from $r$ to $z$.

Numerical evaluation of matrix elements of the running potential is delicate because of the singularity at $r=\rs$. Singular terms were subtracted from $V_{\rm r}$ and their contribution was evaluated separately as follows  ($F(r)$ is a smooth trial wave function or a product of basis functions):
\be
\int_0^{2\rs}dr\, V_{\rm r}(r) F(r) =\int_0^{2\rs}dr\, V_{\rm reg}(r)F(r) +
\int_0^{2\rs}dr\, V_{\rm PSR}(r) F(r),
\quad V_{\rm reg} = V_{\rm r} -V_{\rm PSR}\,.
\ee
The first integral on the right hand side was done numerically, the second analytically. The regularized potential $V_{\rm reg}$ is finite but develops at larger masses ($m>2$) a deep trough around $\rs$ as a sort of premonition of the double pole in the CE-I model, which slows numerical integration. Distributional aspects in the analytic evaluation can be taken care of in various ways, (\ref{defdistr2}), or
\be
\left(\int_0^{\rs-\ep} dr + \int_{\rs + \ep} ^{2\rs} dr \right)
\frac{\rs}{r-\rs} \,F(r),\quad \ep\downarrow 0\,.
\label{defdistr3}
\ee
for the pole, or
\be
\re\left[\int dr\,\frac{1}{(r-\rs +i\ep)^n}\, F(r) \right]_{\ep\downarrow 0}
, \qquad n=1,2\,,
\label{defdistr4}
\ee
(assuming real $F(r)$ in the `$i\ep$ method').
Numerically, the principal-value in the symmetric integration $\int_0^{2\rs}$ around $\rs$ can be obtained conveniently by a subtraction in the integrant, $F(r)\to F(r)-F(\rs)$. The methods lead to identical results.

\subsection{Orthogonality under transposition and variational method}
\label{appnormal}

The interpretation of the singular potential as a distribution becomes implemented when evaluating matrix elements  of the Hamiltonian, $\langle \ph|H|\ps\rangle = \int dr\, \ph(r) H \ps(r)$. Starting formally, consider basis functions $b_n(r)$ forming a complete set, and
\be
H_{mn}=\int_0^\infty dr\, b_m^*(r)\, (K+V)\, b_n(r) \equiv K_{mn} + V_{mn} \, .
\ee
The basis functions can be the s-wave Hydrogen eigenfunctions (including the unbound states), or Fourier-sine functions (the $b_n(r)$ have to vanish at the origin). We assume them to be {\em real} and orthonormal,
\be
\int_0^\infty dr\, b_m(r) b_n(r) =\dl_{mn},\quad \sum_n b_n(r) b_n(r^\prime) = \dl(r-r^\prime)\,.
\ee
For simplicity we use a notation in which the labels $m$ and $n$ are discrete and which has to be suitably adapted in case of continuous labeling. When the $K_{mn}$ integrals diverge at infinite $r$ we assume them to be regularized by $b_n(r)\to b_n(r) \exp(-\ep r)$ with the limit $\ep\downarrow 0$ taken at a suitable place. Then $K_{mn}$ and $V_{mn}$ are symmetric in $m\leftrightarrow n$.

Since the potential is complex for $r<\rL$, $V_{mn}\neq V_{nm}^*$, the Hamiltonian is not Hermitian and its eigenvalues $E$, eigenvectors $f_{En}$ and eigenfunctions $f_E(r)$ are complex.
The eigenvalue problem takes the form
\be
f_E(r) = \sum_n f_{E n}\, b_n(r), \quad \sum_n H_{mn} f_{E n} = E f_{E m}.
\ee
The symmetry $H_{mn} = H_{nm}$ invites an inner product under transposition, without complex conjugation. Using matrix notation $f_{E'}^T H f_E = E' f_{E'}^T f_E = E f_{E'}^T f_E \to (E'-E) f_{E'}^T f_E = 0$, where we used $H^T=H$. Eigenvectors belonging to different eigenvalues are still orthogonal and normalizing them to 1 (under transposition), we have in more explicit notation\footnote{Characters $j,k$ refer to eigenvectors of the Hamiltonian, characters $m,n$ refer to basis vectors.}
\bea
f_j(r) &=&
\sum_n f_{j n}\, b_n(r)\,,
\quad
\sum_n H_{mn}\,f_{j n} = E_j\, f_{j m}\,,
\label{eigeneq}
\\
\sum_{n} f_{jn} f_{kn} &=& \dl_{jk} \,, \quad
\sum_j f_{j m} f_{j n} =\dl_{mn}\,,\quad
b_n(r) = \sum_j f_{j n}\, f_j(r)\,,\quad
\label{normCj}
\\
\int_0^\infty dr\, f_j(r)f_k(r) &=& \dl_{jk} \,,\quad
\sum_j f_j(r) f_j(r') = \dl(r-r')\,.
\eea
For finite matrices $f_{jn}$ (which will be the case in our approximations) the second equation in (\ref{normCj}) follows from the first
($f f^T=\unity \to f^T f = \unity$ since a right-inverse is also a left-inverse);
at the formal level with infinitely many basis functions it is an assumption.
We also have
\be
\sum_j E_j f_{jm} f_{jn} = H_{mn}\,, \quad\,
\sum_j E_j\, f_j(r) f_j(r')= H(r,r')\,.
\label{decomphamil}
\ee
In model-I the $f_{jn}$ and $f_j(r)$ are complex; they are real for model-II and the other models with a real potential. In the discrete part of the spectrum the labels $j$ on the eigenvectors will be assigned according to
\be
\re[E_1] < \re[E_2]<\re[E_3] <\cdots\, ,
\label{orderE}
\ee
assuming no degeneracy at zero angular momentum.
An arbitrary wave function $\ps(r)$ in radial Hilbert space can be decomposed as\footnote{Note that in Dirac notation $\langle n|\ps\rangle = \ps_n$ but $\langle j|\ps\rangle=\int_0^\infty dr\, f_j(r)^*\,\ps(r)\neq \ps_j$, in model-I.}
\be
\ps(r) = \sum_n \ps_n\, b_n(r) = \sum_j \ps_j\, f_j(r)\,,\quad
\ps_j = \int_0^\infty dr\, f_j(r)\,\ps(r)\,.
\ee

Conventionally, the functional depending on a variational trial function $\ps$ is
\be
\mE[\ps] = \frac{\langle\ps|H|\ps\rangle}{\langle\ps|\ps\rangle} =\frac{\sum_{mn}\ps^*_m H_{mn}\ps_n}{\sum_{n}\ps_n^*\ps_n}= \frac{\sum_{mn}(\rh_m H_{mn}\rh_n+\sg_m H_{mn}\sg_n)}{\sum_{n}(\rh_n\rh_n+\sg_n \sg_n)}\,,
\ee
where $\ps_n = \rh_n + i \sg_n$ (real $\rh$ and $\sg$) and the symmetry of $H_{mn}$ is used. The variational equations become
\be
\sum_n H_{mn} \rh_n = \mE \rh_n,\quad \sum_n H_{mn}i \sg_n = \mE i\sg_n\,.
\ee
The sum of these equations appears equivalent to (\ref{eigeneq}), their difference to the complex conjugate of  (\ref{eigeneq}) {\em without} conjugating $E$. Hence they are not equivalent to (\ref{eigeneq}) unless $E$ and $\mE$ are real, i.e.\  only for real potentials. On the other hand,
\be
\mE[\ps] =
%\frac{\langle\bar\ps|H|\rangle\ps}{\langle\ps|\ps\rangle} =
\frac{\sum_{mn}\ps_m H_{mn}\ps_n}{\sum_{n}\ps_n\ps_n}
\ee
leads to the correct equation
\be
\sum_{n} H_{mn}(\rh_n + i \sg_n) = \mE\, (\rh_m + i\sg_m)\,,
\ee
implying that $\mE$ is an eigenvalue.
In variational estimates we shall minimize the real part of $\mE[\ps(a,\ldots)]$ with respect to variational parameters $a,\ldots$. However the corresponding theorem in case of a Hermitian Hamiltonian, $\mE \geq E_1$, does not appear to hold true with a complex symmetric Hamiltonian: using a transpose-normalized trial function $\ps$
($\sum_n\ps_n^2 = 1$, $\sum_j\ps_j^2 = 1$) gives
\be
\re[\mE]-\re[E_1] = \re\left[\sum_j \left(E_j - \re[E_1]\right)\, \ps_j^2\right]\,,
\ee
from which one cannot conclude positivity since the individual $\ps_j^2$ are complex. In the conventional case with a real and symmetric $H_{mn}$,  eigenvalues are real, transpose-normalized eigenvectors are real and with a real $\ps$ trial function $\ps_j^2 \geq 0$; then, since $E_j-E_1>0$ for $j\geq 2$, the r.h.s.\ is positive.\footnote{It is comforting that with variational functions $\ps(r)$ lying entirely in the subspace spanned by the Fourier-sine $b_n(r,L)$ (finite $N$) we {\em did} find $\re[\mE]> \re[E_1]$ in model-I.}

A finite discrete set of basis function can be used for approximations that diagonalize $H_{mn}$. For eigenfunctions $f_j(r)$ which are negligible when $r>L$ (typically those near the ground state $j=1$), Fourier-sine functions in finite volume $r<L$ should be able to give a good approximation,
\be
b_n(r,L)=\sqrt{\frac{2}{L}}\, \sin\left(\frac{n\pi r}{L}\right)\theta(L-r)\,,
%,\quad 0\le r\le L,
\quad n=1,\,\ldots,\, N
\label{sines2}
\ee
($\theta$ is the unit-step function), which form a complete set in $r\in(r,L)$ with Dirichlet boundary conditions when $N\to \infty$. Their simplicity is useful in numerical computations with finite $N$, in which $L$ controls finite-size effects and $p_{\rm max} = N\pi/L$, is a cutoff on the mode momenta.  Such a UV cutoff can be avoided in variational calculations.

\subsection{H-like trial function}
\label{appHtrial}

Here follow a few variational calculations using $u_1(r,a)$ in (\ref{unswave}) as a normalized trial wave function with variational parameter $a$ and variational energy
\be
\mE(a) = \int_0^\infty dr\, u_1(r,a)\left(K + V(r) \right)u_1(r,a) =
\langle K \rangle + \langle V\rangle\,,
\ee
and similar with $K\to\Krel$.
The first concerns the relativistic model with the classical Newton potential $V_{\rm N} (r) = -m^2/r$ (units $G=1$).  The potential energy in the state $u_1$ equals
\be
\langle V_{\rm N}\rangle = -\frac{m^2}{a}\,.
\ee
Using the Fourier-Sine representation
\be
u_1(r,a)=\frac{2}{\pi}\int_0^\infty dp\, \sin(p r)\, \frac{4 p a^{3/2}}{(1+a^2 p^2)^2}\,,
\ee
the relativistic energy is found to be
\bea
\langle \Krel \rangle &=& \frac{4}{3\pi}\, \left(\frac{4-4a^2 m^2 + 3 a^4 m^4}{a(a^2 m^2 -1)^2}
+ \frac{3 a^3 m^4(a^2 m^2-2)\,{\rm arcsec}(a m)}{(a^2 m^2 -1)^{5/2}}\right)\,.
\\
&=& 2 m + \frac{1}{m a^2}+ \mO(m^{-3}) \,,\quad m\to\infty\,,
\label{kNa}
\\
&=& \frac{16}{3\pi a} + \frac{16 m^2 a}{3\pi}+\mO(a^3)\,,\quad a\to 0\,.
\label{krelNa}
\eea
As $m$ increases from 0, the value of $a$ where
$\mE(a,m) = \langle \Krel \rangle + \langle V_{\rm N}\rangle$
has its minimum, moves from $a\simeq \aB$ towards zero. Keeping the first two terms in (\ref{krelNa}) one finds that the position of the minimum of $\mE(a,m)$, $\amin$, reaches zero when the mass reaches a critical value $\mc$:
\be
\mc=\frac{4}{\sqrt{3\pi}}\,, \quad \amin=\frac{\sqrt{\mc^2-m^2}}{m \mc}\,,
\quad \mE(\amin,m)=2 m \mc\sqrt{\mc^2-m^2}\,.
\label{Newtonrel}
\ee
Since $\amin$ and also the minimal $\mE$ vanish as $m \uparrow\mc$, the limiting variational binding energy $\Eb=2m-\mE= 2 \mc\simeq 1.30$. By the variational theorems $\mE$ is an upper bound to the energy of the ground state. Since $\lim_{a\downarrow 0}\mE(a,m) = -\infty$ for $m>\mc$, the relativistic Hamiltonian with the Newton potential is unbounded from below.

Next calculation: With the potential $V_{\CEI}$ in (\ref{CEImodel}) the variational function (\ref{F1a}) becomes, in terms of $\ab=a/(3m)$,
\be
\frac{\mE_{1,{\CEI}}(a)}{m} = \frac{\mE_K(\ab)}{m} + \frac{1}{3}\left[-\frac{1}{\ab} -\frac{4}{\ab^2}+ \frac{4}{\ab^3} +\left(\frac{12}{\ab^3}-\frac{8}{\ab^4}\right)
e^{-2/\ab}{\rm Ei}\left(\frac{2}{\ab}\right) \right]\,,
\quad \frac{\mE_K(\ab)}{m} = \frac{1}{9 \ab^2 m^4}\,,
\label{FCE-I}
\ee
where $\mE_K$ corresponds to the non-relativistic operator $K$ (the second term in (\ref{kNa})). Neglecting the latter, the above expression is represented by the dashed curve in the left plot of figure \ref{figpF1a}. Its two minima are at $\ab_1=2.85$, $\ab_2=0.23$. In model-I, the positions of the two minima are for $m=2$ already close to these values; using them to estimate the average non-relativistic squared velocity gives
$v^2 = \mE_K(\ab)/m =8.6\times 10^{-3}$ and $0.13$, respectively at $\ab_1$ and $\ab_2$.

Last calculation in this section: Leading small-mass dependence of the imaginary part of the variational energy. The potential gets an imaginary part in $0<r<\rs$ and as mentioned earlier the expansion (\ref{zsqrbas}) converges when keeping only the leading (linear) terms in $b= 3m/2$ as $m\to 0$.
Consider first the term $\propto 1/\sqrt{r-\rs}$ in (\ref{zsqrbas}), which corresponds to the square-root term of the PSR model (\ref{PSRmodel})),
\bea
\int_0^{\rs}d r\, \im[V_{\rm PSR}]\, u_1(r,a)^2 &=&
\frac{m^3 a x}{8\sqrt{2}\, c^{3/2}}\, \left[x(15+8x^2+4x^4) \right.
\nonumber\\&&
\left. -(15+18x^2+12x^4+8x^6){\rm DawsonF}(x)\right]\,,
\nonumber\\
x&=& \sqrt{\frac{2\rs}{a}}\,.
\label{decayrate1}
\eea
Using the small $m$ forms $a=\aB$, $\rs=\sqrt{c}$, and further expansion to leading order in $m$ leads to the decay rate
\be
\Gm_{\rm b} \equiv -2\,\im[\Emin]\approx \frac{32\sqrt{2}}{105}\,d \sqrt{c}\, m^{12} = 1.48\, m^{12}\,,
\label{Gm1}
\ee
where $d=2 b/m=3$ indicates the perturbative order in the parameters of model-I. Continuing the expansion (\ref{zsqrbas}) up to $\mO((\sqrt{r-\rs})^{11})$ and keeping again only terms linear in $b$, gives instead of (\ref{Gm1}) (avoiding quoting fractions of excessively large integers)
\be
\Gm_{\rm b} =1.38329\, m^{12}\,.
\ee
(The $\mO((\sqrt{r-\rs})^{11})$ contribution in only about $10^{-5}$ of (\ref{Gm1}). )

Above, instead of finding the minimum $\amin$ of the variational integral $\mE(a)$, we simply used the Bohr radius, which means the calculation is really a perturbative evaluation of imaginary part of the Hamiltonian in the ground state wave function. The result (\ref{Gm1}) is quoted in (\ref{Gm}).

\subsection{CE-I model with the Fourier-sine basis}
\label{appCEIsine}

The kinetic energy matrix is diagonal in the basis of sine functions (\ref{sines2}),
\be
\frac{K_{mn}}{m} = \frac{1}{m^2 \rs^2} \left(\frac{n\pi}{\Lb}\right)^2 \dl_{mn}\,,
\quad \Lb\equiv \frac{L}{\rs}\,.
\ee
In the CE-I model $\rs=3m$. Using $\rb = r/\rs$ as integration variable, the potential matrix becomes
\be
\frac{V_{\CEI,mn}}{m} =
-\frac{2}{3\Lb}\int_0^{\Lb} d\rb\, \frac{\rb}{(\rb-1)^2}\,\sin\left(\frac{m\pi\rb}{\Lb}\right)
\sin\left(\frac{n\pi\rb}{\Lb}\right)\,,
\label{VCEImn}
\ee
which can be evaluated analytically into a host of terms (using the $i\ep$ method to implement the distributional interpretation of the double pole), too many to record here. The explicit dependence on the mass has canceled in (\ref{VCEImn}). The binding energy ratio $\Eb/m$ can now be considered a function of $1/m^4$ coming from $K_{mn}$, of $N/\Lb=2\rs/\lmmin\equiv \rh$, and of $\Lb$. Assuming $\Lb$ is large enough such that finite-size effects may neglected, and that $m$ is large enough to neglect the kinetic energy contribution, there remains only the dependence on $\rh$.
This was tested twice  (t1, t2) by three computations (c1, c2, c3):
\begin{itemize}
\item[c1] computed the mass dependence of $\Eb/m$ at fixed $\lmmin=1$, for $m=1$, 2, \ldots, 10; %using data at $N=64$ and 128 and varying $\Lb$ (to keep $\lmmin$ fixed);
\item[c2] computed the $\rh$ dependence of $\Eb/m$ at $m=2$; data ranging from $\rho=2$ to 512;
\item[c3] computed the $\rh$ dependence of $\Eb/m$ while leaving out the contribution from $K_{mn}$ $\propto 1/m^4$; data ranging from $\rho=2$ to 64,
%the binding energy at fixed $\Lb=8$ ($N$ such that $\rh=2$, 4, 8, 16, 32, 64), while leaving out the $K_{mn}$ contribution (quenching the $1/m^4$ dependence),
which were fitted by $\Eb/m = 1.339\, \rh^2$;
\item[t1] The data in c3\ are consistently 3\% higher than those in c2\, which indicates that  already at $m=2$ the effect of the kinetic energy is only 3\%;
\item[t2] Substituting $\rh=2\rs/\lmmin = 6m$ in the fit from c3 gives $\Eb/m=48.2\, m^2$, which describes the data in c1 well within a few \% for $m\geq 2$.
\end{itemize}

\subsection{Bounds on $\Eb/m$}
\label{appbounds}

\FIGURE[t]{
\includegraphics[width=7cm]{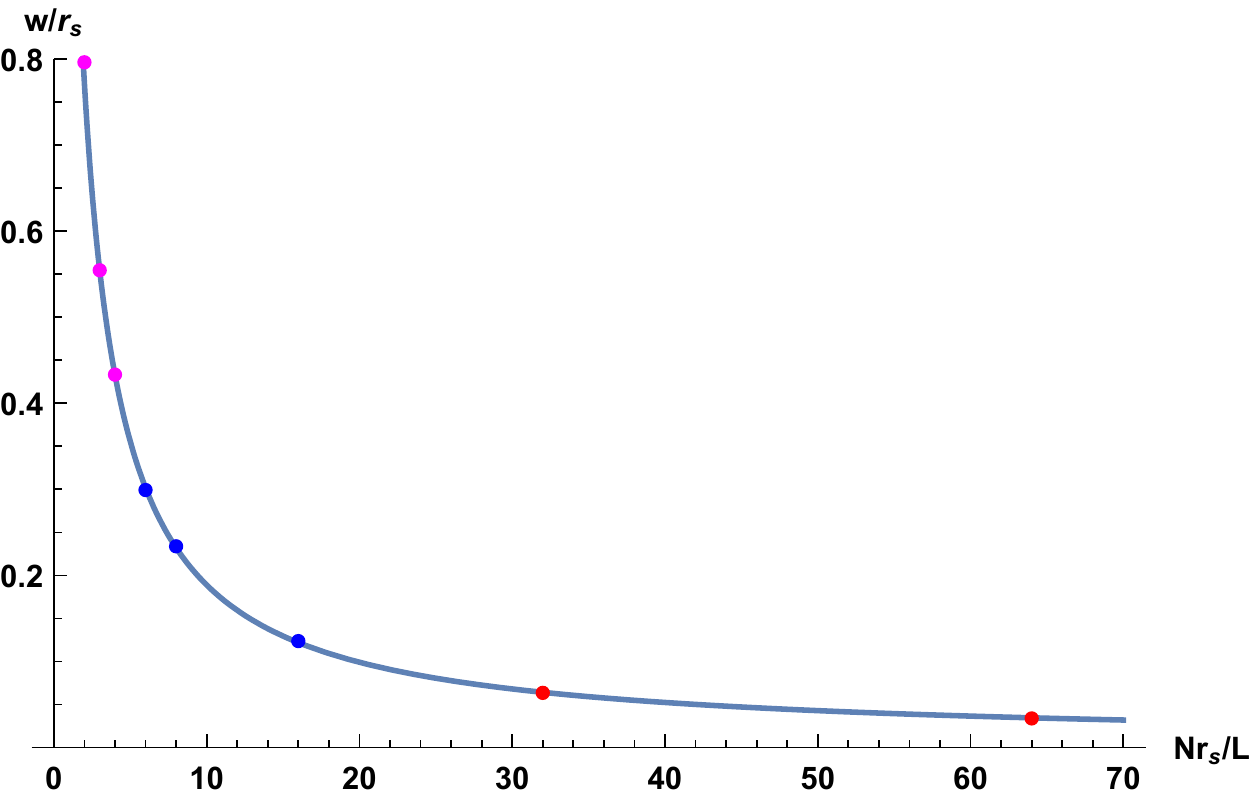} %plots
\includegraphics[width=7cm]{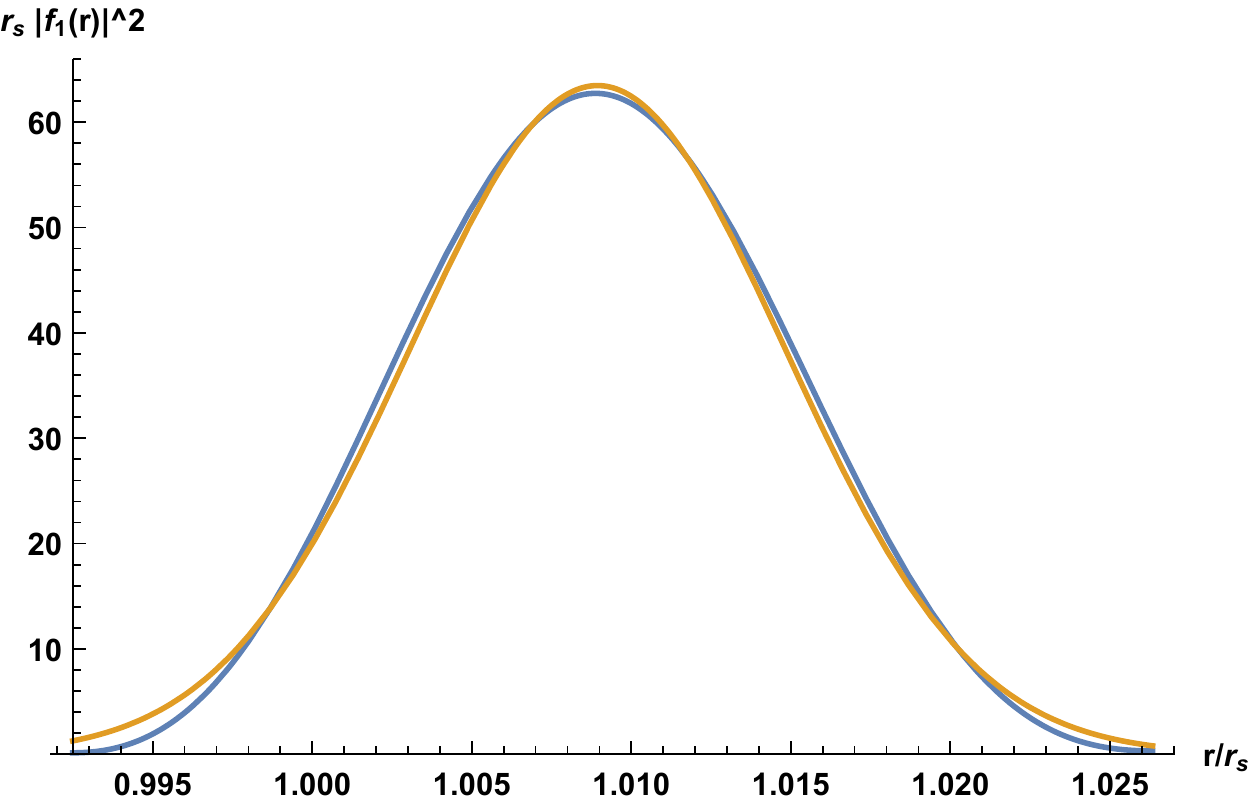} %plots
\caption{Left: results for $w/\rs$ and fitting function (\ref{widths}); $L/\rs=\{32,8,2\}\leftrightarrow {\rm\{Magenta,Blue,Red\}}$.
Right: $\rs |f_1(r)|^2$ (blue) versus $\rb=r/\rs$ for $L/\rs=2$, $N=128$, $m=2$, fitted by a by a Gaussian $\propto \exp[-(r-a)^2/(2s^2)]$, $a=1.0089\, \rs$, $s= 0.0059\, \rs$ (brown). In this case
%$\{\rb_-,\, \rb_+\}=\{0.992,\, 1.026\}$, $\rs\int_{\rb_-}^{\rb_+} d\rb\, |f_1(\rb\rs)|^2 = 0.94$,
%$\rs\int_{\rb_-}^{\rb_+} d\rb\, \re[f_1(\rb\rs)]^2 = 0.91$\,.
$\{r_-,\, r_+\}=\{0.992\, \rs,\, 1.026\,\rs\}$, $w=0.034\,\rs$, $\int_{r_-}^{r_+} dr\, |f_1(r)|^2 = 0.94$,
$\int_{r_-}^{r_+} dr\, \re[f_1(r)]^2 = 0.91$\,.
}
\label{figpwidth}
}
The question whether the binding energy is bounded was followed up using the basis of sine functions, which then helped to choose improved trial functions for the variational method.  With the sine functions the UV cutoff was raised by reducing $L$, since going beyond $N=128$ was numerically impractical. Results were obtained for $N=16$, 24, 32, 48, 64, 96, 128, and $L/\rs=2$, 8, 32.
To make sure that the ground state $f_1(r)$ fitted-in easily in the smaller $L$ domains, we studied its width. A convenient measure of the width is the distance between the two minima of $|f_1(r)|^2$ closest to $\rs$. For example, in figure \ref{figpfcnm20L32N128} these minima are at $r_-=0.938\,\rs$ and $r_+=1.371\,\rs$, giving a width $w=(r_+ -r_-)\rs= 0.433\,\rs$. The left plot in figure \ref{figpwidth} shows results for the width as a function of $N\rs/L=\rh$, with data at each $\rh$ selected to correspond to the largest available $L$. The curve is a fit to $w/\rs$ by  a rational function
\be
R_w(\rh) = \frac{6.41 + 0.0146\, \rh}{1 + 0.534\rh},
\qquad \rh = N\rs/L=2\rs/\lmmin
\label{widths}
\ee
(the first point was left out of the fit to improve agreement with the data at larger $\rh$). The fit indicates a finite width as $\lmmin\to 0$: $R_w(\infty) = 0.044$. As $\rs/\lmmin$ increases, $|f_1(r)|^2$ looks more and more like a Gaussian, narrowing in width and the position of its maximum approaching $\rs$. The right plot in figure \ref{figpwidth} shows an example. The fit gives a standard deviaton $s\simeq 0.00588\, \rs$, from which we deduce a conversion factor between $w$ and the standard deviation $s$:
\be
w/s \simeq 5.76\, .
\label{ratiows}
\ee

The left plot in figure \ref{figpfebNdata} shows $\re[E_1]/m$ obtained from
%$L= 32 \, \rs$, $N=64$, 128, from $L= 8\, \rs$,
%$N\rs/L =2$, 3, 4, 6, 8, 16, and from $L/\rs=2$, $N\rs/L=16$, 32, 64.
the same selected $\{N,\,L/\rs\}$ values.
The results are fitted well (using all data points for $\re[E_1]/m$ and omitting the first three for $\im[E_1]/m$) by the rational functions
\bea
R_{\re[E]}(\rh) &=&
-\frac{0.375322 + 0.727212\, \rh + 0.349396\, \rh^2}{1+0.0410593\, \rh + 0.000172826\, \rh^2}\,,
\label{febLN} \\
R_{\im[E]}(\rh) &=&-\frac{3.20098 + 0.901784\, \rh + 0.0977789\, \rh^2}{1 + 0.0448932 \, \rh + 0.00112354\, \rh^2}\,.
\label{febLNi}
\eea
The second derivative $R_{\re[E]}^{\prime\prime}(\rh)$ is negative at the smaller $\rh$, changes sign at $\rh\simeq 35$, reaches a maximum at $\rh\approx 73$
--- properties almost within the data region ---
and then slowly falls to zero while the function becomes constant. This suggests that $\re[E_1]/m$ is finite; extrapolation gives
$R_E(\infty) = -2022$.
The corresponding fit to the imaginary part of $E_1$ has similar properties with a relatively moderate limit $R_{\im[E]}(\infty) = -87$.
Extrapolation to, say, within 20\% of the infinite $\rh$ limits would involve values of $\rh$ into the many hundreds, which still might seem preposterously far from the computed results. To substantiate the finiteness of the binding energy we need data in this region, but going beyond $\rh=64$ is numerically difficult.

\FIGURE[t]{
\includegraphics[width=7cm]{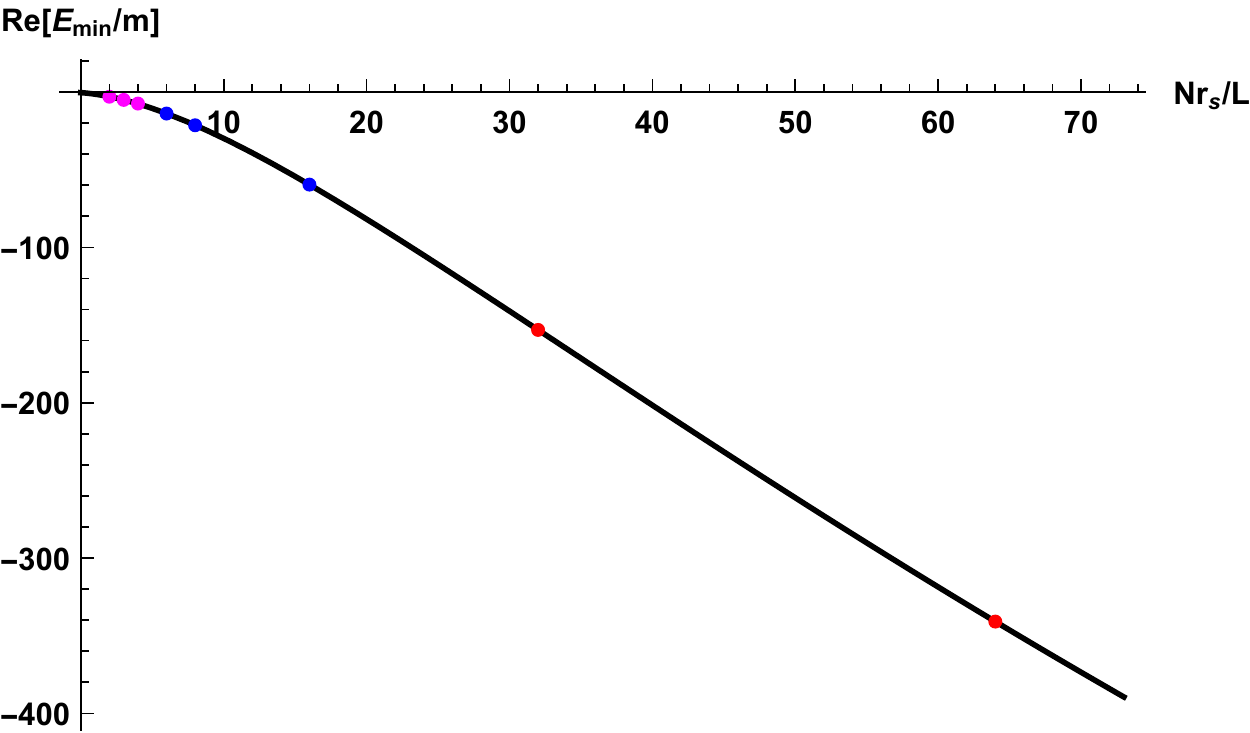} %plots
\includegraphics[width=7cm]{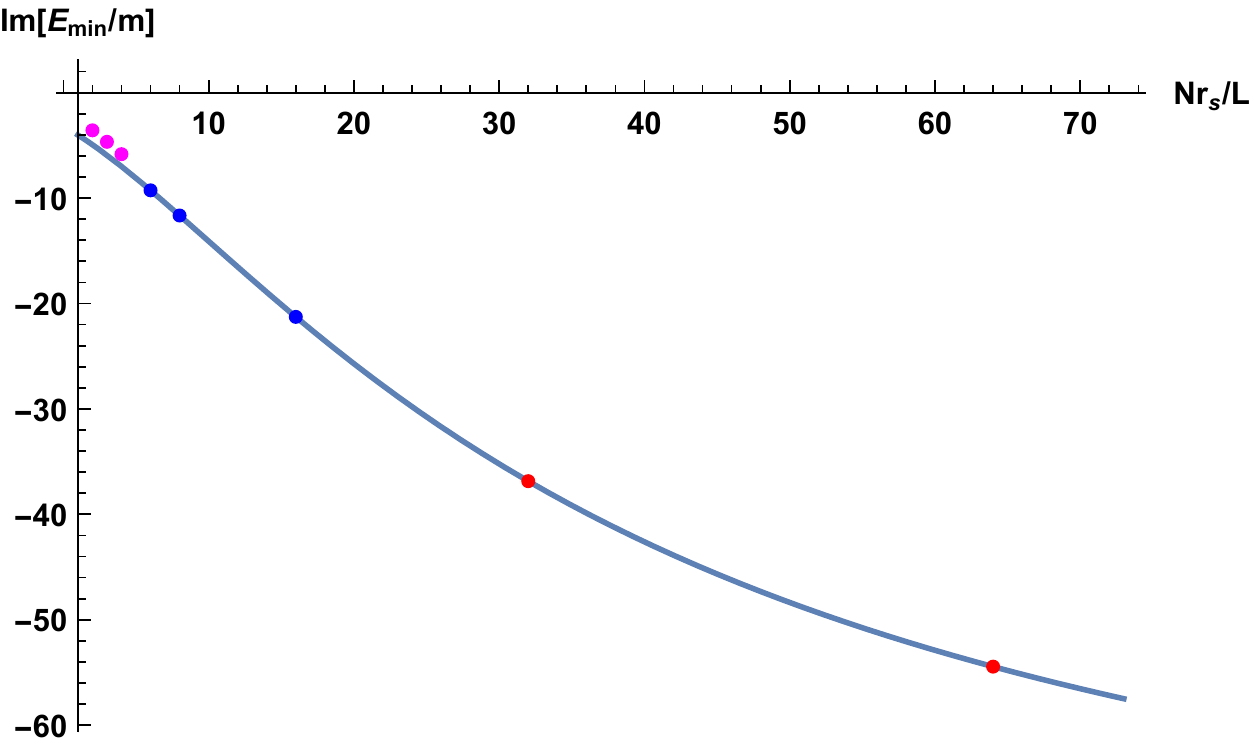} %plots
\caption{Data for $\re[E_1/m]$ (left) and $\im[E_1]/m$ (right) at with fits by the functions in (\ref{febLN}), (\ref{febLNi}). Same data with color coding as in figure \ref{figpwidth}.
}
\label{figpfebNdata}
}

The lowest-energy eigenfunction $f_1(r)$ receives most of its normalization integral from the region $\rs\lesssim r\lesssim \rs+w$ and the small ratios $w/\rs$ in figure \ref{figpwidth} suggest that the large binding energies found thus far are caused by the singularity at $\rs$.
Changing tactics, we focus in appendices \ref{appbounds} and \ref{appG} on the region around $\rs$ by studying simpler models: the pole model (P), the pole+square-root model (PSR).
The good approximation of the Gaussian to $|f_1(r)|^2$ in figure \ref{figpwidth} suggests using a Gaussian for a variational approximation in the large-mass region:
\bea
f_{\rm G}(r)&=&\mu^{-1/2}_{\rm G} \left(\exp\left[-\frac{(r-a)^2}{2 s^2}\right]\right)^{1/2}\, , \quad \int_0^{2\rs}dr\, f_{\rm G}(r)^2 = 1\,,
\label{fG}\\
\mE_{\rm GP}(a,s) &=& \int_0^{2\rs} dr\,f_{\rm G}(r) H_{\rm P} f_{\rm G}(r),
\label{FGP}
\eea
for the P-model; the normalization integral determines $\mu_{\rm G}$.  With upper integration limit $2\,\rs$ we can compare with results using the sine basis functions with $L=2\,\rs$. Extending the integration range to $-\infty<r<\infty$ facilitates analytical evaluation of the resulting variational integral---let's denote it by $\mE(a,s)$. This extension is permitted if $f_{\rm G}(r)$ is at $r=\{0,\, 2\rs\}$ small enough for satisfying the boundary conditions to sufficient accuracy when $\{a,\,s\}$ is near the minimum of $\mE(a,s)$, which may replace $\mE_{\rm GP}(a,s)$ under these circumstances.
%$\{\amin\, ,s_{\rm min}\}$

The PSR-model potential contains also a square root in the potential; this appears to inhibit analytic evaluation. A rational form of $f(r)^2$,
\bea
f_{\rm BW}(r) &=&
\mu^{-1/2}_{\rm BW}\, r(2\rs-r) \left(\frac{1}{(r-a)^2 + s^2}\right)^{1/2},
\quad \int_0^{2\rs} dr\, f_{\rm BW}(r)^2 = 1\, ,
\label{BW}\\
\mE_{\rm BWPSR}(a,s) &=& \int_0^{2\rs} dr\,f_{\rm BW}(r) H_{\rm PSR} f_{\rm BW}(r),
\label{FWBPSR}
\eea
allows analytic evaluation of the variational integral $\mE_{\rm BWPSR}$ (the factor $r(2\rs-r)$ has been added to satisfy the boundary conditions even at the lower end of the large-mass region where $f_{\rm BW}^2$ without this factor would be rather broad).
We dub $f_{\rm BW}$  the Breit-Wigner (BW) trial function. Note that $f_{\rm G}(r)$ and $f_{\rm BW}(r)$ approach the square root of a Dirac delta function as $s\to 0$.

\FIGURE[t]{
\includegraphics[width=7cm]{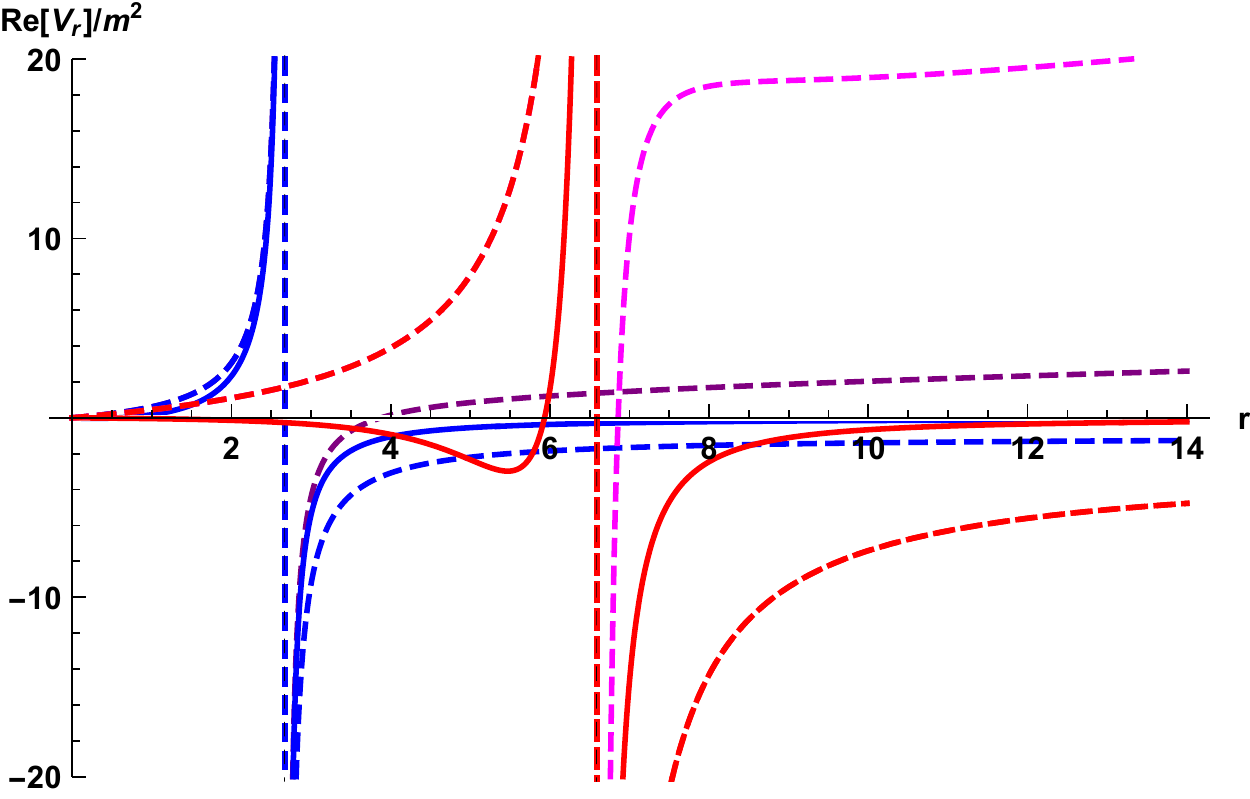} %see plots
\includegraphics[width=7cm]{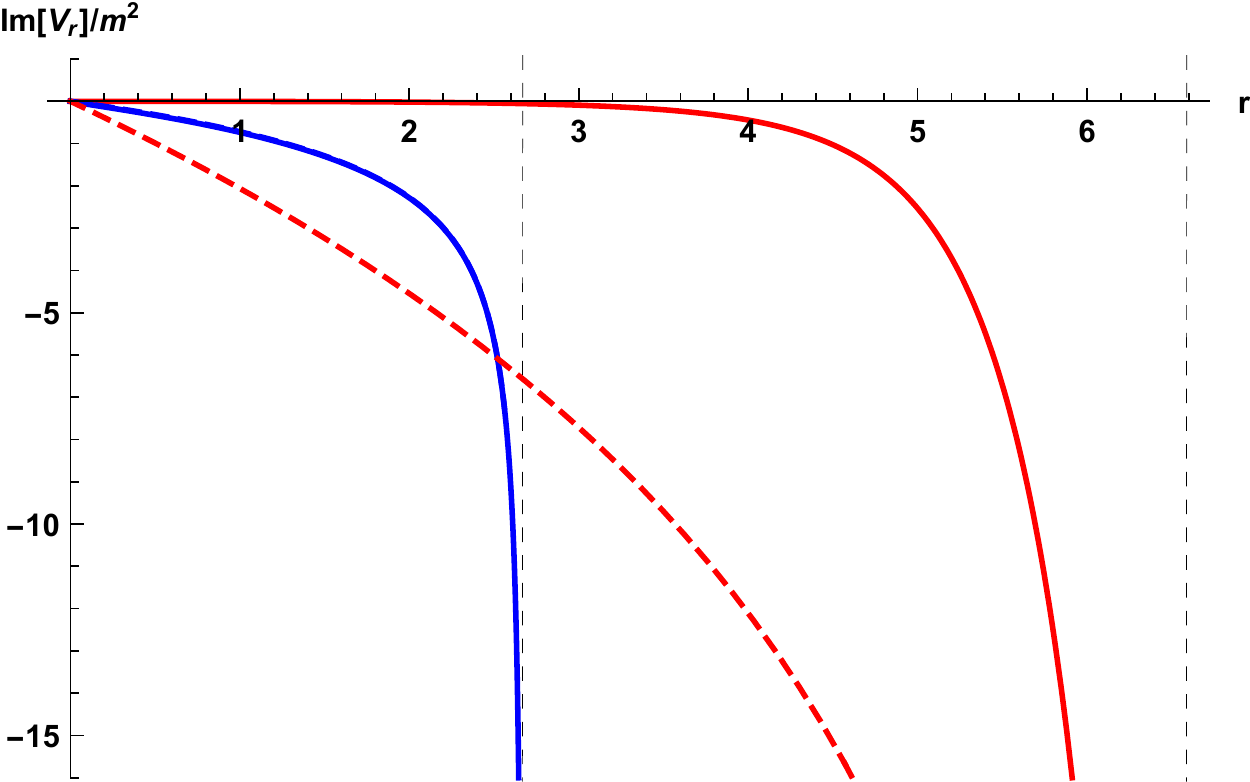} %see plots
\caption{
Running potentials $V_{\rm r}/m^2$: model-I, the P-model and the PSR-model, for $m=0.6$ (blue, dashed-blue and dashed-purple) and for $m=2$ (red, dashed-red and dashed-magenta).
Left: real part; Right: imaginary parts, in which to the eye the blue and dashed-purple curves overlap.
}
\label{figpVrm06m2tot}
}
Figure \ref{figpVrm06m2tot} shows  again the potential in the critical region, here with the potentials of the P-model and the PSR-model included for comparison. In the left plot, the dashed curve for the PSR-model is above that of model-I, hence, its variational energy is definitely above $\re[E_1]$ of model-I, it will produce a {\em lower} bound on its binding-energy. The dashed curve for the P-model lies below that of model-I. Assuming the  Gaussian variational energy to be accurate for large masses for the P-model we may expect its variational energy to lie below $\re[E_1]$ of model-I, hence to produce -- or to be close to -- an {\em upper} bound on its binding energy. This putative upper bound and the lower bound are shown in figure \ref{figpevarI}. The large mass asymptotes  are given by (see appendix \ref{appG} for their evaluation)
\bea
-\re[\mE]/m &\simeq& 6.96\, m^8\,, \qquad\qquad\qquad\qquad\qquad\quad\;\;\,  \mbox{(GP)}
\\
-\re[\mE]/m &\simeq& 4.17\, m^8\,, \quad -\im[E]/m \simeq 4.17\, m^6\,.\quad \mbox{(BWPSR)}
\label{EbasGPBWSQR}
\eea
The asymptote with the Gaussian trial function is shown dashed between the P-model Gauss curve and the PSR-model BW-curve.

Turning to the estimates for $m=2$ obtained with the basis of sine functions, $-R_{\re[E]}(\infty)=2022$ lies indeed between the variational 1427 (BWPSR) and 2596 (GP). Furthermore, the conversion factor 5.76 in (\ref{ratiows}) from the width $w$ of the wave function to the fitted Gaussian standard-deviation $s$, gives, when applied to the extrapolated width, $R_w(\infty)/5.76 = 0.00077$, remarkably close to the GP value $s_{\rm min}/\rs = 0.00075$ (cf.\ below (\ref{Kummerfcns})).

The importance of the singularity for the binding energy helps understanding the change of sign of the second derivative of $R_{\re[E]}(\rh)$ in (\ref{febLN}).
Using the sine basis for the P-model with $L=2\,\rs$, a rational function of the form (\ref{febLN}) fitted to its numerical data of $E_1/m$ has a {\em positive} second  derivative for all $2<\rh<\infty$, with a finite $R_E(\infty)=-4336$.  In similar fashion the PSR-model yields $R_{\re[E]}(\infty) = -3666$ and  $R_{\im[E]}(\infty)=-205$.
In the classical-evolution model CE-I, the data clearly indicate a diverging limit $E_1/m\to -\infty$: a purely quadratic form with
$R_{E}^{\prime\prime}(\rh)= -1.30$
gives a good fit over the whole range $2\leq \rh\leq 512$.
This divergence reflects the stronger singularity of the double pole in this model. In the quantum model-I the diverging and converging behaviors compete: since $\lmmin=(2/\rh)\,\rs$, the smallest wavelength modes still average the double-pole behavior of the potential
%(cf.\ figure \ref{figpVrm06m2})
when $\rh\ll 35$, thus the classical-evolution behavior
($R_{\re[E]}^{\prime\prime}(\rh)<0$) wins,
whereas at larger $\rh\gg 35$ the true single-pole+square-root singularity ($R_{\re[E]}^{\prime\prime}(\rh)>0$) wins with $R_{\re[E]}^{\prime\prime}(\rh)\downarrow 0$,  $R_{\re[E]}^{\prime}(\rh)\uparrow 0$ as $\rh\to\infty$.

\subsection{Gaussian and Breit-Wigner trial functions}
\label{appG}

For the P-model (\ref{Pmodel}) and the Gaussian in (\ref{fG}) we can use the implementation (\ref{defdistr2}) of the principle value in the variational integral:
\bea
\mE_{\rm GP} &=& \mE_{K} + \mE_{V}\,,
\\
\mE_{K, \rm GP} &=& -\frac{1}{m} \int_{-\infty}^{\infty} dr\, f_{\rm G} \frac{\partial^2 }{\partial r^2}\, f_{\rm G} =
\frac{1}{4 m s^2}\,,
\\
\mE_{V, {\rm GP}} &=&\frac{m^2 \rs}{2c} \int _{-\infty}^{\infty} dr\, \ln(|r-\rs|)\, \frac{\partial}{\partial r} \left(r f_{\rm G}^2\right) \,.
\eea
After writing
\be
s=\sbar \,\rs, \quad a=(1+\sbar y)\rs\,,
\label{introy}
\ee
the potential part can be worked into the form
\bea
\mE_{V, {\rm GP}}&=& -\frac{m^2 \rs}{2 c}\left(1 + \frac{h(y)}{\sbar}\right)\,,
\\
h(y) &=& \frac{y}{2}\left[2+ M^{(1,0,0)}\left(0,\half,-\frac{y^2}{2}\right)
- M^{(1,0,0)}\left(0,\frac{3}{2},-\frac{y^2}{2}\right)\right]\,,
\label{Kummerfcns}
\eea
where $M$ is the Kummer confluent hypergeometric function
and its superscript denotes differentiation with respect to its first argument.
The odd function $-h(y)$ has a minimum at $y=y_{\rm min}=1.307$, $h(y_{\rm min})=0.765$.
The pair of variational equations $\{\partial_{\sbar} \mE=0,\, \partial_y \mE=0\}$ was solved numerically and the resulting binding energy is plotted in figure \ref{figpevarI}. For $m=2$,  $\sbar = 0.000745$, $y=1.31$,  and $\mE_{\rm GP}/m = - 2596$.
When $m$ increases $\sbar$ approaches zero.
The asymptotic form for $\sbar\to 0$,
\be
\frac{\mE_{\rm GP}^{\rm as}}{m} =
\frac{1}{4 m \rs^2 \sbar^2} - \frac{m \rs\, h_{\rm min}}{2 c \sbar}\,.
%+ \mO(1)\,.
\label{FGPas}
\ee
gives with $\rs\to 3 m$ a finite minimum at large $m$:
\be
\sbar \to 0.0632\, m^{-6},\quad
\frac{\mE_{\rm GP}^{\rm as}}{m} \to -6.96 \, m^8\, .
\label{FGPasnum}
\ee
Numerical results in $m\gtrsim 1$ are  plotted in figure \ref{figpevarI}.

With the Breit-Wigner trial function (\ref{BW}) the principle-value in the P-model was treated with the definition (\ref{defdistr3}).
In this case the resulting expressions for $\mE_K$ and $\mE_V$ do not involve functions more sophisticated than logarithms but they are too long to record here. A representation in terms of the analogue $\sbar$ and $y$ for this trial function is also useful here. At $m=2$, the minimum is at $y=1$ (machine precision), $\sbar=0.000761$, with $\mE_{\rm BWP}/m = -2220$, which is somewhat higher than the above value $-2596$ for GP.  Indeed, the BW trial function is less accurate than the Gaussian one. The leading asymptotic form for $\sbar\to 0$ of $\mE_{\rm BWP}$ simplifies to
 \be
\frac{\mE_{\rm BWP}^{\rm as}}{m}=
\frac{1}{8 m^2 \rs^2 \sbar^2} - \frac{m \rs y}{2c(1+y^2)\sbar}\,,
\label{FBWPas}
\ee
in which the dependence on $1/\sbar$ and $y$ has again decoupled in the potential term, which has its minimum at $y=1$. Using $\rs=3m$ for large $m$ the solution for the minimum becomes
\be
%\sbar=\frac{c}{m^3 \rs^3}
y \to 1\,, \quad\sbar=0.0483\,m^{-6}\,, \quad \mE_{\rm BWP}^{\rm as}/m = -5.94\, m^8\,.
\label{BWPsoln}
\ee

Turning to the square-root term in the PSR-potential (\ref{PSRmodel}), Mathematica does not give an analytic form for the variational integral with a Gaussian, but the Breit-Wigner form poses no further difficulty for the integral
\be
%\mE_{{\rm BW}}=
-\frac{\sqrt{2}\,m^3}{c^{3/2}}\int_0^{2\rs}dr\, \sqrt{\frac{\rs}{r-\rs}}
\,r f_{\rm BW}(r)^2\,.
\ee
The real part of the above expression is added to $\mE_{\rm BWP}$ to make $\re[\mE_{\rm BWPSR}]$, the imaginary part is evaluated at the minimum of the latter.
Numerical results in $m\geq 1$ are  plotted in figure \ref{figpevarI}.
For $m=2$, $\re[\mE_{\rm min}]/m \simeq -1427$, $\im[\mE_{\rm min}]/m \simeq - 312$. As $m$ increases, $y_{\rm min}$ and
$\sbar_{\rm min}$ rapidly approach 1 and 0 respectively. The asymptotic form for $\sbar\to 0$ can be worked in the form
\bea
\frac{\re[\mE_{\rm BWPSR}^{\rm as}]}{m}&=& \frac{1}{8 m^2 \rs^2 \sbar^2} - \frac{m \rs y}{2c(1+y^2)\sbar}
+ \frac{m^2\rs}{c^{3/2}\sqrt{\sbar}}\sqrt{\frac{y+\sqrt{1+y^2}}{1+y^2}}\,,
\label{FBWPSRsy}\\
\frac{\im[\mE_{\rm BWPSR}^{\rm as}]}{m} &=&-\frac{m^2\rs}{c^{3/2}\sqrt{\sbar}}\, \sqrt{\frac{-y+\sqrt{1+y^2}}{1+y^2}}\,.
\eea
The square-root term in $\mE_{{\rm BWPSR}}^{\rm as}$ is of order $m\sqrt{\sbar}$ relative to the pole term and (\ref{BWPsoln}) also gives the leading behavior of $\mE_{{\rm BWPSR}}^{\rm as}/m$ as $m\to \infty$.
The imaginary part behaves as
\be
\im[\Emin]/m \to  -4.17\, m^6\,.
\ee

In the {\em relativistic} version of model-I with the Gaussian trial function the first term in (\ref{FGPas}) is replaced by
\bea
\frac{\langle \Krel\rangle}{m} &=&
\frac{\sqrt{2}}{m \rs \sbar}\; U\left(-\half,\,0,\, 2 m^2\rs^2 \sbar^2\right)
= \frac{\sqrt{2}}{\sqrt{\pi}\,m \rs \sbar} + \mO(\ln \sbar)
\nonumber\\
\frac{\mE_{\rm GPrel}^{\rm as}}{m} &=& \frac{\sqrt{2}}{\sqrt{\pi}\,m \rs \sbar}
- \frac{h_{\rm min} m \rs}{2 c \sbar} \,,
\label{FGPasrel}
\eea
where $U$ is Kummer's confluent hypergeometric function. The relativistic kinetic-energy contribution scales like $1/\sbar$, in contrast to the non-relativistic $1/\sbar^2$. The potential contribution is unchanged, of order $1/\sbar$. Comparing coefficients of $1/\sbar$ it follows that
the variational integral goes to negative infinity as $\sbar\to 0$ for masses greater than 0.612 (the asymptotic forms  (\ref{FGPas}) and (\ref{FGPasrel}) used only $\sbar\to 0$, they hold for all $m$).
In the Breit-Wigner case we can use a simpler trial function without the factor $r(2\rs-r)$ in (\ref{BW}) when focussing on the limit $\sbar\to 0$. This simplifies analytical evaluation of $\langle \Krel \rangle$, it can be expressed in a Meijer G-function, and
\be
\frac{\mE_{\rm BWPSRrel}^{\rm as}}{m}=\frac{\mE_{\rm BWPrel}^{\rm as}}{m}=
\frac{4}{\pi^2 m \rs \sbar} - \frac{m \rs }{4c \sbar}\,
\label{fBWPasrel}
\ee
(we inserted $y_{\rm min=1}$ in the potential part of (\ref{FBWPas})). Comparing coefficients of $1/\sbar$ again indicates  no lower bound on $\Eb$ already for $m>0.565$.

\FIGURE[t]{
\includegraphics[width=7cm]{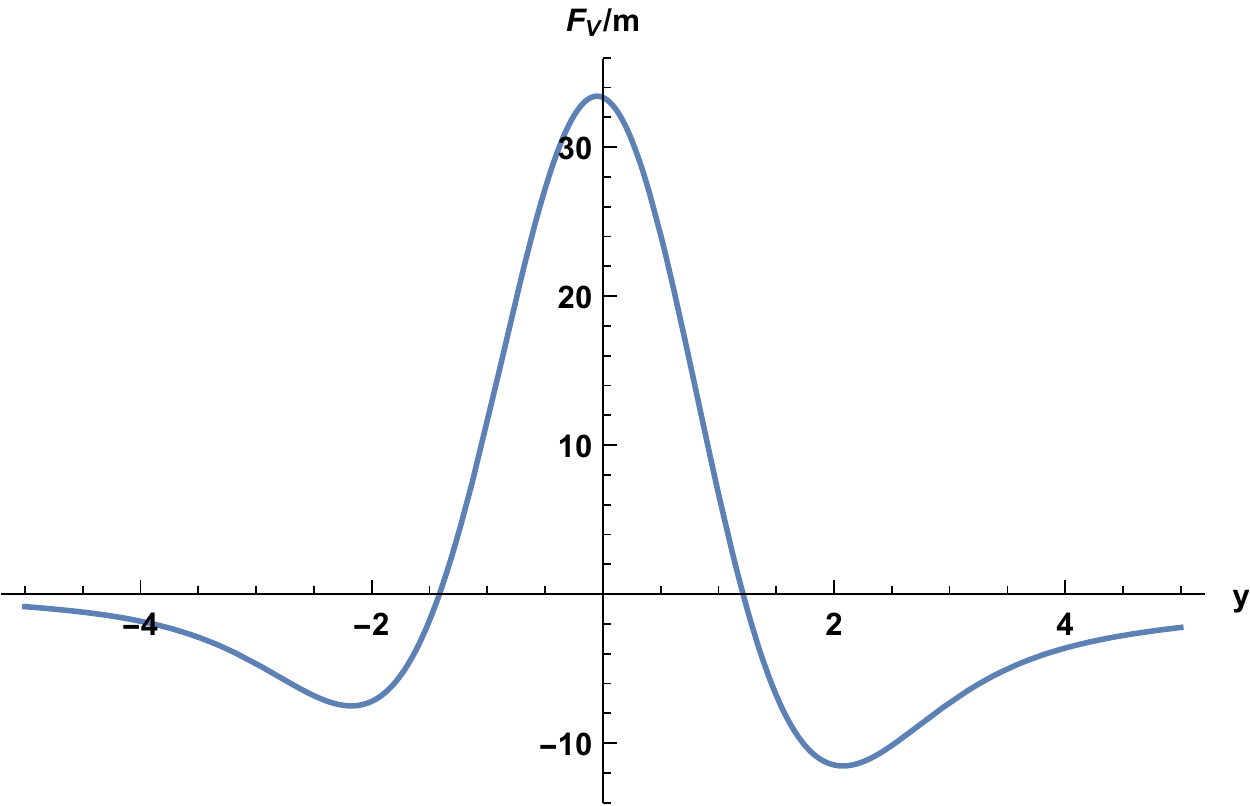} %see bindingGauss.nb
\includegraphics[width=7cm]{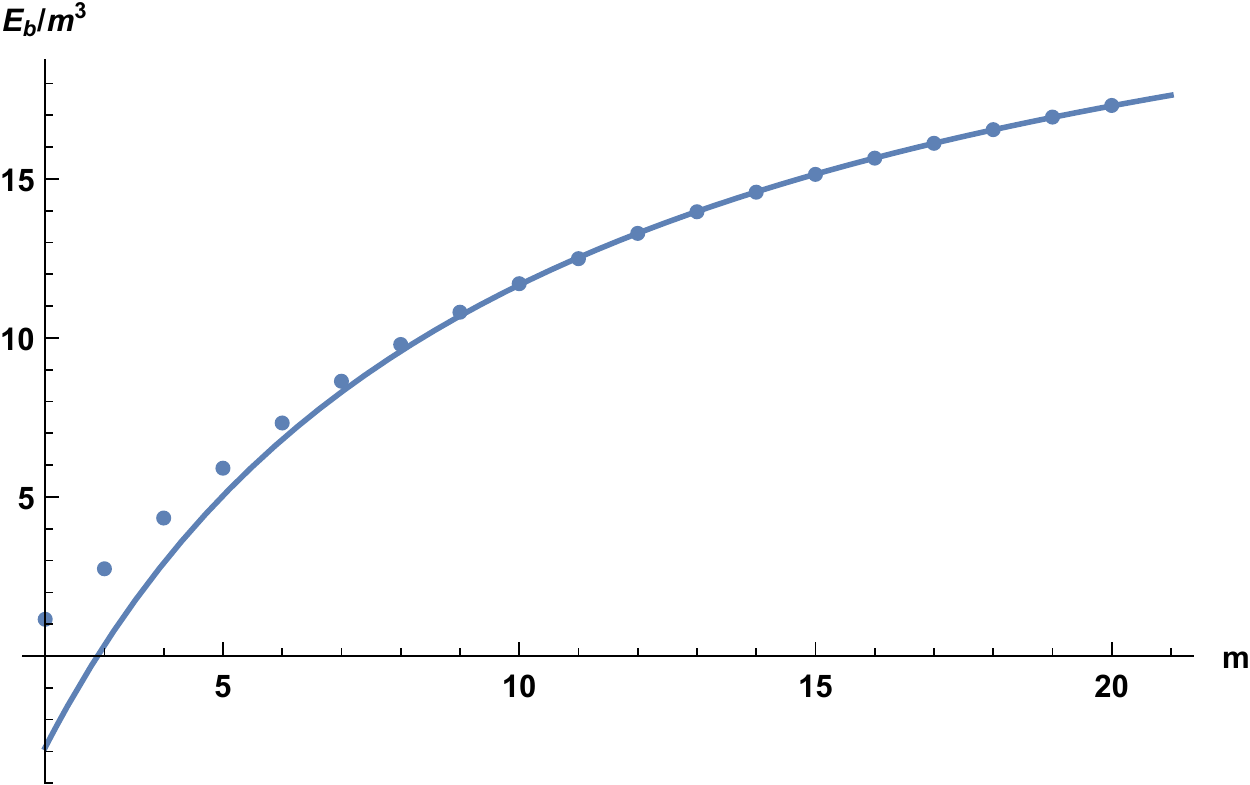} %see binding1loopNew2bIPV7Gauss.nb
\caption{
Left: $\mE_{\GCEI}/m$ in (\ref{FGCEIsy}) for $\sbar=0.1$.
Right: Numerically evaluated $\Eb/m^3$ of model-I with Gaussian trial wave function (dots) fitted at
$m=\{10, 11, \ldots, 20\}$ by the function $(p_0 + c_{h2}\, q_1\, m)/(1+q_1 m)$ (curve), $p_0=-12.51$, $q_1=0.1647$.
}
\label{figFVCEI}
}

The classical-evolution model (\ref{CEImodel}) can be treated in similar fashion. With the Gaussian trial function, the potential part
\be
\mE_{V,\GCEI}= m^2\int_{-\infty}^\infty dr\,\ln(|r-\rs|) \frac{\partial^2}{\partial r ^2}\left(r f_G(r^2\right)\,,
\quad \rs=3 m,
\ee
has after the substitution (\ref{introy}) the form
\be
\frac{\mE_{V,\GCEI}}{m} = \frac{h_1(y)}{\sbar} + \frac{h_2(y)}{\sbar^2}\,,
\label{FGCEIsy}
\ee
where $h_1(y)$ and $h_2(y)$ are are odd and even in $y\to - y$, of similar size, and expressible in Kummer functions, as in (\ref{Kummerfcns}).
The left plot in figure \ref{figFVCEI} shows $\mE_{\GCEI,V}$ for $\sbar=0.1$.
The value of its left minimum approaches that of the right minimum as $\sbar\to 0$. We could use a trial function with two maxima, its square would have turned into two delta functions each with prefactor 1/2 as $\sbar\to 0$. For finite $\sbar$ the single Gaussian picks out the right (lowest) minimum and would still approach the same limiting $\Eb/m$ of order $m^2$.

The leading $s$-dependence in $\mE_V$ is of order $s^{-2}$, and
\be
\frac{\mE_{\GCEI}^{\rm as}}{m} \simeq \frac{1}{36 m^4\sbar^2} - \frac{h_2(y_{\rm min})}{\sbar^2}\,,\quad
y_{\rm min} =
2.12\,, \quad h_2(y_{\rm min})=0.0949\,,
\, ,
\label{FGCEas}
\ee
which has no finite minimum in $\sbar >  0$ for large $m$ (in fact for $m>0.736$). This confirms the lack of ground state found for this model with the basis of sine functions.

The idea of setting  a maximum on the derivative of wave functions (realized at the end of section \ref{secbeI} by a minimal wavelength $\lmmin$) can be implemented also by a mass-independent width parameter $s$. Then $\sbar$ can diminish with increasing mass only slowly, $\sbar = s/\rs\simeq s/(3m)$, and the P and PSR models cannot be used for estimating the binding energy: the terms in the potential part of (\ref{FBWPSRsy}) are of order $m^3$ and $m^{7/2}$, implying that the square-root term inevitably overtakes the pole term (resulting in a negative binding energy). But all other terms in the expansion (\ref{zsqrbas}) would have become relatively large as well---the expansion would not converge. The CE-I model {\em can} be used because (\ref{FGCEIsy}) holds for all $\sbar$ such that the boundary condition for $f_G(r,a,s)$ at the origin is fulfilled to sufficient accuracy, which is typically the case for large masses.

The question arises wether the binding energy of model-I approaches that of the CE-I model under these circumstances. This was investigated as follows. A fit with a Gaussian to the ground-state wave function of model-I determined with the Fourier-since basis at $\lmmin=1$ and $m=3$ gives $s=0.1801$ ($\rs=9.49$, $\sbar=0.0190$). Using this $s$ in $\sbar=s/(3m)$, neglecting the kinetic energy contribution and evaluating (\ref{FGCEIsy}) at the minimum of $h_2(y)$  gives
\be
\Eb^{\rm as}/m
= c_{h1}\, m + c_{h2}\, m^2\,, \quad c_{h1} = 3.359\,,\quad c_{h2} = 26.34 \qquad \mbox{(CE-I model)}\,.
\label{ECEIPlanckas}
\ee
This estimate is shown in the left plot of figure \ref{figpevarI} by the black dashed line a little above the numerically evaluated  model-I curve of the variational Gaussian estimate. Numerical $\lmmin=1$ data for $\Eb/m^3$ at $m=10$, 11, \ldots 20 are well fitted by the rational function $(p_0 + p_1 m)/(1 + q_1 m)$. Its extrapolation $m\to\infty$ differs only 2\% from the $c_{h2}$ in (\ref{ECEIPlanckas}) of the CE-I model. An equally good looking fit with the constraint $p_1/q_1=c_{h2}$ is shown in the right plot of figure \ref{figFVCEI}. There is no reason to doubt that in case of a UV cutoff on the wave function the CE-I model gives the correct asymptotic mass dependence of model-I.\footnote{The Gaussian trial binding energy with $s=0.1801$ is higher than $\Eb$ obtained with the Fourier-basis.
% with $\lmmin=1$.
The reason may be the fact that the maximum derivative of the Gaussian, 7.5, is larger than the $2\pi$ of $\sin(2\pi r/\lmmin)$. }

\section{More on model-II}
\label{appII}

In model-II, the running potential has a minimum determined by
\be
V_{\rm r} = -z^2 r m^2,
\quad 0=\frac{\partial V_{\rm r}}{\partial r} =-z^2 m^2 +2 z \bt_z m^2.
\ee
The relevant solution is given by
\be
z_{\rm min} = -\frac{2b+\sqrt{4b^2 -2c}}{2c},
\ee
from which $\rmin$ and $V_{r\,{\rm min}}$ follow using (\ref{rtozII}), and also their asymptotic behavior in (\ref{rminVrminas}).
At large $m$ the potential approaches its classical-evolution form
\be
V_{\CEII}= -m^2r\, \frac{1}{(r+m)^2}
\ee
uniformly.
Matrix elements of the potential can be conveniently computed using the transformation of variables $r\to z$ as given by the solution in (\ref{rtozII}), with Jacobian
$\partial r/\partial z=-r/\bt_z$.

\section{Classical motion in the relativistic CE and Newton models}
\label{appbounceCEN}

The classical motion in the CE models is expected to approximate the motion of the wave packet at very large masses $m\gg 1$, with initial spread $s_0\ll m$ and initial distance $r_0 \gg m$.
Consider the particles released at rest form a large mutual distance $r=r_0$ with the dynamics specified by the Hamiltonians
\bea
H(r,p) &=&2\sqrt{m^2 + p^2} - \frac{m^2 r}{(r-3m)^2}\,,\qquad\qquad \mbox{CE-I model}
\\
H(r,p) &=&2\sqrt{m^2 + p^2} - \frac{m^2 r}{(r+m)^2}\,,\;\qquad\qquad \mbox{CE-II model}
\\
H(r,p) &=& 2\sqrt{m^2 + p^2} - \frac{m^2}{|r|}\,. \quad\qquad\qquad\qquad \mbox{Newton model}\\
\dot r &=& \frac{\partial H}{\partial p}\,,\quad\dot p = -\frac{\partial H}{\partial r}\,,
\label{clasNewton}
\eea
We start with model CE-I. Numerical integration rapidly shows that, starting from $r_0$,  $r\downarrow \rs= 3m$, $p\to-\infty$ and the velocity $\vrel\to -1$,
at  a time $\tc$.
%As mentioned earlier in section \ref{secevol} this happens also with Schwarzschild black holes and for this comparison the time-evolution stops at $\tc$.
%
Similarly, releasing the particles near the origin at a distance $r_0^\prime$ determined by energy conservation
\be
H(r_0^\prime,0) = H(r_0,0),
\ee
gives a motion $r\uparrow \rs$, $p\to +\infty$, $\vrel\to + 1$, in a time $\tc^\prime$. We can extend the first falling-in motion by gluing to it the time-reversed second motion and in this way continue it towards the origin, where it reaches $r_0^\prime$ and reverses (`bounces') back towards $\rs$. The motion can then be extended again by gluing a time-reversed version of the first motion, after which $r$ reaches $r_0$ again. The process can be repeated such that a non-linear oscillating motion emerges. The gluing implies that the limit points of $r(t)$ and $\dot r(t)=2\,\vrel(t)$ at the gluing times $t=\tc$, $\tc+2\tc^\prime$,  $2\tc+2\tc^\prime$, \ldots\, are added, which renders these functions continuous at these times. Figure \ref{figCEIbouncer} shows $r(t)$ over one period for $m = 2$ and $r_0=10\,\rs = 60$. Figure \ref{figCEIbouncev} shows the velocity. The gluing procedure has replaced the `black hole' interval $0\leq t<\tc$ into a black-hole--white-hole cyclic dependence on time.

\FIGURE[t]{
\includegraphics[width=7cm]{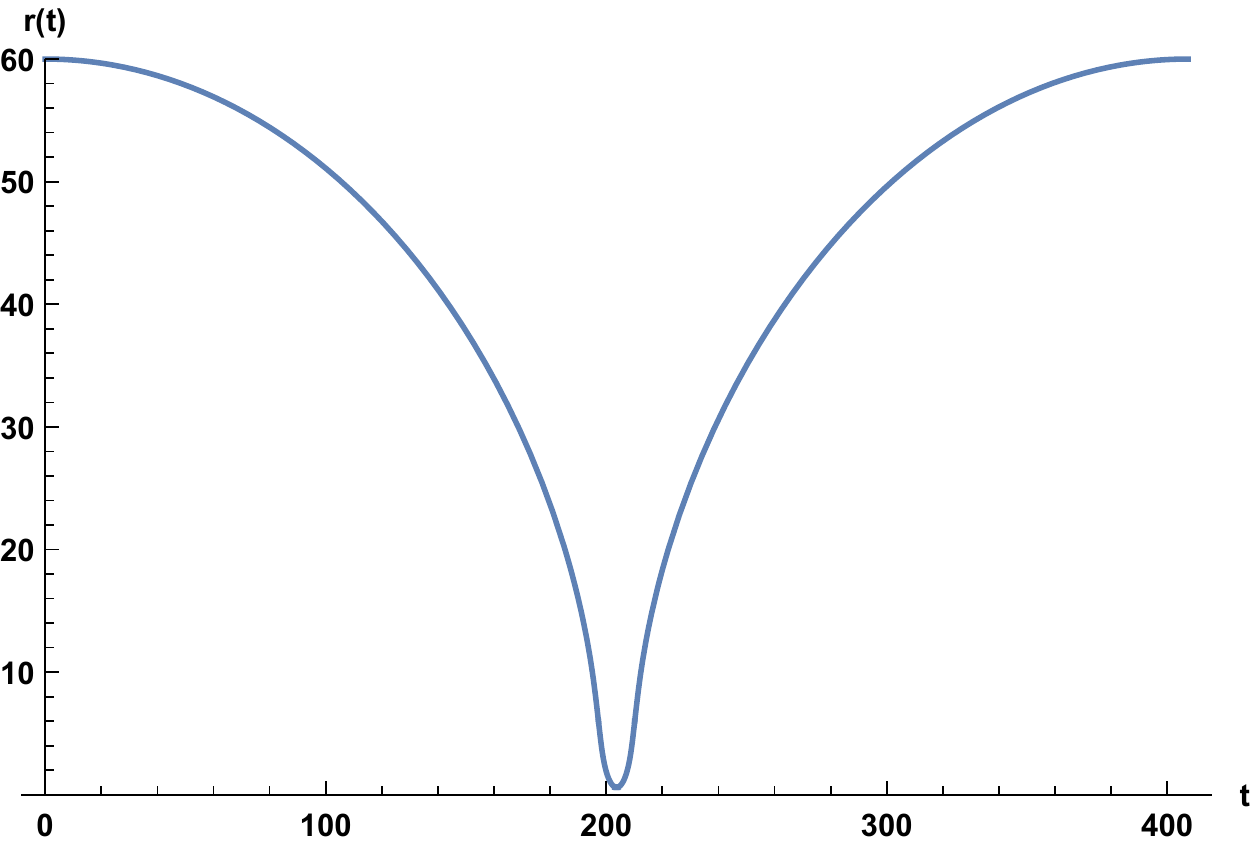} %see classical-I.nb
\includegraphics[width=7cm]{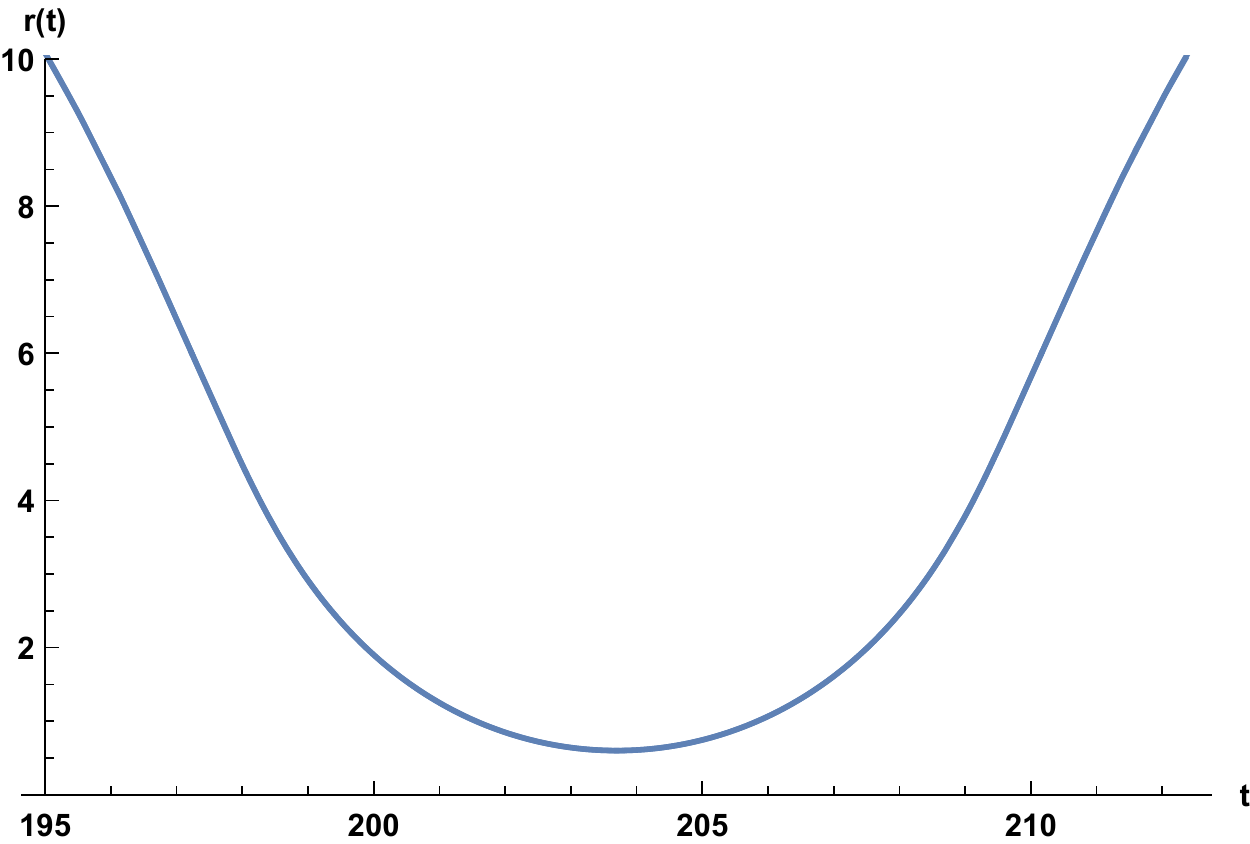} %see classical-I.nb
\caption{Relativistic CE-I model, $m=2$, $r_0=60$. Left: Once cycle of falling in and bouncing back; $\tc=197.2$, $\tc^\prime=6.5$.  The gluing times are $\tc$ and $\tc+2\tc^\prime$.  Right: close-up around $\tc+\tc^\prime=203.7$; the minimum at $t=\tc+\tc^\prime$ is $r_0^\prime=0.6$\,.
}
\label{figCEIbouncer}
}
\FIGURE[h]{
\includegraphics[width=7cm]{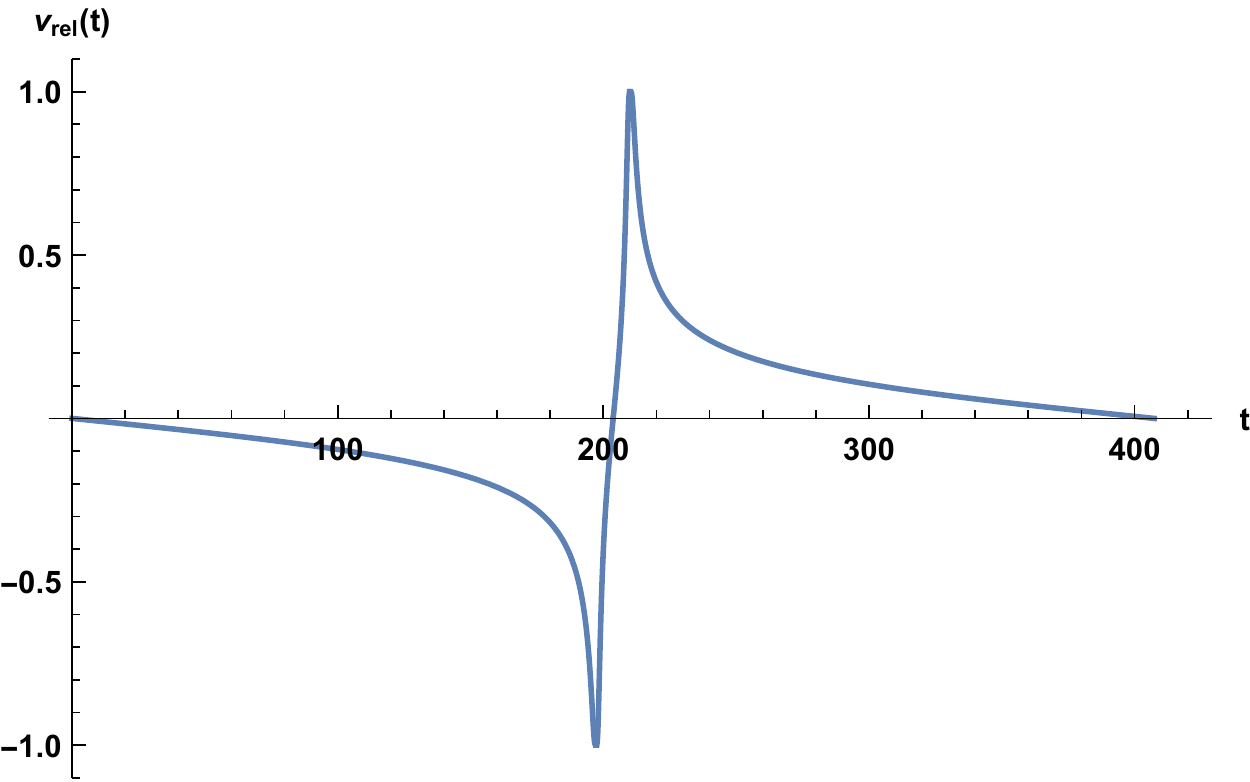} %see classical-I.nb
\includegraphics[width=7cm]{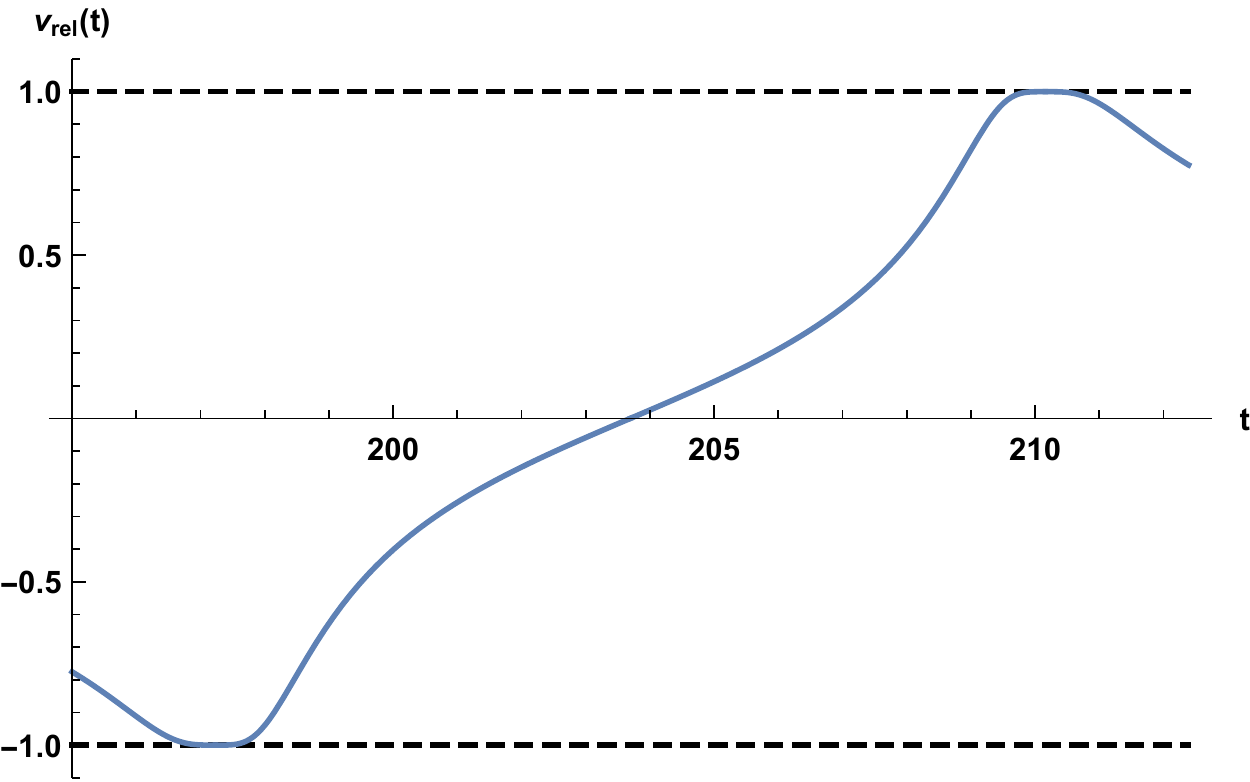} %see classical-I.nb
\caption{As in figure \ref{figCEIbouncer} for $\vrel(t)$.
}
\label{figCEIbouncev}
}

\FIGURE[h]{
\includegraphics[width=7cm]{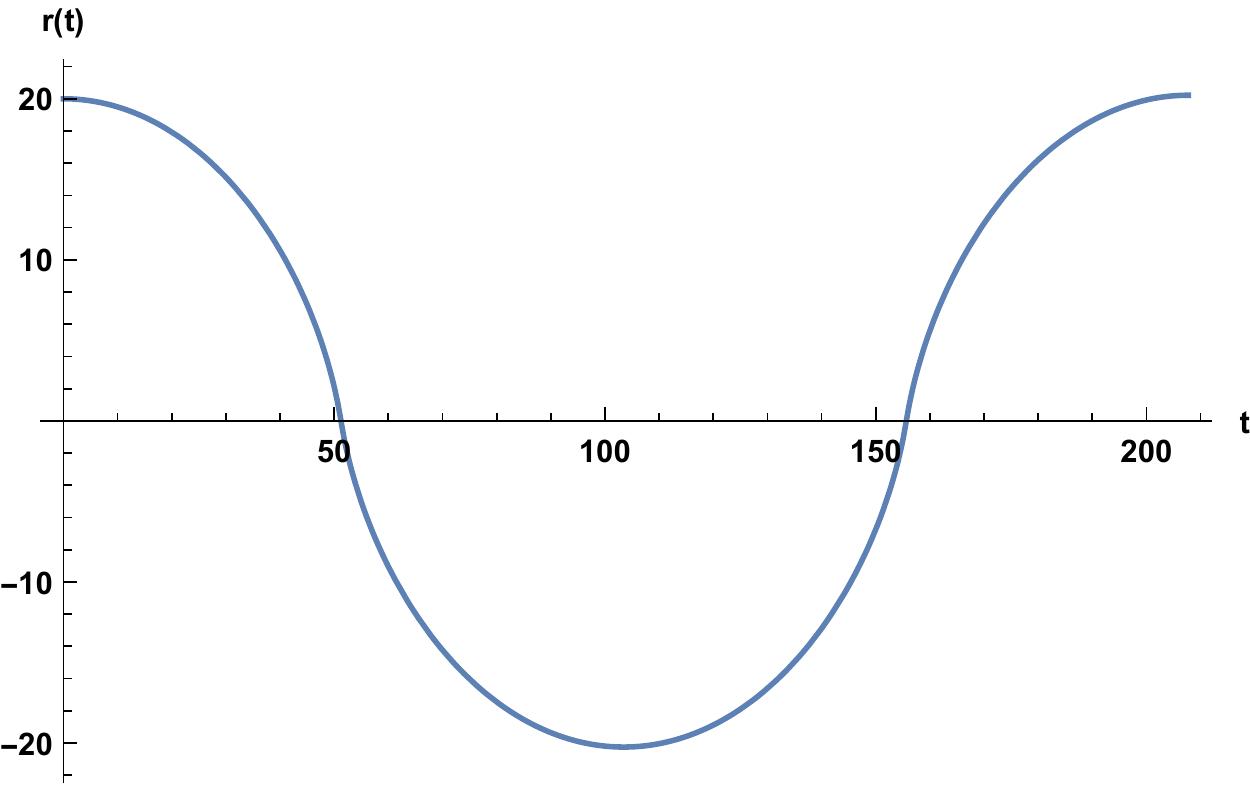}  %see classicalNewtonreg.nb
\includegraphics[width=7cm]{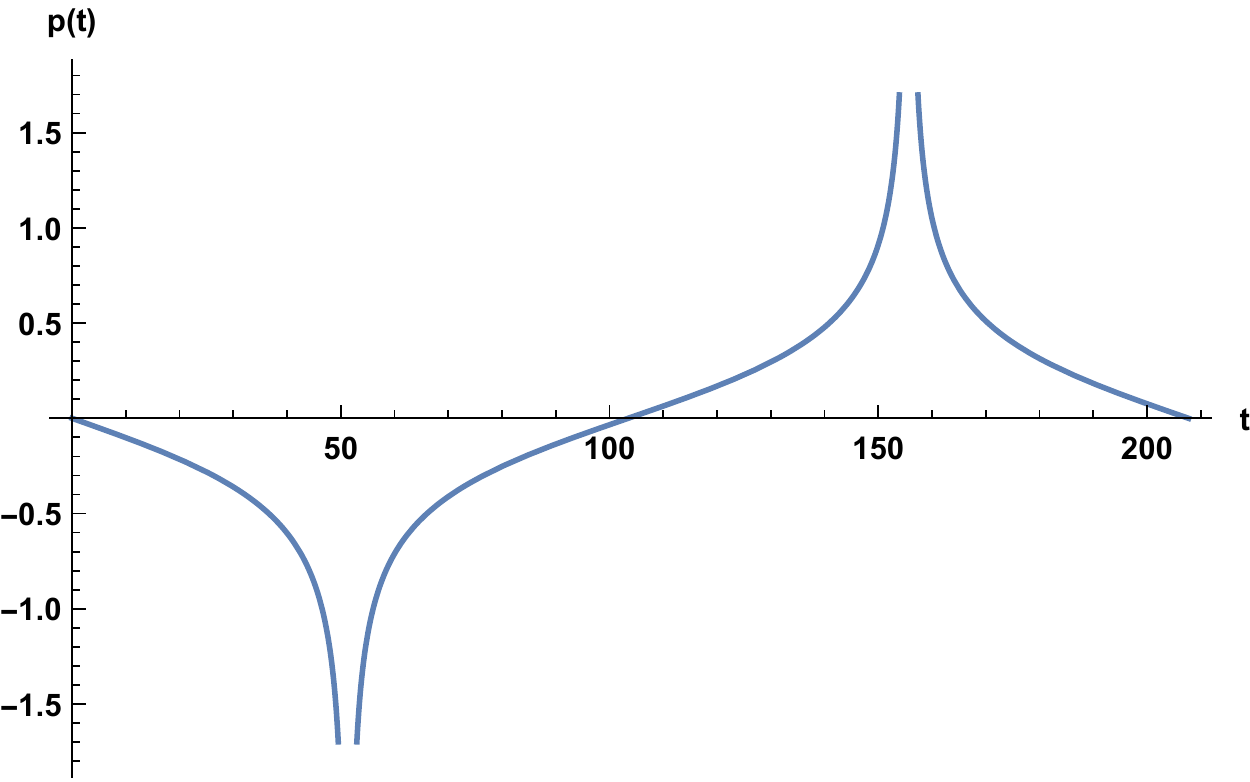}
\caption{Relativistic Newton model with {\em Cartesian} coordinates $r$ and $p$; $m=2$, $r_0=60$. Left: Once cycle of $r(t)$; the velocity $\vrel(t) = r'(t)/2= -1$, $+1$ at the gluing times $\tc=51.25$, $3\tc$.  Right: momentum $p$ diverging at gluing times.
}
\label{figrpclasNewton}
}

In the CE-II model the plots look similar, except that the maximum velocity does not reach $\pm 1$, even when falling in from infinity, and the flattening of $\vrel$ in the CE-I model at $\vrel=\pm 1$ is rounded off in the CE-II model. Using $2m=H(\infty,0)=H(r,p)$ gives $|p|$ as a function of $r$ which is maximal at $r=m$;
$|p_{\rm max}|= m \sqrt{17}/8$, $|v_{\rm rel, \, max}|= \sqrt{17}/9 \simeq 0.46$.

In the Newton model the potential is singular at the origin and this point is reached at a finite time $\tc$ with any initial distance $0<r_0<\infty$. The infinite force at the origin is attractive, not repelling as needed for a bounce, the evolution seemingly has to stop at $\tc$,
However, re-interpreting $r$ as a Cartesian coordinate that may become negative, as anticipated by the absolute value in the denominator in (\ref{clasNewton}), and assuming that the point particles may occupy the same point, the motion can be continued through the origin with continuous $r(t)$ and $\vrel(t)$, as shown in figure \ref{figrpclasNewton}.

\section{Perturbative mass renormalization}
\label{appmm0}

\FIGURE[h]{
\includegraphics[width=12cm]{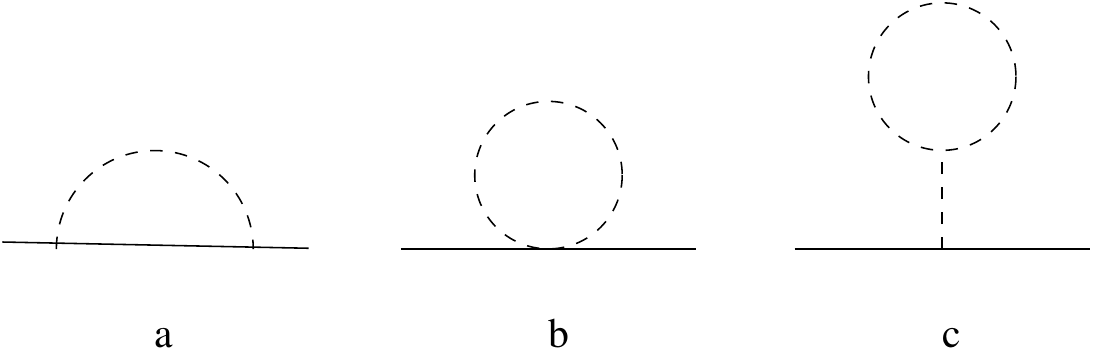}
\caption{Diagrams for the scalar selfenergy $\Sg_0$, the dashed lines represent gravitons. The graviton loop in the {\em tadpole diagram} $c$ is to be accompanied by a ghost loop (not shown).
}
\label{figselfenergydiagrams}
%see contSelfenergy2.nb for evaluation
}

In the perturbative vacuum $\langle g_{\mu\nu}\rangle=\et_{\mu\nu}={\rm diag}(-1,1,1,1)$, $\langle\ph\rangle=0$,    the renormalized selfenergy $\Sg(p^2)$ of the scalar field $\ph$, Wick-rotated to Euclidean momentum space, is related to the renormalized scalar-field propagator $G(p)$ by
\be
G(p)^{-1}= m^2 + p^2 + \Sg(p^2).
\label{Gp}
 \ee
Relevant one-loop scalar selfenergy diagrams are shown in figure \ref{figselfenergydiagrams} (a ghost loop should be added to the tadpole). A possible scalar field tadpole closed loop is left out since in the comparison with SDT we are interested in effects caused by the pure gravity model without `back reaction' of the scalar field---the quenched approximation. To 1-loop order pure-gravity can be renormalized in itself \cite{'tHooft:1974bx}, here we assume it to be done such that $\langle g_{\mu\nu}\rangle=\et_{\mu\nu}$. We use the graviton propagator and vertex functions in harmonic gauge given in \cite{Bjerrum-Bohr:2002kt} and dimensional regularization. The closed graviton loops in diagrams $b$ and $c$ (and its ghost companion) are often declared zero with dimensional regularization. However, for the tadpole diagram $c$ this leads to the ambiguous result $0/0$ (the zero in the denominator comes from the zero mass of the graviton in its propagator), which was analyzed in \cite{deWit:1977av,Antoniadis:1985ub,Johnston:1987jb,deWit:1991sc}. A graviton mass parameter $\lm$ regulates infrared divergencies. It induces a violation of gauge invariance that disappears in infrared-safe quantities where the limit $\lm\to 0$ can be taken.

The diagrams correspond to the unrenormalized selfenergy $\Sg_0$, which differs from $\Sg$ by the counterterms for mass renormalization and field rescaling, $\dl_m$ and $\dl_Z$,
\be
\Sg(p^2) = \Sg_0(p^2) + \dl_m + \dl_Z\, p^2,
\ee
and which are chosen such that the expansion around the zero of $\Sg(p^2)$ at $p^2=-m^2$ (`on shell') has the form
$\Sg(p^2) = 0 + \mathcal{O}((m^2 + p^2)^2)$\,.\footnote{Part of the notation here follows \cite{Peskin:1995ev}, section 10.2 .} This implies
\be
\Sg_0(-m^2) + \dl_m - \dl_Z m^2 =0, \qquad\Sg'_0(-m^2) + \dl_Z = 0.
\ee
The counterterms $\Delta S$ are introduced by rewriting the bare action $S_0$ in terms of the renormalized field and mass, $S_0 = S+\Delta S$,
$\ph_0 = \sqrt{Z}\,\ph =\sqrt{1+\dl_Z}\,\ph$,
$m_0^2 = Z^{-1}(m^2 + \dl_m)$. In one-loop order,
\be
m_0^2 = m^2 -\dl_Z m^2 + \dl_m = m^2 - \Sg_0(-m^2).
\ee
% In $d<2$ dimensional spacetime the diagrams are convergent and applying
Applying the usual techniques diagram $a$ can be worked into the form
\bea
\Sg_{0a} &=& -8\pi G\, \mu^{4-d}\int_0^1 dx\int \frac{d^d k}{(2\pi)^d}\,
\frac{a_1 k^2 p^2 + a_2 (p^2)^2(1-x)^2 + a_3(1-x)p^2 m^2 + a_4 m^4}
{[k^2 + x(1-x)p^2 + x m^2 + (1-x)\lm^2]^2},
\\
%\quad
a_1&=&2 +\frac{(4-d)(d-2)}{2d},
%\quad
a_2=2+\frac{(4-d)(d-2)}{2},
%\quad
%a_3=-\frac{d^2-4d+4}{4},
a_3=-(d^2-4d+4),
%\quad
%a_4=-\frac{d(d-2)}{8}\,.
a_4=-\frac{d(d-2)}{2}\,.
\nonumber
\eea
Here $\mu$ is the conventional mass parameter that keeps the dimension of $\Sg$ independent of spacetime dimension $d$.
Similarly, diagram $b$ (with symmetry factor 1/2) corresponds to
\bea
\Sg_{0b}&=&-8\pi G\,(b_1 p^2 + b_2 m^2)\, \mu^{4-d} \int \frac{d^d k}{(2\pi)^d}\,
\frac{1}{k^2 + \lm^2}\,,
\\
b_1&=&\frac{3d^2}{4} -\frac{5d}{2}+2,\;
b_2= \frac{3d^2}{4}-\frac{d}{2}.
\eea
The tadpole diagram comes out as
\be
\Sg_{0c} =- 8\pi G\,\left(\frac{2-d}{4}\right)\,\left(\frac{3 d^2}{4}-3d+1 \right)\,
(d m^2 + (d-1) p^2)\,\frac{1}{\lm^2}\,
 \mu^{4-d} \int \frac{d^d k}{(2\pi)^d}\,
\frac{k^2}{k^2 + \lm^2}\,,
\ee
where the factor $1/\lm^2$ comes from the graviton propagator attached to the tadpole tail. As $d\to 4$ the loop integral produces a factor $\lm^4$ and $\Sg_{0c}$ vanishes in the limit $\lm\to 0$. The same should happen in the ghost tadpole, since it cancels un-physical contributions in the graviton loop.
Near $d=4$, $\Sg_{0b}$ is proportional to $\lm^2$, including the residue of the pole at $d=4$, hence also $\Sg_{0b}$ vanishes as $\lm\to 0$.

For generic $p^2$, the limit $\lm\to 0$ of $\Sg_{0a}$ is not zero near $d=4$, and furthermore, the residue of the pole at $d=4$ is proportional to $(m^2 + p^2)$ and vanishes on-shell: $\Sg_{0a}(-m^2)$ is finite. Hence
$\Sg_0(-m^2) = \Sg_{0 a}(-m^2)$; we find
\bea
\Sg_0(-m^2) &=& -\frac{5}{2\pi}\, G\, m^4\,,
\\
m_0^2 &=& m^2 + \frac{5}{2\pi}\, G m^4\,.
\label{mm0cont}
\eea
The cancelation of poles at $d=4$ was noted earlier in \cite{Rodigast:2009zj}, where it was found to occur also in Yukawa models of fermions and scalars coupled to gravity. The question arose if these cancelations occurred only in the harmonic gauge. Gauge independence has been contested in \cite{Mackay:2009cf} where a dependence was found on a gauge parameter $\om$.
Finiteness was mentioned for the harmonic gauge in which $\om=0$ and a second gauge parameter $\al=1$. But one can see from the results in this work that for $\om=0$ the cancelation is in fact in-dependent of $\al$. A similar phenomenon occurs in the work \cite{Capper:1979ej}.
Hence, the on-shell relation (\ref{mm0cont}) is gauge-independent in a restricted class of gauges.
However, in the way we have defined $m^2$ it is the position of the pole as a function of $p^2$ in the renormalized Green function. Such a `pole mass' is a physical gauge-invariant quantity that also describes the position of poles in analytically continued S-matrix elements. Since $m_0$ and Newton's coupling $G$ are gauge invariant, (\ref{mm0cont}) is a gauge-invariant (but regularization-dependent) relation.
%\footnote{\P The definition of $G$ may depend on a particular coordinate system, from which $G$ in other systems should be obtained by a coordinate transformation.}

The derivative $\Sg'_0(p^2)$ is UV-divergent at $d=4$ and it has an IR-divergence on shell as $\lm\to 0$,
\be
\Sg'_0(-m^2) = \frac{8\pi G m^2}{16\pi^2}\left(\frac{8}{4-d} - 4\gm_{\rm E} +4\ln(4\pi)
-\ln\frac{\lm^2}{\mu^2} - 3 \ln\frac{m^2}{\mu^2}\right) + \mathcal{O}((4-d)^2,\lm^2).
\ee
It will not be gauge independent. This IR-divergence is to be resolved similar to the case of QED.

\bibliography{lit}
\end{document}